\documentclass[runningheads]{llncs}
\usepackage[T1]{fontenc}

\renewcommand{\orcidID}[1]{\href{http://orcid.org/#1}{\protect\raisebox{-1.25pt}{\protect\includegraphics{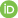}}}}

\usepackage{graphicx}
\usepackage[bookmarks,unicode,pdfusetitle]{hyperref}
\hypersetup{
	colorlinks=true,
	citecolor=blue,
	linkcolor=blue,
	urlcolor=blue,
}
\usepackage{cite}
\usepackage{amsmath,amssymb}
\usepackage{mathtools}
\usepackage{stmaryrd}
\usepackage{listings}
\mathtoolsset{showonlyrefs=true}
\usepackage{tikz,pgfplots}
\usetikzlibrary{cd}
\tikzset{%
	symbol/.style={%
		draw=none,
		every to/.append style={%
			edge node={node [sloped, allow upside down, auto=false]{$#1$}}}
	}
}
\usepackage{mathpartir}
\usepackage{enumitem}
\usepackage{subcaption}
\usepackage{wrapfig}
\usepackage{ifthen}

\newboolean{puplicversion}
\setboolean{puplicversion}{true}
\newboolean{longversion}
\setboolean{longversion}{true}

\ifthenelse{\boolean{puplicversion}}{
\newcommand{\comment}[1]{}

\newcommand{\checkpagelimit}[1]{\ifthenelse{\value{page} > #1}{\errmessage{Error: The main body exceeds the page limit of #1 pages.}}{}}
}{
\newcommand{\comment}[1]{\textcolor{blue}{[Note: #1]}}

\newcommand{\checkpagelimit}[1]{}
}
\renewcommand{\comment}[1]{}

\ifthenelse{\boolean{longversion}}{
\usepackage[bibliography=common]{apxproof}
}{
\usepackage[bibliography=common,appendix=strip]{apxproof}
}

\usepackage[arxiv=arxiv]{macros.apxproof}

\spnewtheorem{mytheorem}{Theorem}{\bfseries}{\itshape}

\spnewtheorem{mylemma}[mytheorem]{Lemma}{\bfseries}{\itshape}

\spnewtheorem{myproposition}[mytheorem]{Proposition}{\bfseries}{\itshape}

\spnewtheorem{mysublemma}[mytheorem]{Sublemma}{\bfseries}{\itshape}
\spnewtheorem{mycorollary}[mytheorem]{Corollary}{\bfseries}{\itshape}

\spnewtheorem{myfact}[mytheorem]{Fact}{\bfseries}{\itshape}
\spnewtheorem{mynotation}[mytheorem]{Notation}{\bfseries}{\rmfamily}
\spnewtheorem{myremark}[mytheorem]{Remark}{\bfseries}{\rmfamily}

\spnewtheorem{myexample}[mytheorem]{Example}{\bfseries}{\rmfamily}

\spnewtheorem{myassumption}[mytheorem]{Assumption}{\bfseries}{\rmfamily}
\spnewtheorem{mydefinition}[mytheorem]{Definition}{\bfseries}{\rmfamily}

\spnewtheorem{myrequirements}[mytheorem]{Requirements}{\bfseries}{\rmfamily}
\spnewtheorem{myproblem}[mytheorem]{Problem}{\bfseries}{\rmfamily}
\spnewtheorem{myconjecture}[mytheorem]{Conjecture}{\bfseries}{\rmfamily}

\usepackage{macros}

\newcommand{\PartialCorrectnessMonad}{\mathbf{Par}}
\newcommand{\CostGradedMonad}{\mathbf{C}}
\newcommand{\ProbMonad}{\mathbf{D}}

\newcommand{\UnionBoundGradedMonad}{\mathbf{UB}}
\newcommand{\SimpleFibredMonad}[1]{s(#1)}
\newcommand{\ExpectationGradedMonad}{\mathbf{EX}}
\newcommand{\WriterLiftingMonad}{\mathbf{Wr}^{\bot}}
\newcommand{\TemporalSafetyGradedMonad}{\mathbf{TS}}
\newcommand{\DependentEffectConstruction}[1]{\mathbf{DE}(#1)}

\newcommand{\GradingMonoid}{\mathcal{M}}
\newcommand{\GradeUnit}{1}
\newcommand{\GradeMult}{\mathbin{\cdot}}

\newcommand{\PowerSet}[1]{2^{#1}}

\newcommand{\BorelClosed}[1]{\mathcal{B}_{\mathrm{cl}}(#1)}

\newcommand{\CategoryOfPreorderedMonoids}{\mathbf{PreMon}}

\title{A Category-Theoretic Framework for Dependent Effect Systems}

\author{Satoshi Kura\texorpdfstring{\inst{1}\orcidID{0000-0002-3954-8255}}{}%
\and Marco Gaboardi\texorpdfstring{\inst{2}\orcidID{0000-0002-5235-7066}}{}%
\and Taro Sekiyama\texorpdfstring{\inst{3,4}\orcidID{0000-0001-9286-230X}}{}%
\and Hiroshi Unno\texorpdfstring{\inst{5}\orcidID{0000-0002-4225-8195}}{}}

\institute{Waseda University, Tokyo, Japan
\and Boston University, Boston, Massachusetts, USA
\and National Institute of Informatics, Tokyo, Japan
\and SOKENDAI, Hayama, Japan
\and Tohoku University, Sendai, Japan}

\begin{document}

\maketitle

\begin{abstract}
  \emph{Graded monads} refine traditional monads using \emph{effect}
  annotations in order to describe quantitatively the computational
  effects that a program can generate.
  They have been successfully applied to a variety of formal systems for reasoning about effectful computations.
  However, existing categorical frameworks for graded monads do not support effects that may depend on program values, which we call \emph{dependent effects}, thereby limiting their expressiveness.
  We address this limitation by introducing \emph{indexed graded monads}, a categorical generalization of graded monads inspired by the
  fibrational ``indexed'' view and by classical categorical
  semantics of dependent type theories. We show how indexed graded monads provide semantics for a refinement type system with dependent effects.
  We also show how this type system can be instantiated with specific choices of parameters to obtain several formal systems for reasoning about specific program properties.
  These instances include, in particular, cost analysis, probability-bound reasoning, expectation-bound reasoning, and temporal safety verification.

\end{abstract}

\section{Introduction}

Refinement types refine the class of values inhabiting types by means
of logical predicates. By using these predicates to quantify over
programs' input and output, users can specify and check properties
that their programs satisfy, combining in this way programming with
program reasoning. Refinement types have found application in a
variety of areas including program verification~\cite{VazouSJ14,SwamyHKRDFBFSKZ16,UnnoPOPL2018,MukaiKS22}, cost
analysis~\cite{GrobauerICFP2001,DanielssonPOPL2008,WangOOPSLA2017,RadicekPOPL2018,HandleyVH20}, probabilistic computations~\cite{KuraICFP2024}, incremental computation~\cite{CicekGA15},
security and privacy~\cite{BengtsonBFGM08,BarthePOPL2015,LiuKLTWSPMDF24}, etc.
In several of these applications domains, computations are associated
with some notion of \emph{computational effects}, which are often
quantitative. For example in cost analysis one usually represents the
computational cost, in terms of time, memory space, or other resources,
as an effect produced by the program while computing the result. In
differential privacy, random noise is used as a way to protect an
individual's data. Aspects of the randomness can be modelled as a
quantitative effect: one can represent the privacy loss as an effect
that a program produces while computing the result, or one can
represent the error incurred because of the noise as an effect
produced during the private computation. As a result, most of the
works from the application areas we mentioned before use some kind of
effect tracking at the type system level, usually either in the form
of a type and effect system~\cite{LucassenPOPL1988,MarioTLDI2009} or in the form of an effect
annotated monad~\cite{Atkey2009,Tate2013,KatsumataPOPL2014,DBLP:journals/corr/HicksBGLS14}.

A unifying theory that has emerged in the last decade to reason about
quantitative effects is the one based on \emph{graded
monads}~\cite{SmirnovJMathSci2008,KatsumataPOPL2014}. Graded monads combines the idea of reasoning
explicitly about effects, as in type and effect systems, with the idea
of modelling computational effects using monads. Graded monads refine
monads using effect annotations which typically are elements of an
effect language which forms a preordered monoid. Similarly to ordinary
monads, graded monads have strong semantical foundations in category
theory. Intuitively, a monad is an endofunctor (on some specific
category) satisfying certain compositionality principles, while a
graded monad is a functor from the effect language (seen as a
category) to the category of endofunctors (on some specific category)
satisfying analogous compositionality laws that additionally track the quantitative effect annotations.
Several works have explored the use of graded monads for reasoning about effects,
e.g.~\cite{KatsumataPOPL2014,OrchardP14,BarthePOPL2015,FujiiKM16,GaboardiKOBU16,OrchardICFP2019,IvaskovicMO20,MoonEO21,GaboardiESOP2021,AguirreICFP2021,RajaniG0021,KatsumataMUW22,AbuahDN22,BreuvartMU22,KellisonH24,Lobo-VesgaOOPSLA2024,Liell-CockPOPL2025,SannierPOPL2026},
and some of them have focused on their semantics
foundation~\cite{KatsumataPOPL2014,FujiiKM16,GaboardiKOBU16,KatsumataMUW22,BreuvartMU22}.

Combining two different ways of refinement, namely refinement types and graded monads, is a natural step towards more expressive systems for reasoning about effectful computations, as explored in several works~\cite{DanielssonPOPL2008,BarthePOPL2015,RajaniG0021,AguirreICFP2021,Lobo-VesgaOOPSLA2024}.
Most of these works have focused on the practical aspects of this combination.
In particular, in order to have precise quantitative information about
effects, it is crucial in these systems to allow effects to depend on
some of the computation's inputs. For example, in cost analysis it is
natural to express the cost of a computation in terms of the size or
value of the input data. So, a type system for cost analysis has to be
able to express this dependency. Indeed, most of the works combining
refinement types and graded monads support this form of
dependency. This is usually achieved by allowing the grade to depend
on the (refined) typing environment.  However, this is not the case
for the semantics models of graded monad studied so far. In all these
models, grades are defined independently of the typing environment and
they cannot express dependences on the input data. One can recover a
limited form of dependency at the meta-level, as done for example
in~\cite{GaboardiESOP2021,AguirreICFP2021}, but one cannot use these models to interpret effects which
can truly depend on the program input as the one used for example in~\cite{BarthePOPL2015,RajaniG0021}.

To overcome this situation, in this paper we focus on the semantics of
effects which may depend on the program's input. We call this class of
effects \emph{dependent effects}. We consider dependent effects in
combination with refinement types and we study their categorical
semantics. Traditionally, categorical models of refinement types which
allow dependencies on the context (similar to models for dependent
types) are based on some form of indexed category. The indexing allows
one to express naturally the different dependencies on context that
different program components may have. While grading may also be seen
as a form of indexing of monads, reason why graded monads are
usually considered refinements of monads,
it is actually of a
semantically different nature from the indexing used to capture the
dependency on the context. In order to combine these two forms of
indexing in a principled way, we define a new notion of \emph{indexed
graded monad} where also effects in the effect language are indexed.
Using this notion we build an interpretation for a refinement type
system with dependent effects. Soundness in systems with graded monads
usually guarantees that the grade is a quantitative bound on the
runtime effect of the program. In our model, as we wanted, soundness
for indexed graded monads extends these quantitative guarantees also
to situations where the runtime effect of the program depends on some
input value. In order to help with the design of models for indexed
graded monads we provide a construction which build a ``naturally
chosen'' indexed graded monad out of a traditional graded monad. This
construction can be used to build a model of dependent effects out of
a model for simple effects. We prove the usefulness of indexed graded
monads and of our constructions by showing that several existing
approaches based on graded monad can be obtained by properly
instantiating our framework. These include cost analysis~\cite{HoffmannPOPL2017}, union bound
reasoning~\cite{BartheICALP2016}, expectation-bound reasoning~\cite{AguirreICFP2021}, and temporal safety verification~\cite{NanjoLICS2018}.
In summary, our contributions include:
\begin{itemize}
	\item We consider a generic refinement type system with dependent effects modelled by indexed graded
	monads. Different dependent effect systems can be obtained by
	instantiating it with different axioms.
        \item We provide categorical semantics for our dependent effect system. Specifically, we
	formalize the notion of indexed preordered monoids and the
	notion of indexed graded monads, which are the indexed versions
	of the usual ones.
	We prove soundness of our
	type system with respect to this semantics.
	\item We provide a construction that builds canonical models
	of indexed graded monads out of models of traditional graded
	monads. This allows us to give a semantics to dependent
	effects starting from the semantics of simple effects.
        \item We instantiate our framework to capture several examples
        of dependent effects that have been studied in the
        literature. 
\end{itemize}

\paragraph{Related work}
Value-dependent systems for reasoning about certain specific effectful behavior
have been broadly studied for a variety of problems.
Cost analysis (especially, in terms of time) is one of the actively studied
problems, and many value-dependent type and effect systems for bound or
amortized analysis have been
proposed~\cite{CicekPOPL2017,RadicekPOPL2018,HandleyVH20,KnothICFP2020,NiuPOPL2022,NiuPOPL2024}.
Cost analysis can be generalized to linear-time temporal safety verification,
which aims to reason about sequences of events, by encoding a cost as a finite
number of events to be raised.
For linear-time temporal verification, value-dependent effect systems for
temporal verification, called \emph{temporal effect systems}, have been
studied~\cite{NanjoLICS2018,SekiyamaPOPL2023}.
As a further generalization, Gordon~\cite{GordonTOPLAS2021} studied \emph{effect
quantales}, which are an algebraic structure behind effects sensitive to their
order, and gave a general value-dependent effect system parameterized by effect
quantales.
Gordon~\cite{GordonTOPLAS2021} showed that the proposed effect system can be
instantiated to support linear-time temporal safety verification.
However, it remains unclear whether effect quantales are expressive enough to
address probabilistic reasoning.
In contrast, our dependent effect system can accommodate probabilistic reasoning
like the union bound logic, thanks to its semantic foundations based on the
category theory.
Another generic approach to dependent effect systems is the use of Dijkstra
monads~\cite{SwamyPLDI2013}, which enable effective reasoning about effectful
higher-order programs through weakest pre-conditions.
The reasoning based on Dijkstra monads can address arbitrary monadic effects,
but it is left open whether they can also be used for reasoning about
probabilistic programs~\cite{MaillardICFP2019}.

\section{Informal Overview}\label{sec:overview}

In this section we will informally introduce our semantic framework. We
will use cost analysis as a running example, and defer other
example instances of our framework to Section~\ref{sec:instances}.

\subsection{Graded Monads}
Cost analysis aims at providing upper bounds on the running time, or
any other computational resource, of programs. A standard approach to
cost analysis is based on the use of a cost
monad~\cite{DanielssonPOPL2008}. This is a monad annotated with a
natural number expressing an upper bound on the number of \verb|Tick|
(representing one unit of cost) the program execution encounters.

The cost monad can be seen as an instance of a \emph{graded monad}, a
concept which emerged in the context of graded
algebra~\cite{SmirnovJMathSci2008}, and which was related to
computational effects in~\cite{KatsumataPOPL2014}.
Given a preordered monoid $\GradingMonoid$, an $\GradingMonoid$-graded monad
$(T, \eta, \mu)$ consists of a functor $T : \GradingMonoid \times \category{C} \to
\category{C}$ (we often write the first argument $m \in \GradingMonoid$ of $T$ as a subscript, like $T_m$), a unit $\eta_X : X \to T_\GradeUnit X$,
where $\GradeUnit$ is the unit of the monoid $\GradingMonoid$, and a
multiplication $\mu_{m_1, m_2, X} : T_{m_1} T_{m_2} X \to
T_{m_1 \GradeMult m_2} X$, where $m_1$ and $m_2$ are elements of
$\GradingMonoid$ and $\GradeMult$ is the multiplication of the monoid, satisfying the graded version of the monad laws.
When using graded monads to reason about effects, we think about the
elements $m\in\GradingMonoid$ as
\emph{effects}, and we use them to \emph{grade} the functor $T$. A graded
monad $T_m A$, for some type $A$, can be seen as a \emph{quantitative refinement} of $T A$
in the sense that it is inhabited by those computations in $T A$ that generate effects upper bounded by $m$.
Here, the preorder on $\GradingMonoid$ expresses the subsumption relation: if $m_1 \le m_2$, then $T_{m_1} A$ is subsumed by $T_{m_2} A$.
In programs, graded monads can be
manipulated using the traditional monadic return and let binding typed with the
following graded rules.
\begin{mathpar}
	\inferrule{
		\Gamma \vdash V : A
	}{
		\Gamma \vdash \return{V} : T_{\GradeUnit} A
	}
	\and
	\inferrule{
		\Gamma \vdash M : T_{m_1} A \\
		\Gamma, x : A \vdash N : T_{m_2} B
	}{
		\Gamma \vdash \letin{x}{M}{N} : T_{m_1 \GradeMult m_2} B
	}
\end{mathpar}
To capture cost analysis we can instantiate the framework of graded
monads using the additive monoid $(\mathbb{N}_{\infty}, 0, {+}, {\le})$ of
extended natural numbers and adding the
primitive $\mathtt{Tick} :T_1\, \UnitType$ which accounts for one unit of
cost (notice that the unit of the monoid used in the type of return is
now $0$).
The semantic soundness of the graded monad
framework~\cite{KatsumataPOPL2014} will then guarantee that the type
$T_m A$ can be assigned only to computations that generate the number of
ticks (i.e., the cost) upper bounded by $m$.
It is worth stressing that it is the semantic soundness of the graded
monad framework that justifies the ``quantitative refinement'' view we
informally discussed above.

\subsection{Motivation for Dependent Effects}
One of the limitations of existing frameworks for graded
monads is that they cannot express effects that depend on program
values, and so they are limited in their foundational scope. To
explain this, let us consider the following example.
\[ \mathtt{loop\_cost} \quad\coloneqq\quad \recfun{f}{x}{\ifexpr{\ x < n\ }{\ \mathtt{Tick}\, ;\ f(x + 1)\ }{\ \return{()}}} \]
The function $\mathtt{loop\_cost}$ with simple type $\IntType \to \UnitType$ takes an
integer $x$ and increments it by $1$ until it reaches $n$.  It is easy
to see that the cost of this function is $n - x$ if $x < n$ and $0$ otherwise.
A natural way to express the cost of this function
is to allow effects to depend on values in the program.  We call
such effects \emph{dependent effects}. If we allow dependent effects, the natural typing judgment for the function is
\[ n : \IntType \quad\vdash\quad \mathtt{loop\_cost} \quad:\quad (x : \IntType) \to T_{\max\{ 0, n - x \}} \UnitType \]
Here, $\max\{ 0, n - x \}$ is a dependent effect, which depends on $n
: \IntType$ and $x : \IntType$.

Most of existing framework for graded monads cannot express dependent effects because effects are restricted to constant elements of a monoid, which we call \emph{simple effects}.
Simple effects limit the expressiveness of these frameworks.
If we reason about the example above with simple effects, then a constant $m \in \mathbb{N}_{\infty}$ must be chosen so that $m$ uniformly bounds the cost of the function $\mathtt{loop\_cost}$ for all $n : \IntType$ and $x : \IntType$.
In this case, the only possible choice of $m$ is the trivial upper bound $m = \infty$. So, the best type those framework can give to the function $\mathtt{loop\_cost}$ is the following:
\[ n : \IntType \quad\vdash\quad \mathtt{loop\_cost} \quad:\quad \IntType \to T_{\infty} \UnitType \]

This is in sharp contrast with current practice, where systems combining specific instances of graded monads actually also support dependent effects. For example, the cost analyses by~\cite{DanielssonPOPL2008,RajaniG0021,HandleyVH20}, all support dependent effects. Similarly, the privacy analyses by \cite{BarthePOPL2015,AbuahDN22,Lobo-VesgaOOPSLA2024}, the incremental computation analysis by \cite{CicekGA15}, the temporal verification by~\cite{NanjoLICS2018} all support dependent effects.
Unfortunately, as far as we know, there is no existing semantic framework based on graded monads that can express these dependent effects.

\subsection{Our Idea: Indexed Graded Monads}

To address the limitation we outlined above, we extend the definition
of graded monads to allow dependencies on program values.  In order to
do this, we take inspiration from (1) the categorical semantics of
dependent type theory and (2) the 2-category theory point of view.

In dependent type theory, types $\Gamma \vdash A$ are indexed by
contexts $\Gamma$.  Thus, a model of dependent type theory is given by
an indexed category $\{ \category{E}_{\interpret{\Gamma}} \}_{\Gamma
: \text{ctx}}$ indexed by contexts $\Gamma$
(Fig.~\ref{fig:models-comparison}, top row).  More precisely, if
$\category{B}$ is a category for interpreting contexts $\Gamma$, then
an indexed category $\category{E}_{(-)}$ is a contravariant functor
$\category{B}^{\op} \to \mathbf{Cat}$, which is known to be equivalent
to a Grothendieck fibration $\category{C} \to \category{B}$.
Below, we keep writing an indexed category $\category{E}_{(-)}$ as $\{ \category{E}_{I} \}_{I \in \category{B}}$.

\begin{figure}[tb]
	\centering
	\setlength{\tabcolsep}{5pt}
	\begin{tabular}{c|cc}
		& Simple effect system & Dependent effect system \\
		\hline
		\begin{tabular}{l} Type $\Gamma \vdash A$ \\ Term $\Gamma \vdash V : A$ \end{tabular} & \begin{tabular}{c}
			category $\category{C}$ \\ $\interpret{A}$: object in $\category{C}$ \\ $\interpret{V}$: morphism in $\category{C}$
		\end{tabular} & \begin{tabular}{c} indexed category $\{ \category{E}_I \}_{I \in \category{B}}$ \\ $\interpret{A}$: object in $\category{E}_{\interpret{\Gamma}}$ \\ $\interpret{V}$: morphism in $\category{E}_{\interpret{\Gamma}}$ \end{tabular} \\
		\hline
		Effect $\Gamma \vdash \mathcal{E}$ & \begin{tabular}{c} monoid $\GradingMonoid$ \\ $\interpret{\mathcal{E}} \in \GradingMonoid$ \end{tabular} & \begin{tabular}{c} indexed monoid $\{ \category{M}_I \}_{I \in \category{B}}$ \\ $\interpret{\mathcal{E}} \in \category{M}_{\interpret{\Gamma}}$ \end{tabular} \\
		\hline
		Computational effect & \begin{tabular}{c} graded monad \\ $T : \GradingMonoid \times \category{C} \to \category{C}$ \end{tabular} & \begin{tabular}{c} indexed graded monad \\ $T = \{ T_I : \category{M}_I \times \category{E}_I \to \category{E}_I \}_{I \in \category{B}}$ \end{tabular} \\
	\end{tabular}
	\caption{Models of simple and dependent effect systems.}
	\label{fig:models-comparison}
\end{figure}

If we take a closer look at the cost analysis example above, we can
see that dependent effects also have a similar ``indexed'' structure.
In the example, we have a dependent effect $n : \IntType, x
: \IntType \vdash \max \{ 0, n - x \} : \mathbf{Effect}$ (using a
well-formedness judgment for effects, formally introduced in
Section~\ref{sec:type-system}), for which the ``index'' is given by
the context $n : \IntType, x : \IntType$.
Hence, we can interpret this dependent effect as a function
$\interpret{\max \{ 0, n -
x \}} \in \Set(\mathbb{Z}^2, \mathbb{N}_{\infty})$, where
$\Set(\mathbb{Z}^2, \mathbb{N}_{\infty})$ is the set of functions
from $\mathbb{Z}^2$ to the preordered monoid $\mathbb{N}_{\infty}$.
Following this intuition, to interpret dependent
effects over a preordered monoid $\GradingMonoid$, we can consider families
of functions $\{ \Set(I, \GradingMonoid) \}_{I \in \Set}$. Since $\GradingMonoid$ is a monoid, also each
set $\Set(I, \GradingMonoid)$ is a monoid, and hence it can be used
as a ``grade'' in the traditional sense. 
More generally, we have that $\Set({-}, \GradingMonoid)$ is a contravariant functor
$\Set^{\op} \to \CategoryOfPreorderedMonoids$
from the category $\Set$ of sets to the category $\CategoryOfPreorderedMonoids$ of preordered monoids.
So, generalizing one step further we consider functors  $\category{B}^{\op} \to \CategoryOfPreorderedMonoids$ and we call them \emph{indexed preordered monoid} (Fig.~\ref{fig:models-comparison}, middle row).

Following the same intuition we define \emph{indexed graded monads} as families
of graded monads $\{ T_{I} : \category{M}_I \times \category{E}_I \to \category{E}_I \}_{I \in \category{B}}$ whose
grading monoids are given by an indexed monoid
$\{ \category{M}_I \}_{I \in \category{B}}$ (Fig.~\ref{fig:models-comparison}, bottom row).  More formally, an
indexed graded monad is a contravariant functor from a category
$\category{B}$ to the category of graded monads. We can now use indexed graded monads to give semantics to dependent effect systems.

\begin{example}[Indexed graded cost monad on the family fibration]\label{ex:indexed-graded-cost-monad-on-fam}
	Consider the family fibration $\mathrm{fam}_{\Set} : \mathbf{Fam}(\Set) \to \Set$, which is a typical model of dependent type theory.
	This corresponds to the indexed category given by a family of $I$-indexed sets: $\{ \mathbf{Fam}(\Set)_I \}_{I \in \Set} = \{ \{ X_i \}_{i \in I} \}_{I \in \Set}$.
	For example, the type $n : \NatType \vdash \mathtt{Vec}\ n$ of integer vectors of length $n$ is interpreted as $\{ \mathbb{Z}^n \}_{n \in \mathbb{N}}$ in this setting.
	Now, an indexed graded monad $\{ T_I : \Set(I, \mathbb{N}_{\infty}) \times \mathbf{Fam}(\Set)_I \to \mathbf{Fam}(\Set)_I \}_{I \in \Set}$ for cost analysis is defined as follows.
	\begin{equation}
		(T_I)_f \{ X_i \}_{i \in I} \quad\coloneqq\quad \{\ X_i \times \{\ m \in \mathbb{N}_{\infty} \mid m \le f(i)\ \} \ \}_{i \in I}
		\label{eq:indexed-graded-cost-monad}
	\end{equation}
	This is an indexed graded version of the cost monad $({-}) \times \mathbb{N}_{\infty}$, which captures the situation where the cost is upper bounded by a dependent effect $f : I \to \mathbb{N}_{\infty}$.
	\qed
\end{example}

The correspondence shown in Fig.~\ref{fig:models-comparison} can also
be interpreted by the 2-category point of view. It is broadly
understood that to pass from simple type theory to dependent type
theory one has to go from the 2-category $\mathbf{Cat}$ of categories to
the 2-category
$\mathbf{ICat}(\mathbb{B})$ of $\mathbb{B}$-indexed
categories (or equivalently, to the 2-category $\mathbf{Fib}(\mathbb{B})$ of fibrations over base category $\category{B}$).
Now, monoids and graded monads are definable in any
2-category (with finite products). If we define them in
$\mathbf{Cat}$, we obtain monoids and graded monads. If we define
them in $\mathbf{ICat}(\mathbb{B})$, then we obtain indexed monoids
and indexed graded monads.

\subsection{From Graded Monads to Indexed Graded Monads}

Constructing models for dependent effects as outlined before can be challenging,
hence we also provide a method to construct a model for dependent effects  out of
a model for simple effects.  More specifically, we
show that given an ordinary graded
monad, we can construct an indexed graded monad.

\paragraph{Construction of indexed preordered monoids}
Suppose we have a preordered monoid $\GradingMonoid$ as a model of
simple effects.  As we discussed above, a natural model to interpret
dependent effects can be the $\Set$-indexed preordered monoid
$\{ \Set(I, \GradingMonoid) \}_{I \in \Set}$.  For
example, $n : \IntType, x : \IntType \vdash \max \{ 0, n - x \}
: \mathbf{Effect}$ can be interpreted as $\interpret{\max \{ 0, n -
x \}} \in \Set(\mathbb{Z}^2, \mathbb{N}_{\infty})$ as we have already
discussed above. So, we could try to have a method to construct this interpretation
from $\mathbb{N}_{\infty}$. However, in general, we may have contexts $\Gamma$ that are
refined by predicates. So, we actually need a $\mathbf{Pred}$-indexed
preordered monoid where $\mathbf{Pred}$ is the category of predicates and predicate-preserving functions, i.e., an object is a pair $(I, P \subseteq I)$ of a set $I$ and a predicate $P$ on $I$, and a morphism $(I, P \subseteq I) \to (J, Q \subseteq J)$ is a function $f : I \to J$ such that $f(i) \in Q$ for each $i \in P$.
One could try to use the change-of-base construction along the
forgetful functor $\mathbf{Pred} \to \Set$, but this does not work because it
simply forgets the predicates, and as a result, the subeffecting
relation becomes weaker than we expect.  For example, the subeffecting relation $x : \{ x
: \IntType \mid x \ge 1 \} \vDash |x + 1| \le |2 x|$, which compares two effects $|x + 1|$ and $|2 x|$ under the assumption $x \ge 1$, should be valid.
However, the change-of-base construction does not allow us to take the
predicate $x \ge 1$ into account, in which case the subeffecting
relation becomes invalid, as $|x + 1| \le |2 x|$ is not true for $x =
0$.
To address this issue, our construction instead defines a
$\mathbf{Pred}$-indexed preordered monoid as follows.
\[ \mathbf{Pred} \ni (I, P \subseteq I) \mapsto \big( \Set(I, \GradingMonoid), {\le}_P \big) \quad\qquad f \le_P g \stackrel{\text{def}}{\iff} \forall i \in P, f(i) \le g(i) \]
A key point in this definition is that we use the pointwise order on $\Set(I, \GradingMonoid)$ restricted to the predicate $P$ instead of the unrestricted pointwise order.

\paragraph{Construction of indexed graded monads}
We now want to construct an indexed graded monad graded by the
$\mathbf{Pred}$-indexed preordered monoid above.

Let $T$ be a monad on $\Set$ that models computational effects.
We can build a model of simple effect systems, by using an \emph{$\GradingMonoid$-graded monad lifting} $\dot{T}$ of $T$ along the predicate fibration $\mathrm{pred} : \mathbf{Pred} \to \Set$.
\[ \dot{T}: \qquad \GradingMonoid \times \mathbf{Pred} \ni (m, (I, Q \subseteq I)) \qquad\mapsto\qquad (T I, \dot{T}_m Q \subseteq T I) \in \mathbf{Pred} \]
As in the case of ordinary monad liftings, the functor $\mathrm{pred}$ maps the graded monad structure of $\dot{T}$ to the monad structure of $T$.
This ensures that interpreting computations in the total category $\mathbf{Pred}$ yields liftings of their interpretations in the base category $\Set$, thereby establishing the soundness of the simple effect system.

We would like to extend this construction to indexed graded monads.
Since we now consider contexts and types refined by predicates, a context $\Gamma$ is interpreted as a predicate $\interpret{\Gamma} = (I, P \subseteq I) \in \mathbf{Pred}$, and a type $\Gamma \vdash A$ is interpreted as a family of predicates $\{ (X_i, Q_i \subseteq X_i) \}_{i \in P}$.
If we apply $\dot{T}$ index-wise, then a graded monadic type $\Gamma \vdash T_{\mathcal{E}} A$ where $\Gamma \vdash \mathcal{E} : \mathbf{Effect}$ would be interpreted as a family of predicates $\{ (T X_i, \dot{T}_{f(i)}
Q_i) \}_{i \in P}$ where
$\interpret{\mathcal{E}} = f \in \Set(I, \GradingMonoid)$ is the
interpretation of the dependent effect $\mathcal{E}$.  This intuition
brings us to  the construction of a $\mathbf{Pred}$-indexed graded monad $\dot{T}' = \{ \dot{T}'_{(I, P)} \}_{(I, P) \in \mathbf{Pred}}$ (graded by $\Set({-}, \GradingMonoid)$) from the graded monad lifting $\dot{T}$ (graded by $\GradingMonoid$) as follows.
\[ \dot{T}'_{(I, P)}: \quad \big(f : I \to \GradingMonoid,\quad \{ (X_i, Q_i \subseteq X_i) \}_{i \in P}\big) \qquad\mapsto\qquad \{ (T X_i, \dot{T}_{f(i)} Q_i) \}_{i \in P} \]
We formalize this construction in Section~\ref{sec:indexed-graded-monad}, allowing slightly more general predicate fibrations, and use it to provide concrete instances of our type system for cost
analysis, union bound reasoning, expectation reasoning, and temporal safety verification in Section~\ref{sec:instances}.

\section{A Dependent Effect System with Refinement Types}\label{sec:type-system}
We consider two layers of type systems.
The first layer is a simply typed language based on fine-grain call-by-value (FGCBV)~\cite{LevyInformationandComputation2003}.
The language is parameterized by a set of generic effects~\cite{PlotkinApplCategStruct2003}, which allows us to instantiate the language for various applications such as cost analysis and union bound.
Then, we define the second layer, a dependent effect system on top of the simply typed FGCBV.
This is a novel dependent type system that allows effects to depend on values.
Combined with refinement types, the dependent effect system allows us to reason about higher-order functional programs similarly to graded Hoare logics~\cite{GaboardiESOP2021}.
We explain these type systems using cost analysis and union bound as running examples.
We will show other instances later in Section~\ref{sec:instances}.

\subsection{Simply Typed Fine-Grain Call-by-Value}\label{sec:stfgcbv}

In FGCBV, values and computations are syntactically separated as follows.
\begin{definition}\label{def:terms}
	Let $\mathbf{Op}$ be a set of effect-free operations and $\mathbf{GenEff}$ a set of generic effects.
	\emph{Value terms} $V, W$ and \emph{computation terms} $M, N$ of the fine-grain call-by-value are defined as follows.
	\begin{align}
		V, W &\coloneqq\quad x \mid \lambda x. M \mid () \mid (V, W) \mid \leftinj{V} \mid \rightinj{V} \mid \recfun{f}{x}{M} \mid \mathtt{op}(V) \\
		M, N &\coloneqq\quad \return{V} \mid \letin{x}{M}{N} \mid \patternmatch{V}{x}{y}{M} \mid V\ W \\
		&\qquad \mid \mathtt{gef}(V) \mid \caseof{V}{\casepattern{\leftinj{x}}{M}, \casepattern{\rightinj{y}}{N}}
		\tag*{\qed}
	\end{align}
\end{definition}
Here, a recursive function written as $\recfun{f}{x}{M}$ is defined (semantically) as the fixed point of $f \mapsto \lambda x. M$.
Effect-free operations (or pure functions) $\mathtt{op} \in \mathbf{Op}$ and generic effects $\mathtt{gef} \in \mathbf{GenEff}$ can take multiple arguments by using tuples.
The language has a simple type system, which we will later extend to a dependent effect system in Section~\ref{sec:dependent-effect-system}.
Since the typing rules for simple types are standard, we only give the definition of simple types here and omit lengthy typing rules.
\begin{definition}\label{def:simple-types}
	Let $\mathbf{Base} = \{ \UnitType, \IntType, \RealType, \NatType, \dots \}$ be a set of base types.
	We assume that $\mathbf{Base}$ contains at least the unit type $\UnitType$.
	\emph{Simple value types} $A, B$ and \emph{simple computation types} $C, D$ are defined as follows.
	\begin{align}
		A, B\ &\coloneqq\ b \mid A \times B \mid A \to C \mid A + B \quad \text{where $b \in \mathbf{Base}$}\quad &
		C, D\ &\coloneqq\ T A \tag*{\qed}
	\end{align}
\end{definition}

Since we have sum types $A + B$, the type $\BoolType$ of boolean values is defined as a shorthand for $\BoolType \coloneqq \UnitType + \UnitType$, and the if-expression is defined as a shorthand for case analysis on $\mathtt{true} \coloneqq \leftinj{()}$ and $\mathtt{false} \coloneqq \rightinj{()}$.
We say that a value type is a \emph{ground type} if it does not contain arrow types.
Specifically, a ground type is in the fragment of simple value types defined by $A, B \coloneqq b \mid A \times B \mid A + B$.
We assume that each effect-free operation $\mathtt{op} \in \mathbf{Op}$ and each generic effect $\mathtt{gef} \in \mathbf{GenEff}$ have a simple type signature $\mathtt{op} : A \rightarrowtriangle B$ and $\mathtt{gef} : A \rightarrowtriangle B$, respectively, where the input type $A$ and the output type $B$ are ground types.

\begin{example}[simple type signatures of effect-free operations]\label{ex:effect-free-operations-simple-type-signatures}
	Effect-free operations contain operations such as arithmetic operations, comparison operations, type conversion functions, and constant values for $\IntType$, $\RealType$, and $\NatType$.
	For example, the addition operation for integers has the simple type signature ${+} : \IntType \times \IntType \rightarrowtriangle \IntType$, and the constant value $0$ for integers has the simple type signature $0 : \UnitType \rightarrowtriangle \IntType$.
	\qed
\end{example}

\begin{example}[cost analysis]\label{ex:cost-analysis-generic-effect}
	For cost analysis, we use the following generic effect:
	$\mathbf{GenEff} = \{ \mathtt{Tick} : \NatType \rightarrowtriangle \UnitType \}$.
	Here, we slightly change the type signature of the generic effect $\mathtt{Tick}$ from Section~\ref{sec:overview} so that it takes a natural number $n$ and incurs a cost of $n$.
	\qed
\end{example}

\begin{example}[union bound]
	For union bound logic, we consider the following generic effect:
	$\mathbf{GenEff} = \{ \mathtt{Lap} : \RealType \times \RealType \rightarrowtriangle \RealType \}$.
	The generic effect $\mathtt{Lap}(\epsilon, m)$ takes two real numbers as parameters for the Laplace distribution, where $\epsilon$ is the scale parameter and $m$ is the location parameter, and returns a real number sampled from the Laplace distribution if $\epsilon > 0$.
	Here, the probability density function of the Laplace distribution is given by $L_{\epsilon, m}(x) = \frac{1}{2 \epsilon} \exp(- \frac{|x - m|}{\epsilon})$.
	If $\epsilon \le 0$, then $\mathtt{Lap}(\epsilon, m)$ returns an arbitrary value.
	\qed
\end{example}

Well-typed terms in the simple type system are denoted as $\Gamma \vdash V : A$ and $\Gamma \vdash M : C$ where $\Gamma$ is a \emph{simple context}, which is defined as a finite list of pairs of variables and simple value types.
Typing rules for the simple type system are standard and omitted.

\subsection{A Dependent Effect System with Refinement Types}\label{sec:dependent-effect-system}

\subsubsection{Types}

We extend simple value/computation types in three ways: (1) by using dependent types, (2) by refining base types with predicates, and (3) by grading computation types with dependent effects.
The extension is called \emph{graded refinement value/computation types}.
Meta-variables $\dot{A}, \dot{C}$ for these types are written with a dot above them to distinguish them from simple types.
\begin{definition}\label{def:refinement-types}
	Let $\mathbf{BEff}$ be a set of \emph{basic effects}, which are symbols that represent dependent effects.
	\emph{Graded refinement value types} $\dot{A}, \dot{B}$; \emph{graded refinement computation types} $\dot{C}, \dot{D}$; and \emph{dependent-effect terms} $\mathcal{E}$ are defined as follows.
	\begin{align}
		\dot{A}, \dot{B} &\coloneqq\ \{ x : b \mid \phi \} \mid (x : \dot{A}) \times \dot{B} \mid (x : \dot{A}) \to \dot{C} \mid \dot{A} + \dot{B} \\
		\dot{C}, \dot{D} &\coloneqq\quad T_{\mathcal{E}} \dot{A} \qquad\qquad\qquad
		\mathcal{E} \coloneqq\quad \GradeUnit \mid \mathcal{E}_1 \GradeMult \mathcal{E}_2 \mid \mathtt{be}(V) \qquad \text{where $\mathtt{be} \in \mathbf{BEff}$}
	\end{align}
	Here, $\phi$ ranges over formulas.
	We write $\underlying{\dot{A}}$ and $\underlying{\dot{C}}$ for the simple types obtained by forgetting predicates and dependent-effect terms.
	We call them the \emph{underlying type} of $\dot{A}$ and $\dot{C}$.
	For example, the underlying type of $(x : \{ x : \IntType \mid n \ge x \}) \to T_{n - x} \{ u : \UnitType \mid \mathbf{true} \}$ is $\IntType \to T \UnitType$.
	\qed
\end{definition}

\begin{toappendix}
\begin{definition}
	The \emph{underlying type} of $\dot{A}$ and $\dot{C}$ is defined as follows.
	\begin{gather}
		\underlying{\{ x : b \mid \phi \}} = b \qquad
		\underlying{(x : \dot{A}) \times \dot{B}} = \underlying{\dot{A}} \times \underlying{\dot{B}} \qquad
		\underlying{(x : \dot{A}) \to \dot{C}} = \underlying{\dot{A}} \to \underlying{\dot{C}} \\
		\underlying{\dot{A} + \dot{B}} = \underlying{\dot{A}} + \underlying{\dot{B}} \qquad\qquad
		\underlying{T_{\mathcal{E}} \dot{A}} = T \underlying{\dot{A}}
	\end{gather}
\end{definition}
\end{toappendix}

\paragraph{Dependent Types}
We consider \emph{dependent pair types} $(x : \dot{A}) \times \dot{B}$ and \emph{dependent function types} $(x : \dot{A}) \to \dot{C}$, in which the type of the second component may depend on $x$.
Note that types are allowed to depend only on value terms, but not on computation terms.
This approach to combining dependent types with computational effects follows EMLTT~\cite{AhmanFoSSaCS2016}, a dependently typed call-by-push-value calculus.
When $x$ does not occur in $\dot{B}, \dot{C}$, we often write as $\dot{A} \to \dot{C} = (x : \dot{A}) \to \dot{C}$ and $\dot{A} \times \dot{B} = (x : \dot{A}) \times \dot{B}$.

\paragraph{Refinement Types}
To specify preconditions and postconditions of programs, we use a type of the form $\{ x : b \mid \phi \}$, which we call a \emph{refinement base type}.
Intuitively, a value term $V$ has type $\{ x : b \mid \phi \}$ if $V$ is of base type $b$ and satisfies the formula $\phi$.
When the formula $\phi$ is always true, we write $b$ as a shorthand of $\{ x : b \mid \mathbf{true} \}$.
\emph{Formulas} $\phi, \psi$ in refinement base types are constructed from atomic formulas (including $\mathbf{true}$ and $\mathbf{false}$) using boolean connectives $\land, \lor, \implies$.
Here, an \emph{atomic formula} $\mathtt{a}(V)$ is constructed from a predicate symbol $\mathtt{a}$ and a value term $V$.
We assume that the argument type $A$ of each predicate symbol $\mathtt{a} : A \to \mathbf{Fml}$ is a (simple) ground type and do not consider predicate symbols for computation terms.
This implies that any value term occurring in a formula does not contain lambda abstractions and recursive functions.
On the other hand, it is convenient to extend the syntax of value terms $V, W$ in formulas to include projections (recall that pattern matching is defined as a \emph{computation} term in Definition~\ref{def:terms}), which allows us to write $(x : \dot{A}) \times (y : \dot{B}) \to \dot{C}$ as a syntactic sugar for $(z : \dot{A} \times \dot{B}) \to \dot{C}[\leftproj{z}/x, \rightproj{z}/y]$.
Thus, we use \emph{ground value terms} defined below as value terms occurring in formulas.
\begin{equation}
	V, W \quad\coloneqq\quad x \mid () \mid (V, W) \mid \leftinj{V} \mid \rightinj{V} \mid \leftproj{V} \mid \rightproj{V} \mid \mathtt{op}(V)
	\label{eq:ground-value-terms}
\end{equation}
Note that effect-free operations are allowed here, but generic effects are not.
A formula is well-formed and written as $\Gamma \vdash \phi : \mathbf{Fml}$ if any value terms occurring in $\phi$ is well-typed in $\Gamma$ and predicate symbols are applied to value terms of appropriate types.

\paragraph{Dependent Effects}
The main novelty of our dependent effect system is the use of dependent-effect terms $\mathcal{E}$.
Intuitively, a dependent-effect term represents an element of a grading monoid that depends on value terms.
To convert value terms to dependent-effect terms, we use \emph{basic effects} in $\mathbf{BEff}$.
We write $\mathtt{be} : A \to \mathbf{Effect}$ if $\mathtt{be}$ is a basic effect that takes a value term of type $A$ and returns an effect (i.e., an element of a preordered monoid).
We assume that the argument type $A$ of each basic effect $\mathtt{be} : A \to \mathbf{Effect}$ is a (simple) ground type.
Similarly to formulas, we use ground value terms~\eqref{eq:ground-value-terms} in dependent-effect terms.

\begin{example}[cost analysis]
	Consider the cost of the following function:
	\[ M \quad\coloneqq\quad \recfun{f}{x}{\ifexpr{x < n}{\letin{z}{\mathtt{Tick}\ 1}{f(x + 1)}}{\return{()}}} \]
	Here, we consider $(\mathbb{N}, 0, {+}, {\le})$ as a grading monoid.
	The program $M$ is a function that takes an integer $x$ and increments it by $1$ until it reaches $n$.
	It is easy to see that the cost of this function is upper bounded by $\max \{ 0, n - x\}$.
	We can express this using our type system as follows:
	\[ n : \IntType \quad\vdash\quad M : (x : \IntType) \to T_{\mathtt{nat2eff} (\max \{ 0, n - x\})} \UnitType \]
	Here, we use the function $\mathtt{nat2eff} : \NatType \to \mathbf{Effect}$ as a basic effect that converts a value term $\max \{ 0, n - x\} : \NatType$ to an effect.
	\qed
\end{example}

\begin{example}[simple effect]
	If a set of basic effects $\mathbf{BEff}$ consists only of a constant effect $\underline{m} : \UnitType \to \mathbf{Effect}$ for each $m \in \GradingMonoid$ in a grading monoid $\GradingMonoid$, then we obtain a simple effect system.
	\qed
\end{example}

\subsubsection{Well-formed Types}

We define a \emph{graded refinement context} $\dot{\Gamma}$ as a finite list of pairs of variables and graded refinement value types.
The \emph{underlying context} $\underlying{\dot{\Gamma}}$ is a simple context obtained by applying $\underlying{{-}}$ to each type in $\dot{\Gamma}$, that is, $\underlying{x_1 : \dot{A}_1, \dots, x_n : \dot{A}_n} \coloneqq x_1 : \underlying{\dot{A}_1}, \dots, x_n : \underlying{\dot{A}_n}$.

As is common in dependent type systems, we consider judgements for well-formed contexts $\dot{\Gamma} \vdash $ and well-formed types $\dot{\Gamma} \vdash \dot{A}$, $\dot{\Gamma} \vdash \dot{C}$.
In addition, we need a judgement for well-formed dependent-effect terms $\dot{\Gamma} \vdash \mathcal{E} : \mathbf{Effect}$.
The intuition of well-formedness is that free variables in types and dependent-effect terms are bound in the context $\dot{\Gamma}$.
Rules for well-formedness are mostly straightforward.
The only rule that requires some attention is the rule for basic effects: for each basic effect $\mathtt{be} : A \to \mathbf{Effect}$, the dependent-effect term $\dot{\Gamma} \vdash \mathtt{be}(V) : \mathbf{Effect}$ is well-formed if $\underlying{\dot{\Gamma}} \vdash V : A$ is well-typed in the simple type system, and $\dot{\Gamma} \vdash$ is well-formed.
One may notice that the predicates in the context $\dot{\Gamma}$ are not used when constructing dependent-effect terms; however, as we will see in Example~\ref{ex:subtyping}, they play a role in sub-effecting.

\subsubsection{Typing Rules for Terms}
Well-typed value terms and well-typed computation terms are denoted as $\dot{\Gamma} \vdash V : \dot{A}$ and $\dot{\Gamma} \vdash M : \dot{C}$, respectively.
Most typing rules are straightforward (Fig.~\ref{fig:typing-rules}).
In the rules for function application and generic effects, the argument value term $V$ is substituted for the variable $x$ in dependent-effect terms.
For $\mathtt{return}$ and $\mathtt{let}$, we use the monoid structure of effects.
The rule for $\mathtt{let}$ does not allow $\mathcal{E}_2$ to depend on $x$.
This restriction stems from the difficulty of combining computational effects with dependent types, as discussed in \cite[Section~2.6]{PedrotPOPL2020} and \cite{AhmanFoSSaCS2016}.
At first glance, this restriction may appear to be a serious limitation of our dependent effect system.
However, even with this restriction, our type system can reason about interesting examples by using refinement types to describe the possible outcomes of computations.
\begin{example}
	Suppose that $f : (n : \mathtt{int}) \to T_{n} \{ m : \mathtt{int} \mid m = 2 n \}$ is a function that returns $2 n$ at cost $n$, and consider the cost of the following simple program:
	\begin{equation}
		x : \mathtt{int} \quad\vdash\quad \letin{y}{f\ x}{f\ y} \quad:\quad T_{3 x} \{ m : \mathtt{int} \mid m = 4 x \} \label{eq:let-dep-example}
	\end{equation}
	Here, we omit $f$ from the context for brevity.
	By the typing rule for function application, the second computation $f\ y$ has the following type:
	\[ x : \mathtt{int},\quad y : \{ m : \mathtt{int} \mid m = 2 x \} \quad\vdash\quad f\ y \quad:\quad T_y \{ m : \mathtt{int} \mid m = 2 y \} \]
	We cannot directly apply the typing rule for $\mathtt{let}$ because the result type contains the variable $y$, which is bound to the result of the first computation.
	However, since we know $y = 2 x$ from the refinement type, the variable $y$ in the result type of $f\ y$ can be eliminated using the following subtyping relation.
	\[ T_y \{ m : \mathtt{int} \mid m = 2 y \} \quad<:\quad T_{2 x} \{ m : \mathtt{int} \mid m = 4 x \} \]
	Therefore, our type system can still verify that the cost of the program in \eqref{eq:let-dep-example} is upper bounded by $3 x$, which, in this case, coincides with the exact cost.
	\qed
\end{example}

This approach may not work in general, as we show in \referappendix{Example}{40}{ex:union-bound-let-problem} for union bound logic, but a better treatment of sequencing is left as future work.

\begin{toappendix}
\begin{example}\label{ex:union-bound-let-problem}
	We show an example where the restriction of the typing rule for $\mathtt{let}$ can be problematic.
	Consider reasoning about the following program using the union-bound logic:
	\[ \nvdash \letin{x}{\mathtt{Bern}(0.2)}{f\ x} \quad:\quad T_{0.08} \{ y : \IntType \mid y \ge 0 \} \]
	Here, $\mathtt{Bern}(p)$ is a generic effect that returns $\mathtt{true}$ with probability $p$ and $\mathtt{false}$ with probability $1 - p$; and $f : (b : \BoolType) \to T_{\ifexpr{b}{0.4}{0}} \{y : \IntType \mid y \ge 0\}$ is a function that returns non-negative integer with probability more than $0.6$ when $b$ is $\mathtt{true}$ and with probability $1$ when $b$ is $\mathtt{false}$.
	In this example, the true failure probability is upper bounded by $0.2 \times 0.4 + 0.8 \times 0 = 0.08$.
	However, since the type of the first computation $\mathtt{Bern}(0.2) : T_0 \BoolType$ does not contain sufficient information about its output distribution, our dependent effect system yields only a very loose overapproximation:
	\[ \nvdash \letin{x}{\mathtt{Bern}(0.2)}{f\ x} \quad:\quad T_{0.4} \{ y : \IntType \mid y \ge 0 \} \]
\end{example}
Nonetheless, this does not mean that our dependent effect system is useless for the union bound case.
\begin{example}
	Continue from Example~\ref{ex:union-bound-let-problem}.
	Consider the following program:
	\begin{align}
		x : \BoolType \quad&\vdash\quad \letin{y}{f\ x}{\letin{z}{f\ (\mathtt{not}\ x)}}{\return{(y, z)}} \\
		&\qquad:\quad T_{0.4} \{ (y, z) : \IntType \times \IntType \mid y \ge 0 \land z \ge 0 \}
	\end{align}
	If we use a simple effect system, then it would estimate the failure probability as $0.4 + 0.4 = 0.8$.
	In contrast, our dependent effect system can verify that the failure probability is actually at most $0.4$.
\end{example}
\end{toappendix}

To reason about effect-free operations and generic effects in our dependent effect system, we assume that their \emph{refinement type signatures} are given.
An effect-free operation has a refinement type signature of the form $\mathtt{op} : (x : \dot{A}) \rightarrowtriangle \dot{B}$ where $\vdash \dot{A}$ and $x : \dot{A} \vdash \dot{B}$ are well-formed, and $\underlying{\dot{A}} \rightarrowtriangle \underlying{\dot{B}}$ coincides with the simple type signature of $\mathtt{op}$.
For example, the constant $0$ has refinement type signature $0 : \UnitType \rightarrowtriangle \{ x : \IntType \mid x = 0 \}$, and the addition operator has $({+}) : (x : \IntType) \times (y : \IntType) \rightarrowtriangle \{ z : \IntType \mid z = x + y \}$.
Similarly, a generic effect has a refinement type signature of the form $\mathtt{gef} : (x : \dot{A}) \stackrel{\mathcal{E}}{\rightarrowtriangle} \dot{B}$ where $\vdash \dot{A}$, $x : \dot{A} \vdash \dot{B}$, and $x : \dot{A} \vdash \mathcal{E} : \mathbf{Effect}$ are well-formed, and $\underlying{\dot{A}} \rightarrowtriangle \underlying{\dot{B}}$ is the simple type signature of $\mathtt{gef}$.
For example, the generic effect $\mathtt{Tick}$ for cost analysis (Example~\ref{ex:cost-analysis-generic-effect}) has refinement type signature $\mathtt{Tick} : (n : \NatType) \stackrel{n}{\rightarrowtriangle} \UnitType$, which means that the cost of $\mathtt{Tick}\ n$ is upper bounded by $n$.

\begin{figure}[tb]
	\begin{mathpar}
		\inferrule{
			\dot{\Gamma} \vdash V : (x : \dot{A}) \to T_{\mathcal{E}} \dot{B} \\
			\dot{\Gamma} \vdash W : \dot{A}
		}{
			\dot{\Gamma} \vdash V \ W : T_{\mathcal{E}[W/x]} \dot{B}[W/x]
		}
		\and
		\inferrule{
			\dot{\Gamma} \vdash V : \dot{A} \\
			\mathtt{gef} : (x : \dot{A}) \stackrel{\mathcal{E}}{\rightarrowtriangle} \dot{B}
		}{
			\dot{\Gamma} \vdash \mathtt{gef}(V) : T_{\mathcal{E}[V/x]} \dot{B}[V/x]
		}
		\and
		\inferrule{
			\dot{\Gamma} \vdash V : \dot{A}
		}{
			\dot{\Gamma} \vdash \return{V} : T_{\GradeUnit} \dot{A}
		}
		\and
		\inferrule{
			\dot{\Gamma} \vdash M : T_{\mathcal{E}_1} \dot{A} \\
			\dot{\Gamma} \vdash T_{\mathcal{E}_2} \dot{B} \\
			\dot{\Gamma}, x : \dot{A} \vdash N : T_{\mathcal{E}_2} \dot{B}
		}{
			\dot{\Gamma} \vdash \letin{x}{M}{N} : T_{\mathcal{E}_1 \GradeMult \mathcal{E}_2} \dot{B}
		}
	\end{mathpar}
	\caption{Selected typing rules. Substitution of a value term $V$ for a variable $x$ is denoted by suffixing $[V/x]$.}
	\label{fig:typing-rules}
\end{figure}

\subsubsection{Subtyping}
Subtyping relation is denoted as $\dot{\Gamma} \vdash \dot{A} <: \dot{B}$ and $\dot{\Gamma} \vdash \dot{C} <: \dot{D}$.
Similarly to the consequence rule in Hoare logic, it allows us to derive $\dot{\Gamma} \vdash V : \dot{B}$ from $\dot{\Gamma} \vdash V : \dot{A}$ and $\dot{\Gamma} \vdash \dot{A} <: \dot{B}$.
To derive subtyping relations $\dot{\Gamma} \vdash \{ x : b \mid \phi \} <: \{ x : b \mid \psi \}$ for refinement base types, we use the semantic validity of formulas $\dot{\Gamma}, x : b \vDash \phi \implies \psi$, which means that the formula $\phi$ implies $\psi$ when all the predicates in $\dot{\Gamma}$ are satisfied.
To derive subtyping relations $\dot{\Gamma} \vdash T_{\mathcal{E}_1} \dot{A} <: T_{\mathcal{E}_2} \dot{B}$ for graded refinement computation types (Fig.~\ref{fig:subtyping}), we use the semantic subeffecting relation $\dot{\Gamma} \vDash \mathcal{E}_1 \le \mathcal{E}_2$, which means that the dependent-effect term $\mathcal{E}_1$ is less than or equal to $\mathcal{E}_2$ when all the predicates in $\dot{\Gamma}$ are satisfied.
We do not provide syntactic derivation rules for these semantic judgements because they are intended to be solved by an external solver, such as an SMT solver~\cite{DeMouraTACAS2008,BarbosaTACAS2022,DutertreCAV2014}, during type checking.

\begin{example}\label{ex:subtyping}
	For cost analysis, we can derive the following subtyping relation.
	For notational simplicity, we omit the basic effect $\mathtt{nat2eff} : \NatType \to \mathbf{Effect}$ for conversion below.
	\[ x : \{ x : \IntType \mid x \ge 1 \} \quad\vdash\quad T_{|x + 1|} \{ y : \IntType \mid y = x \} \quad<:\quad T_{|2 x|} \{ y : \IntType \mid y \ge 1 \} \]
	Here, we need to compare both dependent-effect terms and formulas.
	For the former, $x : \{ x : \IntType \mid x \ge 1 \} \vDash |x + 1| \le |2 x|$ is semantically valid, and for the latter, $x : \{ x : \IntType \mid x \ge 1 \}, y : \IntType \vDash y = x \implies y \ge 1$ is semantically valid.
	Note that $x \ge 1$ in the context is necessary for the validity.
	\qed
\end{example}

\begin{figure}[tb]
	
	\begin{mathpar}
		\inferrule{
			\dot{\Gamma} \vdash M : \dot{C} \\
			\dot{\Gamma} \vdash \dot{C} <: \dot{D}
		}{
			\dot{\Gamma} \vdash M : \dot{D}
		}
		\and
		\inferrule{
			\dot{\Gamma} \vdash \mathcal{E}_1 : \mathbf{Effect} \\
			\dot{\Gamma} \vdash \mathcal{E}_2 : \mathbf{Effect} \\\\
			\dot{\Gamma} \vDash \mathcal{E}_1 \le \mathcal{E}_2 \\
			\dot{\Gamma} \vdash \dot{A} <: \dot{B}
		}{
			\dot{\Gamma} \vdash T_{\mathcal{E}_1} \dot{A} <: T_{\mathcal{E}_2} \dot{B}
		}
	\end{mathpar}
	\caption{Selected rules for subtyping. Here, $\dot{\Gamma} \vDash \mathcal{E}_1 \le \mathcal{E}_2$ is semantic subeffecting in the refinement context $\dot{\Gamma}$.}
	\label{fig:subtyping}
\end{figure}

\begin{toappendix}
Well-formed context:
$\dot{\Gamma} \vdash$
\begin{mathpar}
	\inferrule{ }{
		\diamond \vdash
	}
	\and
	\inferrule{
		\dot{\Gamma} \vdash \dot{A}
	}{
		\dot{\Gamma}, x : \dot{A} \vdash
	}
\end{mathpar}

Well-formed value type:
$\dot{\Gamma} \vdash \dot{A}$
\begin{mathpar}
	\inferrule{
		\dot{\Gamma} \vdash \\
		\underlying{\dot{\Gamma}}, x : b \vdash \phi : \mathbf{Fml}
	}{
		\dot{\Gamma} \vdash \{ x : b \mid \phi \}
	}
	\and
	\inferrule{
		\dot{\Gamma} \vdash \\
		\underlying{\dot{\Gamma}}, x : \UnitType \vdash \phi : \mathbf{Fml}
	}{
		\dot{\Gamma} \vdash \{ x : \UnitType \mid \phi \}
	}
	\and
	\inferrule{
		\dot{\Gamma}, x : \dot{A} \vdash \dot{C}
	}{
		\dot{\Gamma} \vdash (x : \dot{A}) \to \dot{C}
	}
	\and
	\inferrule{
		\dot{\Gamma}, x : \dot{A} \vdash \dot{B}
	}{
		\dot{\Gamma} \vdash (x : \dot{A}) \times \dot{B}
	}
	\and
	\inferrule{
		\dot{\Gamma} \vdash \dot{A} \\
		\dot{\Gamma} \vdash \dot{B}
	}{
		\dot{\Gamma} \vdash \dot{A} + \dot{B}
	}
\end{mathpar}

Well-formed computation type:
$\dot{\Gamma} \vdash \dot{C}$
\begin{mathpar}
	\inferrule{
		\dot{\Gamma} \vdash \mathcal{E} : \mathbf{Effect} \\
		\dot{\Gamma} \vdash \dot{A}
	}{
		\dot{\Gamma} \vdash T_{\mathcal{E}} \dot{A}
	}
\end{mathpar}

Well-formed effect:
$\dot{\Gamma} \vdash \mathcal{E} : \mathbf{Effect}$
\begin{mathpar}
	\inferrule{
		\dot{\Gamma} \vdash
	}{
		\dot{\Gamma} \vdash \GradeUnit : \mathbf{Effect}
	}
	\and
	\inferrule{
		\dot{\Gamma} \vdash \mathcal{E}_1 : \mathbf{Effect} \\
		\dot{\Gamma} \vdash \mathcal{E}_2 : \mathbf{Effect}
	}{
		\dot{\Gamma} \vdash \mathcal{E}_1 \GradeMult \mathcal{E}_2 : \mathbf{Effect}
	}
	\and
	\inferrule{
		\dot{\Gamma} \vdash \\
		\underlying{\dot{\Gamma}} \vdash V : A \\
		\mathtt{be} : A \to \mathbf{Effect}
	}{
		\dot{\Gamma} \vdash \mathtt{be}(V) : \mathbf{Effect}
	}
\end{mathpar}

Well-typed value term
$\dot{\Gamma} \vdash V : \dot{A}$

\begin{mathpar}
	\inferrule[\hypertarget{rule:VT-Var}{VT-Var}]{
		\dot{\Gamma} \vdash \\
		(x : \dot{A}) \in \dot{\Gamma}
	}{
		\dot{\Gamma} \vdash x : \dot{A}
	}
	\and
	\inferrule[\hypertarget{rule:VT-VarSelf}{VT-VarSelf}]{
		\dot{\Gamma} \vdash \\
		(x : \{ x : b \mid \phi \}) \in \dot{\Gamma}
	}{
		\dot{\Gamma} \vdash x : \{ x' : b \mid x' = x \}
	}
	\and
	\inferrule[\hypertarget{rule:VT-EffFreeOp}{VT-EffFreeOp}]{
		\dot{\Gamma} \vdash V : \dot{A} \\
		\mathtt{op} : (x : \dot{A}) \rightarrowtriangle \dot{B}
	}{
		\dot{\Gamma} \vdash \mathtt{op}\ V : \dot{B}[V / x]
	}
	\and
	\inferrule[\hypertarget{rule:VT-Lam}{VT-Lam}]{
		\dot{\Gamma}, x : \dot{A} \vdash M : \dot{C}
	}{
		\dot{\Gamma} \vdash \lambda x. M : (x : \dot{A}) \to \dot{C}
	}
	\and
	\inferrule[\hypertarget{rule:VT-Unit}{VT-Unit}]{
		\dot{\Gamma} \vdash
	}{
		\dot{\Gamma} \vdash () : \{ x : \UnitType \mid \mathbf{true} \}
	}
	\and
	\inferrule[\hypertarget{rule:VT-Pair}{VT-Pair}]{
		\dot{\Gamma} \vdash V : \dot{A} \\
		\dot{\Gamma} \vdash W : \dot{B}[V/x]
	}{
		\dot{\Gamma} \vdash (V, W) : (x : \dot{A}) \times \dot{B}
	}
	\and
	\inferrule[\hypertarget{rule:VT-Inl}{VT-Inl}]{
		\dot{\Gamma} \vdash V : \dot{A}_1 \\
		\dot{\Gamma} \vdash \dot{A}_2
	}{
		\dot{\Gamma} \vdash \leftinj{V} : \dot{A}_1 + \dot{A}_2
	}
	\and
	\inferrule[\hypertarget{rule:VT-Inr}{VT-Inr}]{
		\dot{\Gamma} \vdash \dot{A}_1 \\
		\dot{\Gamma} \vdash V : \dot{A}_2
	}{
		\dot{\Gamma} \vdash \rightinj{V} : \dot{A}_1 + \dot{A}_2
	}
	\and
	\inferrule[\hypertarget{rule:VT-Sub}{VT-Sub}]{
		\dot{\Gamma} \vdash V : \dot{A} \\
		\dot{\Gamma} \vdash \dot{A} <: \dot{B}
	}{
		\dot{\Gamma} \vdash V : \dot{B}
	}
\end{mathpar}

Well-typed computation term
$\dot{\Gamma} \vdash M : \dot{C}$

\begin{mathpar}
	\inferrule[\hypertarget{rule:CT-App}{CT-App}]{
		\dot{\Gamma} \vdash V : (x : \dot{A}) \to T_{\mathcal{E}} \dot{B} \\
		\dot{\Gamma} \vdash W : \dot{A}
	}{
		\dot{\Gamma} \vdash V \ W : T_{\mathcal{E}[V/x]} \dot{B}[V/x]
	}
	\and
	\inferrule[CT-Ret]{
		\dot{\Gamma} \vdash V : \dot{A}
	}{
		\dot{\Gamma} \vdash \return{V} : T_{\GradeUnit} \dot{A}
	}
	\and
	\inferrule[\hypertarget{rule:CT-Let}{CT-Let}]{
		\dot{\Gamma} \vdash M : T_{\mathcal{E}_1} \dot{A} \\
		\dot{\Gamma} \vdash T_{\mathcal{E}_2} \dot{B} \\
		\dot{\Gamma}, x : \dot{A} \vdash N : T_{\mathcal{E}_2} \dot{B}
	}{
		\dot{\Gamma} \vdash \letin{x}{M}{N} : T_{\mathcal{E}_1 \GradeMult \mathcal{E}_2} \dot{B}
	}
	\and
	\inferrule[\hypertarget{rule:CT-GenEff}{CT-GenEff}]{
		\dot{\Gamma} \vdash V : \dot{A} \\
		\mathtt{gef} : (x : \dot{A}) \stackrel{\mathcal{E}}{\rightarrowtriangle} \dot{B}
	}{
		\dot{\Gamma} \vdash \mathtt{gef}\ V : T_{\mathcal{E}[V/x]} \dot{B}[V/x]
	}
	\and
	\inferrule[\hypertarget{rule:CT-Sub}{CT-Sub}]{
		\dot{\Gamma} \vdash M : \dot{C} \\
		\dot{\Gamma} \vdash \dot{C} <: \dot{D}
	}{
		\dot{\Gamma} \vdash M : \dot{D}
	}
	\and
	\inferrule[\hypertarget{rule:CT-PatternMatch}{CT-PatternMatch}]{
		\dot{\Gamma} \vdash V : (x : \dot{A}) \times \dot{B} \\
		\dot{\Gamma}, z : (x : \dot{A}) \times \dot{B} \vdash \dot{C} \\
		\dot{\Gamma}, x : \dot{A}, y : \dot{B} \vdash M : \dot{C}[(x, y)/z]
	}{
		\dot{\Gamma} \vdash \patternmatch{V}{x}{y}{M} : \dot{C}[V / z]
	}
	\and
	\inferrule[\hypertarget{rule:CT-Case}{CT-Case}]{
		\dot{\Gamma} \vdash V : \dot{A} + \dot{B} \\
		\dot{\Gamma}, z : \dot{A} + \dot{B} \vdash \dot{C} \\
		\dot{\Gamma}, x : \dot{A} \vdash M : \dot{C}[\iota_1\ x/z] \\
		\dot{\Gamma}, y : \dot{B} \vdash N : \dot{C}[\iota_2\ x/z]
	}{
		\dot{\Gamma} \vdash \caseof{V}{\casepattern{\leftinj{x}}{M}, \casepattern{\rightinj{y}}{N}} : \dot{C}[V/z]
	}
	\and
	\inferrule[\hypertarget{rule:CT-Rec}{CT-Rec}]{
		\dot{\Gamma}, f : (x : \dot{A}) \to T_{\mathcal{E}} \dot{B}, x : \dot{A} \vdash M : T_{\mathcal{E}} \dot{B}
	}{
		\dot{\Gamma} \vdash \recfun{f}{x}{M} : (x : \dot{A}) \to T_{\mathcal{E}} \dot{B}
	}
\end{mathpar}

Subtyping
$\dot{\Gamma} \vdash \dot{A} <: \dot{B}$, $\dot{\Gamma} \vdash \dot{C} <: \dot{D}$

\begin{mathpar}
	\inferrule{
		\underlying{\dot{\Gamma}}, x : b \vdash \phi : \mathbf{Fml} \\
		\underlying{\dot{\Gamma}}, x : b \vdash \psi : \mathbf{Fml} \\
		\dot{\Gamma}, x : b \vDash \phi \implies \psi
	}{
		\dot{\Gamma} \vdash \{ x : b \mid \phi \} <: \{ x : b \mid \psi \}
	}
	\and
	\inferrule{
		\dot{\Gamma} \vdash \mathcal{E}_1 : \mathbf{Effect} \\
		\dot{\Gamma} \vdash \mathcal{E}_2 : \mathbf{Effect} \\
		\dot{\Gamma} \vDash \mathcal{E}_1 \le \mathcal{E}_2 \\
		\dot{\Gamma} \vdash \dot{A} <: \dot{B}
	}{
		\dot{\Gamma} \vdash T_{\mathcal{E}_1} \dot{A} <: T_{\mathcal{E}_2} \dot{B}
	}
\end{mathpar}
Here, $\dot{\Gamma} \vDash \mathcal{E}_1 \le \mathcal{E}_2$ is the semantic subeffecting, and $\dot{\Gamma} \vDash \phi$ is the (semantic) validity of logical formulas, whose precise meanings are defined later in \eqref{eq:semantic-subeffecting} and \eqref{eq:semantic-validity}.
We do not provide syntactic rules for deriving these judgements because we will use constraint solvers to obtain these judgements.
We allow subeffecting that depends on the predicates in the context $\dot{\Gamma}$, like $x : \{ x : \RealType \mid x = 1 \} \vDash |2 x| \le |x + 2|$.

\begin{mathpar}
	\inferrule{
		\dot{\Gamma} \vdash \dot{A}_2 <: \dot{A}_1 \\
		\dot{\Gamma}, x : \dot{A}_2 \vdash \dot{C}_1 <: \dot{C}_2
	}{
		\dot{\Gamma} \vdash (x : \dot{A}_1) \to \dot{C}_1 <: (x : \dot{A}_2) \to \dot{C}_2
	}
	\and
	\inferrule{
		\dot{\Gamma} \vdash \dot{A}_1 <: \dot{A}_2 \\
		\dot{\Gamma}, x : \dot{A}_1 \vdash \dot{B}_1 <: \dot{B}_2
	}{
		\dot{\Gamma} \vdash (x : \dot{A}_1) \times \dot{B}_1 <: (x : \dot{A}_2) \times \dot{B}_2
	}
	\and
	\inferrule{
		\dot{\Gamma} \vdash \dot{A}_1 <: \dot{A}_2 \\
		\dot{\Gamma} \vdash \dot{B}_1 <: \dot{B}_2
	}{
		\dot{\Gamma} \vdash \dot{A}_1 + \dot{B}_1 <: \dot{A}_2 + \dot{B}_2
	}
\end{mathpar}

\end{toappendix}

\section{Preliminaries on Semantics: Stepping Towards Dependent Effect Systems}\label{sec:preliminary-semantics}
In this section, we gradually develop categorical semantics starting from the simply typed FGCBV.
This development is completed in Section~\ref{sec:indexed-graded-monad} where we introduce the semantics of dependent effects and use it to prove the soundness.
We need careful treatment of circularity of definitions when dealing with dependent type systems.
However, since we focus on a dependent type system that is a refinement of a simple type system, the situation is slightly simpler than general dependent type systems.
The dependency of the definitions of the interpretation is summarized in \referappendix{Fig.}{4}{fig:dependency-interpretation}.
To simplify the presentation, we focus on describing semantics of the recursion-free fragment of FGCBV and leave the details of recursion to \referappendix{Appendix}{C.2}{sec:fgcbv-model-with-recursion}.

\ifthenelse{\boolean{longversion}}{
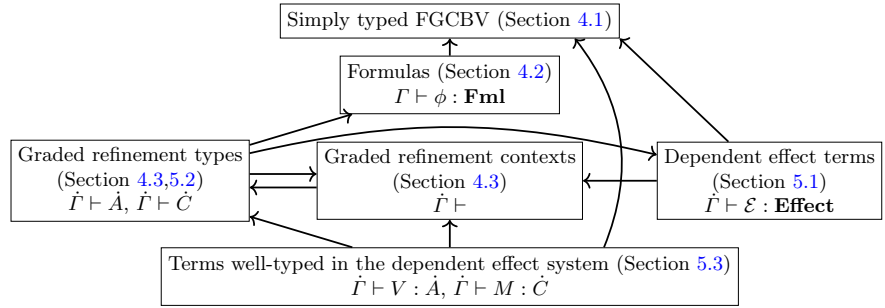
\begin{figure}[tb]
	\centering
	\scalebox{0.85}{%
	\begin{tikzpicture}
		\node[draw] (stfgcbv) at (0, 0) {Simply typed FGCBV (Section~\ref{sec:stfgcbv-semantics})};
		\node[draw, align=center] (formula) at (0, -1) {Formulas (Section~\ref{sec:formula-semantics}) \\ $\Gamma \vdash \phi : \mathbf{Fml}$};
		\node[draw, align=center] (grtype) at (-5, -2.5) {Graded refinement types \\ (Section~\ref{sec:semantics-dependent-refinement},\ref{sec:indexed-graded-monad-definition}) \\ $\dot{\Gamma} \vdash \dot{A}$, $\dot{\Gamma} \vdash \dot{C}$};
		\node[draw, align=center] (grcontext) at (0, -2.5) {Graded refinement contexts \\ (Section~\ref{sec:semantics-dependent-refinement}) \\ $\dot{\Gamma} \vdash$};
		\node[draw, align=center] (depeff) at (5, -2.5) {Dependent effect terms \\ (Section~\ref{sec:indexed-preordered-monoids}) \\ $\dot{\Gamma} \vdash \mathcal{E} : \mathbf{Effect}$};
		\node[draw, align=center] (term) at (0, -4) {Terms well-typed in the dependent effect system (Section~\ref{sec:dependent-effect-system-soundness}) \\ $\dot{\Gamma} \vdash V : \dot{A}$, $\dot{\Gamma} \vdash M : \dot{C}$};
		\draw[->, thick] (formula) -- (stfgcbv);
		\draw[->, thick] (grtype) -- (formula);
		\draw[->, thick] (grtype) edge[transform canvas={yshift=3pt}] (grcontext);
		\draw[->, thick] (grcontext) edge[transform canvas={yshift=-3pt}] (grtype);
		\draw[->, thick] (grtype) edge[bend left=13] (depeff);
		\draw[->, thick] (depeff) -- (grcontext);
		\draw[->, thick] (depeff) -- (stfgcbv.south east);
		\draw[->, thick] (term) -- (grtype);
		\draw[->, thick] (term) -- (grcontext);
		\draw[thick] ([xshift=-60pt]term.north east) edge[->,bend right] ([xshift=-20pt]stfgcbv.south east);
	\end{tikzpicture}}
	\caption{Dependency of the interpretation $\interpret{-}$. The interpretation is defined from top to bottom: (1) the interpretation of simply typed FGCBV is defined first; (2) using the interpretation of value terms in simply typed FGCBV, formulas are interpreted; (3) the interpretation of graded refinement types, graded refinement contexts, and dependent-effect terms are defined by simultaneous induction; and (4) the soundness of the dependent effect system is established by showing the existence of a lifting of the interpretation of terms in simply typed FGCBV.}
	\label{fig:dependency-interpretation}
\end{figure}
}{}

\subsection{Simply Typed FGCBV}\label{sec:stfgcbv-semantics}

Categorical semantics of simply typed FGCBV (Section~\ref{sec:stfgcbv}) is given by a bicartesian closed category with a strong monad \cite{LevyInformationandComputation2003}.

\begin{definition}\label{def:simple-fgcbv-model}
	A \emph{simple FGCBV model} is a tuple $(\category{C}, T, \interpret{{-}})$ where $\category{C}$ is a bicartesian closed category, $T$ is a strong monad on $\category{C}$, and $\interpret{{-}}$ is an interpretation of base types, effect-free operations, and generic effects: $\interpret{b} \in \category{C}$ for each $b \in \mathbf{Base}$, $\interpret{\mathtt{op}} : \interpret{A} \to \interpret{B}$ for each $(\mathtt{op} : A \rightarrowtriangle B) \in \mathbf{Op}$, and $\interpret{\mathtt{gef}} : \interpret{A} \to T \interpret{B}$ for each $(\mathtt{gef} : A \rightarrowtriangle B) \in \mathbf{GenEff}$.
	Here, we implicitly extend the interpretation $\interpret{{-}}$ to ground types $A, B$ by the bicartesian closed structure of $\category{C}$.
	\qed
\end{definition}

Given a FGCBV model $(\category{C}, T, \interpret{{-}})$, we can interpret a well-typed value term $\Gamma \vdash V : A$ as a morphism $\interpret{V} : \interpret{\Gamma} \to \interpret{A}$ in $\category{C}$, and a well-typed computation term $\Gamma \vdash M : T A$ as a morphism $\interpret{M} : \interpret{\Gamma} \to T \interpret{A}$ in $\category{C}$.
The concrete definition of the interpretation $\interpret{{-}}$ is standard and can be found in, e.g., \cite{LevyInformationandComputation2003}.

\begin{example}[cost analysis]\label{ex:cost-analysis-model-wo-recursion}
	A FGCBV model for cost analysis is given as $(\Set, ({-}) \times \mathbb{N}_{\infty}, \interpret{{-}})$.
	Here, $({-}) \times \mathbb{N}_{\infty}$ is the \emph{cost monad} induced by the additive monoid $(\mathbb{N}_{\infty}, 0, {+})$ of extended natural numbers $\mathbb{N}_{\infty} = \mathbb{N} \cup \{ \infty \}$, and it keeps track of the cost of a program.
	The generic effect $\mathtt{Tick} : \NatType \to \UnitType$ is interpreted as a function $\interpret{\mathtt{Tick}} : \mathbb{N} \to 1 \times \mathbb{N}_{\infty}$ that takes $n \in \mathbb{N}$ and returns the pair $((), n)$ of the unit value $()$ and the cost $n$.
	\qed
\end{example}

\subsection{Semantics of Formulas}\label{sec:formula-semantics}

Formulas $\phi$ in refinement types $\{ x : b \mid \phi \}$ are interpreted in a preordered fibration with sufficient structure to interpret logical connectives.

\begin{definition}\label{def:formula-model}
	Let $(\category{C}, T, \interpret{{-}})$ be a simple FGCBV model.
	A \emph{model of formulas} is a pair $(p, \interpret{-})$ of (i) a preordered fibration $p : \category{P} \to \category{C}$ that is fibred bicartesian closed and has simple products $\prod : \category{P}_{I \times X} \to \category{P}_{I}$ satisfying the Beck--Chevalley condition, and (ii) an interpretation $\interpret{a} \in \category{P}_{\interpret{A}}$ for each predicate symbol $\mathtt{a} : A \to \mathbf{Fml}$.
	We often say $p$ is a model of formulas and omit $\interpret{-}$ when it is clear from the context.
	\qed
\end{definition}
Given a model of formulas, a well-formed formula $\Gamma \vdash \phi$ is interpreted as $\interpret{\phi} \in \category{P}_{\interpret{\Gamma}}$.
The definition of the interpretation $\interpret{{-}}$ is straightforward and omitted.
In what follows, we mainly focus on the fibration of $\Omega$-valued predicates defined below, which is sufficient for all instances in Section~\ref{sec:instances}.
However, our dependent effect system and its semantics are not limited to this fibration.

\begin{example}\label{ex:predicates-fibration}
	We define the \emph{predicate fibration} $\mathrm{pred}_{\Set} : \mathbf{Pred} \to \Set$ as follows.
	The total category $\mathbf{Pred}$ is the category of predicates over $\Set$.
	An object in $\mathbf{Pred}$ is a pair $(X, P)$ where $X \in \Set$ and $P \subseteq X$ is a predicate on $X$.
	A morphism from $(X, P)$ to $(Y, Q)$ is a function $f : X \to Y$ that preserves predicates, that is, for any $x \in P$, we have $f(x) \in Q$.
	The forgetful functor $\mathrm{pred}_{\Set} : \mathbf{Pred} \to \Set$ is defined by $\mathrm{pred}_{\Set}(X, P) = X$ and $\mathrm{pred}_{\Set}(f) = f$.
	\qed
\end{example}

\begin{example}\label{ex:omega-valued-predicates-fibration-over-set}
	We consider a generalization of the predicate fibration (Example~\ref{ex:predicates-fibration}) to $\Omega$-valued predicates.
	Let $\Omega$ be a complete Heyting algebra.
	The \emph{$\Omega$-valued predicate fibration} over $\Set$ is defined as follows.
	The total category $\mathbf{Pred}_{\Omega}$ is the category of $\Omega$-valued predicates: an object is a pair $(X, P)$ where $X$ is a set and $P : X \to \Omega$ is a function, and a morphism $f : (X, P) \to (Y, Q)$ is a function $f : X \to Y$ such that $P \le Q \comp f$ with respect to the pointwise order on $\Set(X, \Omega)$.
	Then, the forgetful functor $\mathrm{pred}^{\Omega} : \mathbf{Pred}_{\Omega} \to \Set$ is a fibration that is fibred bicartesian closed and has simple products.
	Note that if $\Omega = \mathbf{2}$ is the two-element Boolean algebra, then this fibration is the same as the predicate fibration in Example~\ref{ex:predicates-fibration}.
	\qed
\end{example}

\begin{example}\label{ex:omega-valued-predicates-fibration}
	We can replace the base category $\Set$ in Example~\ref{ex:omega-valued-predicates-fibration-over-set} with any category $\category{C}$ with terminal object $1$ by applying the change-of-base construction along $\category{C}(1, {-}) : \category{C} \to \Set$.
	This gives us a fibration $\mathrm{pred}^{\Omega}_{\category{C}} : \mathbf{Pred}_{\Omega}(\category{C}) \to \category{C}$ of $\Omega$-valued predicates over $\category{C}$.
	Specifically, an object in $\mathbf{Pred}_{\Omega}(\category{C})$ is a tuple $(I, P)$ where $I \in \category{C}$ and $P : \category{C}(1, I) \to \Omega$ is a function.
	\qed
\end{example}

\subsection{Dependently Typed FGCBV with Refinement Types}\label{sec:semantics-dependent-refinement}
To develop the semantics step by step, we temporarily consider a dependent refinement type system that lies between the simply typed FGCBV (Section~\ref{sec:stfgcbv}) and the dependent effect system (Section~\ref{sec:dependent-effect-system}).
Specifically, we consider types defined as follows.
\begin{align}
	\dot{A}, \dot{B} \ &\coloneqq\ \{ x : b \mid \phi \} \mid (x : \dot{A}) \times \dot{B} \mid (x : \dot{A}) \to \dot{C} \mid \dot{A} + \dot{B} &
	\dot{C}, \dot{D} \ &\coloneqq\ T \dot{A}
\end{align}
We call $\dot{A}$ a \emph{refinement value type} and $\dot{C}$ a \emph{refinement computation type}. 
The difference from Section~\ref{sec:dependent-effect-system} is that computation types $T \dot{A}$ are not graded by dependent effect terms.

We give the semantics of refinement types based on the construction in \cite{KuraFoSSaCS2021}.
The idea is summarized as follows.
Observe that the dependent refinement type system introduced here is a special case of dependently typed FGCBV where types are constructed from $\{ x : b \mid \phi \}$, although we do not spell out the definition of \emph{dependently typed FGCBV}.
Thus, we necessarily use the semantics of dependent type systems as a basis.
Also, simply typed FGCBV (Section~\ref{sec:stfgcbv}) can be regarded as another special case of dependently typed FGCBV where types are constructed from base types $b$.
Thus, the model of the dependent refinement type system should be given as a ``lifting'' of the model of simply typed FGCBV.
The construction in \cite{KuraFoSSaCS2021} gives such a lifting from a model of simply typed FGCBV and a model of formulas (Definition~\ref{def:formula-model}).

More concretely, dependent type systems are interpreted by a \emph{(split) closed comprehension category} (\emph{SCCompC}) \cite[Definition~10.5.3]{Jacobs2001}, which is a fibration $p : \category{E} \to \category{B}$ with certain structures.
Contexts are interpreted as objects in the base category $\category{B}$; and types and terms are interpreted as objects and morphisms in the fibre category $\category{E}_{\interpret{\dot{\Gamma}}}$ over the interpretation of the context.
In the current setting, we need additional structures: strong fibred (binary) coproducts \cite[Exercise~10.5.6]{Jacobs2001} is required to interpret $\dot{A} + \dot{B}$, and a fibred monad (over a fixed base category) is required to interpret computation types $T \dot{A}$ \cite{AhmanFoSSaCS2016,Ahman2017}.
Given a model of simply typed FGCBV $(\category{C}, T, \interpret{-})$, the corresponding SCCompC can be constructed as the simple fibration $s_{\category{C}} : s(\category{C}) \to \category{C}$ \cite[Definition~1.3.1]{Jacobs2001}.
Moreover, a strong monad $T$ on $\category{C}$ bijectively corresponds to a fibred monad $s(T)$ on the simple fibration \cite[Exercise~2.6.10]{Jacobs2001}.
It can be shown that the interpretation of dependently typed FGCBV in the simple fibration $s_{\category{C}} : s(\category{C}) \to \category{C}$ with a fibred monad $s(T)$ corresponds to the standard interpretation of simply typed FGCBV in a bicartesian closed category $\category{C}$ with a strong monad $T$.

The construction in \cite{KuraFoSSaCS2021} combines the simple fibration $s_{\category{C}} : s(\category{C}) \to \category{C}$ with a model of formulas $p : \category{P} \to \category{C}$ and gives us a SCCompC $\{ s(\category{C}) \mid \category{P} \}  \to \category{P}$ with fibred coproducts.
This gives a ``lifting'' of the simple fibration in the sense that there exists a pair of forgetful functors to the simple fibration~\eqref{eq:refinement-fibration-morphism} that strictly preserves the SCCompC structure and fibred coproducts.

\begin{wrapfigure}{r}{0.5\textwidth}
	\vspace{-2.5em}
	\begin{equation}
		\begin{tikzcd}[column sep=large]
			\{ s(\category{C}) \mid \category{P} \} \ar[r] \ar[d, swap, "\{ s_{\category{C}} \mid p \}"] & s(\category{C}) \ar[d, "s_{\category{C}}"] \ar[loop right, looseness=3, "s(T)"] \\
			\category{P} \ar[r, "p"] & \category{C}
		\end{tikzcd}
		\mkern-9mu\label{eq:refinement-fibration-morphism}
	\end{equation}
	\vspace{-2.5em}
\end{wrapfigure}

The construction in \cite{KuraFoSSaCS2021} requires some technical conditions.
Since all the instances in Section~\ref{sec:instances} use $\Omega$-valued predicates (Example~\ref{ex:omega-valued-predicates-fibration}), which automatically satisfy most of the technical conditions, we only state the construction for $\Omega$-valued predicates here.

\begin{proposition}\label{prop:refinement-fibration-omega-valued}
	Let $(\category{C}, T, \interpret{{-}})$ be a simple FGCBV model and $\Omega$ be a complete Heyting algebra such that $\mathrm{pred}^{\Omega}_{\category{C}} : \mathbf{Pred}_{\Omega}(\category{C}) \to \category{C}$ is a model of formulas.
	If\/ $\category{C}(1, {-})$ preserves binary coproducts, then there exists a SCCompC with strong fibred binary coproducts $\{ s_{\category{C}} \mid \mathrm{pred}^{\Omega}_{\category{C}} \} : \{ s(\category{C}) \mid \mathbf{Pred}_{\Omega}(\category{C}) \} \to \mathbf{Pred}_{\Omega}(\category{C})$, which we call the \emph{$\mathrm{pred}^{\Omega}_{\category{C}}$-refinement fibration}.
	Moreover, there exists a morphism of fibrations from $\mathrm{pred}^{\Omega}_{\category{C}}$-refinement fibration to the simple fibration $s_{\category{C}} : s(\category{C}) \to \category{C}$ as in~\eqref{eq:refinement-fibration-morphism}, which strictly preserves the SCCompC structure and fibred binary coproducts \cite{KuraFoSSaCS2021}.
	\qed
\end{proposition}

Consider the situation in Proposition~\ref{prop:refinement-fibration-omega-valued} where $\category{P} = \mathbf{Pred}_{\Omega}(\category{C})$.
If we have a lifting of the fibred monad $s(T)$ along~\eqref{eq:refinement-fibration-morphism} (i.e., a fibred monad on the refinement fibration such that~\eqref{eq:refinement-fibration-morphism} preserves monad structures), then (well-formed) contexts $\dot{\Gamma}$ and types $\dot{A}, \dot{C}$ are interpreted as $\interpret{\dot{\Gamma}} \in \category{P}$ and $\interpret{\dot{\Gamma} \vdash \dot{A}}, \interpret{\dot{\Gamma} \vdash \dot{C}} \in \{ s(\category{C}) \mid \category{P} \}_{\interpret{\dot{\Gamma}}}$, respectively.
A bit more concretely, $\interpret{\dot{\Gamma}} = (\interpret{\underlying{\dot{\Gamma}}}, P)$ gives a pair of the interpretation of the underlying context $\underlying{\dot{\Gamma}}$ and an $\Omega$-valued predicate $P : \category{C}(1, \interpret{\underlying{\dot{\Gamma}}}) \to \Omega$ on the underlying context; and $\interpret{\dot{\Gamma} \vdash \dot{A}} = (\interpret{\dot{\Gamma}}, \interpret{\underlying{\dot{A}}}, Q)$ gives a tuple where $Q : \category{C}(1, \interpret{\underlying{\dot{\Gamma}}} \times \interpret{\underlying{\dot{A}}}) \to \Omega$ is a predicate on the underlying context $\underlying{\dot{\Gamma}}$ and the underlying type $\underlying{\dot{A}}$.
If we apply the functor~\eqref{eq:refinement-fibration-morphism} to these interpretations, then predicates are removed, and we obtain the ``underlying'' or ``simple'' interpretation $\interpret{\underlying{\dot{\Gamma}}} \in \category{C}$ and $(\interpret{\underlying{\dot{\Gamma}}}, \interpret{\underlying{\dot{A}}}) \in s(\category{C})$.
Subtyping relations $\dot{\Gamma} \vdash \dot{A} <: \dot{B}$ are also interpreted in the refinement fibration as a semantic relation $\interpret{\dot{\Gamma} \vdash \dot{A}} <: \interpret{\dot{\Gamma} \vdash \dot{B}}$ defined by $\interpret{\underlying{\dot{A}}} = \interpret{\underlying{\dot{B}}}$ and $Q \le Q'$ where $\interpret{\dot{\Gamma} \vdash \dot{A}} = (\interpret{\dot{\Gamma}},\interpret{\underlying{\dot{A}}}, Q)$ and $\interpret{\dot{\Gamma} \vdash \dot{B}} = (\interpret{\dot{\Gamma}}, \interpret{\underlying{\dot{B}}}, Q')$.
The soundness of the derivation rules for subtyping is straightforward once we define the semantic validity of formulas as follows.
\begin{equation}
	\dot{\Gamma} \vDash \phi \quad\coloneqq\quad \interpret{\dot{\Gamma}} \le \interpret{\phi} \text{ in } \mathbf{Pred}_{\Omega}(\category{C})_{\interpret{\underlying{\dot{\Gamma}}}}
	\label{eq:semantic-validity}
\end{equation}

The interpretation of terms is defined as a lifting of the interpretation of the underlying terms in simply typed FGCBV along~\eqref{eq:refinement-fibration-morphism}.
Since morphisms in the refinement fibration are morphisms in the simple fibration that preserve predicates, the existence of such a lifting implies the soundness of the refinement type system.
Specifically, we have the following.
For any well-typed value term $\dot{\Gamma} \vdash V : \dot{A}$, let $\interpret{V} : \interpret{\underlying{\dot{\Gamma}}} \to \interpret{\underlying{\dot{A}}}$ be the (underlying) interpretation of the term.
Then, $\interpret{V}$ satisfies
$\interpret{\dot{\Gamma}} \le \interpret{\dot{\Gamma}, x : \dot{A}} \comp \category{C}(1, \tupling{\identity{}}{\interpret{V}}) : \category{C}(1, \interpret{\underlying{\dot{\Gamma}}}) \to \Omega$, which means that if the precondition in $\dot{\Gamma}$ holds, then $\interpret{V}$ satisfies the postcondition in $\dot{A}$.
The same holds for computation terms $\dot{\Gamma} \vdash M : \dot{C}$.

\begin{toappendix}
\section{Simply Typed FGCBV as a Special Case of Dependently Typed FGCBV}
We consider a dependently typed version of FGCBV where types are defined as follows.
\begin{align}
	\dot{A}, \dot{B} \quad&\coloneqq\quad \dot{b}(V) \mid (x : \dot{A}) \times \dot{B} \mid (x : \dot{A}) \to \dot{C} \mid \dot{A} + \dot{B} &
	\dot{C}, \dot{D} \quad&\coloneqq\quad T \dot{A}
\end{align}
Here, $\dot{b} : \dot{A} \to \mathbf{Type}$ is a base type constructor.
We also need to modify the definition of effect-free operations, generic effects, and typing rules accordingly, but we omit the details here.

Simply typed FGCBV can be regarded as a special case of dependently typed FGCBV where any base type constructor $\dot{b} : \UnitType \to \mathbf{Type}$ is constant, i.e., $\dot{b}() = b$ for some base type $b$.
There is a semantic counterpart of this situation.
A dependent FGCBV model can be constructed from a simple FGCBV model so that the interpretation in the dependent FGCBV model corresponds to the interpretation in the simple FGCBV model.

\begin{definition}
	Given a category $\category{C}$ with finite products, the \emph{simple fibration} $\mathsf{s}_{\category{C}} : s(\category{C}) \to \category{C}$ is defined as follows.
	An object in $s(\category{C})$ is a pair $(I, X)$ where $I, X \in \category{C}$.
	A morphism from $(I, X)$ to $(J, Y)$ is a pair $(u, f)$ where $u : I \to J$ and $f : I \times X \to Y$ is a morphism in $\category{C}$.
	The functor $s_{\category{C}} : s(\category{C}) \to \category{C}$ is defined by $s_{\category{C}} (I, X) = I$ and $s_{\category{C}} (u, f) = u$.
\end{definition}

\begin{lemma}
	If $\category{C}$ is a cartesian closed category, then the simple fibration $s_{\category{C}} : s(\category{C}) \to \category{C}$ is a SCCompC.
	If $\category{C}$ further has coproducts, then the simple fibration has strong fibred finite coproducts.
\end{lemma}

\begin{lemma}\label{lem:simple-fibration-strong-monad}
	There is a bijection between strong monads on $\category{C}$ and fibred monads on the simple fibration $s(\category{C}) \to \category{C}$ \cite[Exercise~2.6.10]{Jacobs2001}.
	Concretely, given a strong monad $(T, \eta^T, \mu^T)$ on $\category{C}$, we can define a fibred monad $\SimpleFibredMonad{T}$ on the simple fibration as follows.
	\[ \SimpleFibredMonad{T} (I, X) \coloneqq (I, T X) \qquad \eta_{(I, X)} \coloneqq (\identity{I}, \eta^T_X \comp \pi_2) \qquad \mu_{(I, X)} \coloneqq (\identity{I}, \mu^T_X \comp \pi_2) \]
\end{lemma}

By the above lemmas, we can construct a dependent FGCBV model $(s_{\category{C}} : s(\category{C}) \to \category{C}, \SimpleFibredMonad{T}, \interpret{{-}}_d)$ from a simple FGCBV model $(\category{C}, T, \interpret{{-}}_s)$.
It is straightforward to define the interpretation $\interpret{{-}}_d$ of base type constructors, effect-free operations, and generic effects from the interpretation $\interpret{{-}}_s$ in the simple FGCBV model.
For each effect-free operation $\mathtt{op} : A \rightarrowtriangle B$, the interpretation $\interpret{\mathtt{op}}_d \in s(\category{C})_{\interpret{A}}((\interpret{A}, 1), (\interpret{A}, \interpret{B}))$ is defined as $\interpret{\mathtt{op}}_d = (\identity{}, \interpret{\mathtt{op}}_s)$ where the isomorphism $\interpret{A} \times 1 \cong \interpret{A}$ is elided.
The interpretation of generic effects is defined similarly.

\begin{proposition}\label{prop:dependent-fgcbv-simple}
	Suppose that all base type constructors $\dot{b} : \UnitType \to \mathbf{Type}$ are constant.
	Let $(\category{C}, T, \interpret{{-}})$ be a simple FGCBV model, $\interpret{{-}}_s$ be the interpretation of simply typed FGCBV in $(\category{C}, T, \interpret{{-}}_s)$ and $\interpret{{-}}_d$ be the interpretation of dependently typed FGCBV in the simple fibration model $(s_{\category{C}} : s(\category{C}) \to \category{C}, \SimpleFibredMonad{T}, \interpret{{-}}_d)$.
	\begin{itemize}
		\item For any well-formed context $\dot{\Gamma}$, we have
		$\interpret{\dot{\Gamma}}_d = \interpret{\dot{\Gamma}}_s \in \category{C}$.
		\item For any well-formed type $\dot{\Gamma} \vdash \dot{A}$, we have
		$\interpret{\dot{A}}_d = (\interpret{\dot{\Gamma}}_s, \interpret{\dot{A}}_s) \in s(\category{C})_{\interpret{\dot{\Gamma}}_d}$.
		\item For any well-typed value term $\dot{\Gamma} \vdash V : \dot{A}$, we have
		$\interpret{V}_d = (\identity{}, \interpret{V}_s) \in s(\category{C})_{\interpret{\dot{\Gamma}}_d}(1 \interpret{\dot{\Gamma}}_d, \interpret{\dot{A}}_d)$.
	\end{itemize}
\end{proposition}

\end{toappendix}

\section{Graded Monads for Dependent Effect Systems}\label{sec:indexed-graded-monad}

We introduce indexed graded monads, which are a key concept for the semantics of our dependent effect system.
Indexed graded monads have two equivalent presentations in terms of indexed categories and fibrations.
In this section, we present indexed graded monads in terms of indexed categories and leave the fibrational presentation to \referappendix{Appendix}{C.3}{sec:fibrational-presentation}.
Using indexed graded monads, we provide the soundness theorem for the dependent effect system in Section~\ref{sec:type-system}.

\subsection{Indexed Preordered Monoids}\label{sec:indexed-preordered-monoids}

As we have explained in the introduction (Fig.~\ref{fig:models-comparison}), grading monoids for the dependent effect system should be indexed by contexts.
We formalize this idea as \emph{$\category{B}$-indexed preordered monoids}, which is (a specific kind of) a monoidal object in the 2-category $\mathbf{ICat}(\category{B})$ of $\category{B}$-indexed categories.
Below, we borrow notations from fibrations and use them for indexed categories: indexed categories are written as $\category{E}_{({-})} : \category{B}^{\op} \to \mathbf{Cat}$, and for any morphism $u : I \to J$, the functor $\category{E}_u : \category{E}_J \to \category{E}_I$ is written as $u^{*}$ and called the \emph{reindexing functor}.

\begin{definition}[indexed preordered monoid]
	Let $\CategoryOfPreorderedMonoids$ be the category of preordered monoids, i.e., the category whose objects are preordered sets with a monotone monoid structure and whose morphisms are monotone monoid homomorphisms.
	For any category $\category{B}$, a \emph{$\category{B}$-indexed preordered monoid} is a functor $\category{M}_{({-})} : \category{B}^{\op} \to \CategoryOfPreorderedMonoids$.
	That is, for each $I \in \category{B}$, $\category{M}_I = (\category{M}_I, \GradeUnit, ({\GradeMult}), {\le})$ is a preordered monoid, and for each morphism $u : I \to J$ in $\category{B}$, the reindexing functor $u^{*} : \category{M}_J \to \category{M}_I$ is a morphism of preordered monoids.
	\qed
\end{definition}

\begin{toappendix}
	\begin{definition}[indexed monoidal category]
		An \emph{indexed monoidal category} is a monoidal object in the 2-category $\mathbf{Fib}(\category{B})$ of fibrations, fibred functors, and vertical natural transformations.
		Specifically, an indexed monoidal category is a tuple $(m, {\otimes}, I, \alpha, \lambda, \rho)$ where
		\begin{itemize}
			\item $m : \category{M} \to \category{B}$ is a fibration,
			\item $\otimes$ and $I$ are fibred functors, and
			\begin{equation}
				\begin{tikzcd}
					\category{M} \times_{\category{B}} \category{M} \ar[rr, "{\otimes}"] \ar[rd, swap, "m"] & & \category{M} \ar[ld, "m"] \\
					& \category{B}
				\end{tikzcd}
				\qquad
				\begin{tikzcd}
					\category{B} \ar[rr, "I"] \ar[rd, equal] & & \category{M} \ar[ld, "m"] \\
					& \category{B}
				\end{tikzcd}
			\end{equation}
			\item $\alpha : (({-}) \otimes ({-})) \otimes ({-}) \xrightarrow{\cong} ({-}) \otimes (({-}) \otimes ({-}))$, $\lambda : I \otimes ({-}) \xrightarrow{\cong} ({-})$, and $\rho : ({-}) \otimes I \xrightarrow{\cong} ({-})$ are vertical natural isomorphisms such that the triangle and the pentagon identities hold.
			\begin{equation}
				\begin{tikzcd}
					\category{M} \times_{\category{B}} \category{M} \times_{\category{B}} \category{M} \ar[r, "{\otimes} \times_{\category{B}} \category{M}"] \ar[d, swap, "\category{M} \times_{\category{B}} {\otimes}"] & \category{M} \times_{\category{B}} \category{M} \ar[d, "{\otimes}"] \ar[ld, Rightarrow, shorten >=10pt, shorten <=10pt, "\alpha"] \\
					\category{M} \times_{\category{B}} \category{M} \ar[r, swap, "{\otimes}"] & \category{M}
				\end{tikzcd}
				\qquad
				\begin{tikzcd}
					\category{B} \times_{\category{B}} \category{M} \ar[d, ""{name=Iotimes, inner sep=0}, swap, "I \times_{\category{B}} \category{M}"] & \category{M} \ar[l, "\cong"] \ar[d, ""{name=Id, inner sep=0}, equal] \\
					\category{M} \times_{\category{B}} \category{M} \ar[r, "{\otimes}"] & \category{M}
					\ar[Rightarrow, shorten >= 15pt, shorten <= 15pt, from=Iotimes, to=Id, "\lambda"]
				\end{tikzcd}
				\qquad
				\begin{tikzcd}
					\category{M} \times_{\category{B}} \category{B} \ar[d, ""{name=otimesI, inner sep=0}, swap, "\category{M} \times_{\category{B}} I"] & \category{M} \ar[l, "\cong"] \ar[d, ""{name=Id, inner sep=0}, equal] \\
					\category{M} \times_{\category{B}} \category{M} \ar[r, "{\otimes}"] & \category{M}
					\ar[Rightarrow, shorten >= 15pt, shorten <= 15pt, from=otimesI, to=Id, "\rho"]
				\end{tikzcd}
			\end{equation}
		\end{itemize}
		We often say ``$m$ is an indexed monoidal category'' to mean that $(m, {\otimes}, I, \alpha, \lambda, \rho)$ is an indexed monoidal category.
		We say an indexed monoidal category is \emph{strict} if $\alpha, \lambda, \rho$ are identities.
		We also say an indexed monoidal category is \emph{preordered} if the fibration $m$ is preordered.
		Note that $m$ is strict if $m$ is preordered.
	\end{definition}	
\end{toappendix}

\begin{example}[preordered-monoid-valued functions]\label{ex:preordered-monoid-valued-functions}
	For any preordered monoid $\GradingMonoid$, the functor $\Set({-}, \GradingMonoid) : \Set^{\op} \to \CategoryOfPreorderedMonoids$ is a $\Set$-indexed preordered monoid.
	Concretely, for any set $X$, the preordered monoid $\Set(X, \GradingMonoid)$ is the set of functions from $X$ to $\GradingMonoid$ with pointwise order and pointwise multiplication.
	For each morphism $f : X \to Y$ in $\Set$, the reindexing $f^{*} = \Set(f, \GradingMonoid) : \Set(Y, \GradingMonoid) \to \Set(X, \GradingMonoid)$ is defined by precomposition with $f$.
	\qed
\end{example}

\begin{example}[simple effect]\label{ex:simple-effect-indexed-preordered-monoid}
	Let $\GradingMonoid$ be a preordered monoid and $\category{B}$ be a category.
	The constant functor $\Delta \GradingMonoid : \category{B}^{\op} \to \CategoryOfPreorderedMonoids$ is a $\category{B}$-indexed preordered monoid.
	This models the \emph{simple effect} where the grading monoid is constant for all contexts.
	Note that $\Delta \GradingMonoid : \Set^{\op} \to \CategoryOfPreorderedMonoids$ is an indexed preordered \emph{sub}-monoid of $\Set({-}, \GradingMonoid)$ in Example~\ref{ex:preordered-monoid-valued-functions}: $\GradingMonoid \cong \{ f : X \to \GradingMonoid \mid \text{$f$ is constant} \} \subseteq \Set(X, \GradingMonoid)$.
	\qed
\end{example}

The indexed preordered monoid in Example~\ref{ex:preordered-monoid-valued-functions} provides a natural semantics of dependent effect terms if we ignore refinement types.
Consider as an example cost analysis where the grading monoid is $\GradingMonoid = (\mathbb{N}_{\infty}, 0, {+})$.
Let $n : \NatType \vdash \mathtt{nat2eff}(n + 1) : \mathbf{Effect}$ be a dependent effect term.
Then, the interpretation is given by the composite $\interpret{\mathtt{nat2eff}} \comp \interpret{n + 1} \in \Set(\interpret{\NatType}, \GradingMonoid)$ of the interpretation of the term $\interpret{n + 1} : \mathbb{N} \to \mathbb{N}$ and the interpretation of the basic effect $\interpret{\mathtt{nat2eff}} : \mathbb{N} \to \GradingMonoid$.

However, the indexed preordered monoid in Example~\ref{ex:preordered-monoid-valued-functions} has a few problems when we consider refinement types.
The first problem is the mismatch of the indexing categories.
We would like to interpret dependent effect terms $\dot{\Gamma} \vdash \mathcal{E} : \mathbf{Effect}$ as $\interpret{\mathcal{E}} \in \category{M}_{\interpret{\dot{\Gamma}}}$.
This means that the indexing category of indexed preordered monoids $\category{M}_{({-})}$ must be the category of $\Omega$-valued predicates where $\interpret{\dot{\Gamma}}$ lives.
However, the indexing category in Example~\ref{ex:preordered-monoid-valued-functions} is $\Set$.
The second problem is about the semantic subeffecting relation.
We would like to interpret the semantic subeffecting relation $\dot{\Gamma} \vDash \mathcal{E}_1 \le \mathcal{E}_2$ as the order relation $\interpret{\mathcal{E}_1} \le \interpret{\mathcal{E}_2}$ in $\category{M}_{\interpret{\dot{\Gamma}}}$.
It should take the predicates in $\dot{\Gamma}$ into account, as we have explained in Example~\ref{ex:subtyping}.
However, the indexed preordered monoid in Example~\ref{ex:preordered-monoid-valued-functions} does not provide an appropriate interpretation.

We solve these problems as follows.
The first problem can be solved by applying the change-of-base construction along the forgetful functor $\mathbf{Pred}_{\Omega}(\category{C}) \to \mathbf{Pred}_{\Omega} \to \Set$ to the indexed preordered monoid in Example~\ref{ex:preordered-monoid-valued-functions}.
In addition, we replace the order relation with the pointwise order restricted by predicates to solve the second problem.
This gives the following indexed preordered monoid, which we use for the semantics of dependent effect terms.

\begin{definition}\label{def:indexed-graded-monoid-over-omega-valued-predicates}
	Let $\GradingMonoid$ be a preordered monoid.
	We define a $\mathbf{Pred}_{\Omega}$-indexed preordered monoid by $(\Set \sslash_{\Omega} \GradingMonoid)_{(I, P)} \coloneqq (\Set(I, \GradingMonoid), \GradeUnit, ({\GradeMult}), {\le_P})$ for $(I, P) \in \mathbf{Pred}_{\Omega}$ where the unit and the multiplication of $\GradingMonoid$ is extended to $\Set(I, \GradingMonoid)$ pointwise, and the order relation $\le_P$ is defined as follows.
	\[ f \le_P g \quad\coloneqq\quad \forall i \in I, P(i) \neq \bot_{\Omega} \implies f(i) \le g(i) \]
	Here, $\bot_{\Omega} \in \Omega$ is the least element.
	This order relation is a bit ad hoc, but it ensures that the construction in Proposition~\ref{prop:indexed-graded-monad-omega-valued} below gives an indexed graded monad.
	Note that if $\Omega = \mathbf{2}$ is the two-element Boolean algebra, then we have $f \le_P g$ if and only if $f(i) \le g(i)$ for all $i \in P$.
	We also define a $\mathbf{Pred}_{\Omega}(\category{C})$-indexed preordered monoid $(\category{C} \sslash_{\Omega} \GradingMonoid)_{({-})}$ by the change-of-base along the forgetful functor $\mathbf{Pred}_{\Omega}(\category{C}) \to \mathbf{Pred}_{\Omega}$.
	\qed
\end{definition}

\begin{example}
	Let $\Omega = \mathbf{2}$.
	Consider the dependent effect term $x : \{ x : \IntType \mid x \ge 1 \} \vdash |x + 1| : \mathbf{Effect}$ for cost analysis where a grading monoid is given by $\mathbb{N} = (\mathbb{N}, 0, {+})$.
	This dependent effect term can be interpreted as $\interpret{|x + 1|} = \lambda x. |x + 1| \in (\Set \sslash_{\mathbf{2}} \mathbb{N})_{(\mathbb{Z}, \{ x \in \mathbb{Z} \mid x \ge 1 \})}$.
	The semantic subeffecting relation $x : \{ x : \IntType \mid x \ge 1 \} \vDash |x + 1| \le |2 x|$ in Example~\ref{ex:subtyping} is valid in this setting because $|x + 1| \le |2 x|$ holds for any $x \ge 1$.
	\qed
\end{example}

\subsection{Indexed Graded Monads}\label{sec:indexed-graded-monad-definition}

Given an interpretation of $\interpret{\dot{\Gamma} \vdash \dot{A}} \in \{ s(\category{C}) \mid \category{P} \}_{\interpret{\dot{\Gamma}}}$ and $\interpret{\dot{\Gamma} \vdash \mathcal{E}} \in \category{M}_{\interpret{\dot{\Gamma}}}$ where $\interpret{\dot{\Gamma}} \in \category{P} = \mathbf{Pred}_{\Omega}(\category{C})$, we would like to interpret a computation type $\dot{\Gamma} \vdash T_{\mathcal{E}} \dot{A}$ as an object in $\{ s(\category{C}) \mid \category{P} \}_{\interpret{\dot{\Gamma}}}$.
For this purpose, we introduce below a novel notion of \emph{indexed graded monads}.
Then, an indexed graded monad $T : \category{M}_{\interpret{\dot{\Gamma}}} \times \{ s(\category{C}) \mid \category{P} \}_{\interpret{\dot{\Gamma}}} \to \{ s(\category{C}) \mid \category{P} \}_{\interpret{\dot{\Gamma}}}$ indexed by $\interpret{\dot{\Gamma}} \in \category{P}$ naturally gives the interpretation of the computation type $\dot{\Gamma} \vdash T_{\mathcal{E}} \dot{A}$ as $T_{\interpret{\dot{\Gamma} \vdash \mathcal{E}}} \interpret{\dot{\Gamma} \vdash \dot{A}}$.

\begin{toappendix}
\begin{definition}[morphism of graded monads]\label{def:morphism-of-graded-monads}
	Let $(S, \eta^S, \mu^S)$ be an $\mathcal{M}$-graded monad on a category $\category{C}$ and $(T, \eta^T, \mu^T)$ be an $\mathcal{N}$-graded monad on a category $\category{D}$.
	A \emph{morphism of graded monads} from $(S, \eta^S, \mu^S)$ to $(T, \eta^T, \mu^T)$ is a pair of a preordered-monoid morphism $H : \mathcal{M} \to \mathcal{N}$ and a functor $F : \category{C} \to \category{D}$ that preserves the graded monad structure:
	for any $m, m_1, m_2 \in \mathcal{M}$ and $X \in \category{C}$,
	\begin{gather}
		F (S_m X) = T_{H m} F X, \qquad\qquad\qquad
		F \eta^S_X = \eta^T_{F X} \quad:\quad F X \to T_{\GradeUnit} F X, \\
		F \mu^S_{m_1, m_2, X} = \mu^T_{H m_1, H m_2, F X} \quad:\quad F S_{m_1} S_{m_2} X \to T_{H m_1 \GradeMult H m_2} F X.
	\end{gather}
\end{definition}
\end{toappendix}

\begin{definition}[indexed graded monad]
	A \emph{$\category{B}$-indexed graded monad} is a functor from $\category{B}^{\op}$ to the category of graded monads.
	Concretely, a $\category{B}$-indexed graded monad consists of the following data.
	\begin{itemize}
		\item An indexed preordered monoid $\category{M}_{({-})} : \category{B}^{\op} \to \mathbf{Cat}$.
		This is the grading indexed preordered monoid of the indexed graded monad.
		We sometimes say ``indexed $\category{M}_{({-})}$-graded monad'' to make the grading indexed preordered monoid explicit. 
		\item An indexed category $\category{E}_{({-})} : \category{B}^{\op} \to \mathbf{Cat}$ on which the indexed graded monad is defined.
		\item For each object $I \in \category{B}$, an $\category{M}_I$-graded monad $(T I, \eta_I, \mu_I)$ on the category $\category{E}_I$.
		We often omit the index $I \in \category{B}$ in $(T I, \eta_I, \mu_I)$ when clear from the context.
		\item For each morphism $u : I \to J$ in $\category{B}$, the pair of reindexing functors $u^{*} : \category{M}_J \to \category{M}_I$ and $u^{*} : \category{E}_J \to \category{E}_I$ that is a morphism of graded monads (defined in \referappendix{Definition}{47}{def:morphism-of-graded-monads}) from $(T J, \eta_J, \mu_J)$ to $(T I, \eta_I, \mu_I)$.
		That is, for any $m, m_1, m_2 \in \category{M}_J$ and $X \in \category{E}_J$, we have $u^{*} T_m X = T_{u^{*} m} u^{*} X$, $u^{*} \eta_{X} = \eta_{u^{*} X}$, and $u^{*} \mu_{m_1, m_2, X} = \mu_{u^{*} m_1, u^{*} m_2, u^{*} X}$. \qed
	\end{itemize}
\end{definition}

\begin{example}[indexed graded monad for indexed sets]\label{ex:indexed-graded-monad-indexed-sets}
	Indexed sets (or equivalently, the family fibration~\cite[Definition~1.2.1]{Jacobs2001}) give a typical model of dependent type systems \emph{without} refinement types.
	We will not use this model for our dependent effect system, but there is a simple example of an indexed graded monad defined on indexed sets.
	Now, suppose that we have an $\GradingMonoid$-graded monad $T = (T, \eta^T, \mu^T)$ on $\Set$.
	For each $f \in \Set(I, \GradingMonoid)$ and $\{ X_i \}_{i \in I} \in \mathbf{Fam}(\Set)_I$, we define $\hat{T}_f (\{ X_i \}_{i \in I}) \in \mathbf{Fam}(\Set)_I$ as $\hat{T}_f (\{ X_i \}_{i \in I}) \coloneqq \{ T_{f(i)} X_i \}_{i \in I}$.
	Then, $\hat{T}$ is an indexed $\Set({-}, \GradingMonoid)$-graded monad.
	The unit and the multiplication for $\hat{T}$ are defined as expected.
	Example~\ref{ex:indexed-graded-cost-monad-on-fam} is an instance of this construction.
	\qed
\end{example}

The following indexed graded monad defined on the refinement fibration $\{ s_{\category{C}} \mid \mathrm{pred}^{\Omega}_{\category{C}} \} : \{ s(\category{C}) \mid \mathbf{Pred}_{\Omega}(\category{C}) \} \to \mathbf{Pred}_{\Omega}(\category{C})$ will be used for the semantics of our dependent effect system.
\begin{proposition}\label{prop:indexed-graded-monad-omega-valued}
	Suppose that we have a strong $\GradingMonoid$-graded monad lifting $\ddot{T}$ of a strong monad $T$ along $\mathrm{pred}^{\Omega}_{\category{C}} : \mathbf{Pred}_{\Omega}(\category{C}) \to \category{C}$.
	Then, we can construct an indexed $(\category{C} \sslash_{\Omega} \GradingMonoid)_{({-})}$-graded monad $\DependentEffectConstruction{\ddot{T}}$ over the $\mathrm{pred}^{\Omega}_{\category{C}}$-refinement fibration $\{ s_{\category{C}} \mid \mathrm{pred}^{\Omega}_{\category{C}} \} : \{ s(\category{C}) \mid \mathbf{Pred}_{\Omega}(\category{C}) \} \to \mathbf{Pred}_{\Omega}(\category{C})$ as follows.
	\[ \DependentEffectConstruction{\ddot{T}}_f ((I, P), X, Q) \quad\coloneqq\quad ((I, P), T X, \lambda \tupling{i}{x}. P(i) \land (\ddot{T}_{f(i)} Q \tupling{i}{{-}})(x)) \]
	Here, $I, X \in \category{C}$, $P \in \mathbf{Pred}_{\Omega}(\category{C})_I$, $Q \in \mathbf{Pred}_{\Omega}(\category{C})_{I \times X}$, $f \in (\category{C} \sslash_{\Omega} \GradingMonoid)_{(I, P)}$, $i \in \category{C}(1, I)$, and $x \in \category{C}(1, T X)$.
	Moreover, $\DependentEffectConstruction{\ddot{T}}$ is a lifting of the fibred monad $\SimpleFibredMonad{T}$ along the morphism~\eqref{eq:refinement-fibration-morphism}.
	\qed
\end{proposition}
\begin{appendixproof}[Proof of Proposition~\ref{prop:indexed-graded-monad-omega-valued}]
	We define
	\[ \DependentEffectConstruction{\ddot{T}}_{(I, P, f)} (I, X, P, Q) \quad\coloneqq\quad (I, T X, P, \lambda \tupling{i}{x}. P(i) \land (\dot{T}_{f(i)} Q \tupling{i}{{-}})(x)) \]
	\begin{itemize}
		\item (Functoriality): We show that for each $(I, P) \in \mathbf{Pred}_{\Omega}(\category{C})$, the mapping $\ddot{T}$ defines a functor $\ddot{T} : \{ s(\category{C}) \mid \mathbf{Pred}_{\Omega}(\category{C}) \}_{(I, P)} \times (\category{C} \sslash_{\Omega} \GradingMonoid)_{(I, P)} \to \{ s(\category{C}) \mid \mathbf{Pred}_{\Omega}(\category{C}) \}_{(I, P)}$.
		It suffices to show that for any morphism $(\identity{}, h) : (I, X, P, Q) \to (I, Y, P, R)$ and $f \le_P g$, we have $\ddot{T}_{f \le_P g} h = (\identity{}, T h \comp \strength^T) : \ddot{T}_f (I, X, P, Q) \to \ddot{T}_g (I, Y, P, R)$ as a morphism in $\{ s(\category{C}) \mid \mathbf{Pred}_{\Omega}(\category{C}) \}_{(I, P)}$.
		It suffices to prove that for any $\tupling{i}{x} \in \category{C}(1, I \times T X)$, we have
		\[ P(i) \land (\dot{T}_{f(i)} Q \tupling{i}{{-}})(x) \quad\le\quad P(i) \land (\dot{T}_{g(i)} R \tupling{i}{{-}})(T h \comp \strength^T \comp \tupling{i}{x}). \]
		If $P(i) = \bot_{\Omega}$, then both sides are $\bot_{\Omega}$.
		Suppose $P(i) \neq \bot_{\Omega}$.
		Since the strength $\strength^T$ satisfies $\strength^T \comp \tupling{i}{x} = T \tupling{i \comp {!}}{\identity{}} \comp x$, it suffices to show
		\[ (\dot{T}_{f(i)} Q \tupling{i}{{-}})(x) \quad\le\quad (\dot{T}_{g(i)} R \tupling{i}{{-}})(T (h \comp \tupling{i \comp {!}}{\identity{}}) \comp x). \]
		Observe that we have
		\[ h \comp \tupling{i \comp {!}}{\identity{}} : Q \tupling{i}{{-}} \to R \tupling{i}{{-}} \]
		because for any $x : 1 \to X$,
		\[ Q \tupling{i}{x} \le R (\tupling{\pi_1}{h} \comp \tupling{i}{x}) = R \tupling{i}{h \comp \tupling{i \comp {!}}{\identity{}} \comp x} \]
		by definition of $(\identity{}, h) : (I, X, P, Q) \to (I, Y, P, R)$ and $f \le_P g$.
		Since $\dot{T}$ is a graded monad lifting, we have
		\[ T (h \comp \tupling{i \comp {!}}{\identity{}}) : \dot{T}_{f(i)} Q \tupling{i}{{-}} \to \dot{T}_{g(i)} R \tupling{i}{{-}} \]
		in $\mathbf{Pred}_{\Omega}(\category{C})$.
		\item (Unit): We show
		\[ (\identity{}, \eta^T \comp \pi_2) : (I, X, P, Q) \to \ddot{T}_0 (I, X, P, Q) \]
		is a morphism.
		It suffices to prove
		\[ Q \tupling{i}{x} \le P(i) \land (\dot{T}_{0} Q \tupling{i}{{-}})(\eta^T \comp x) \]
		for any $\tupling{i}{x} \in \category{C}(1, I \times X)$.
		We have $Q \tupling{i}{x} \le P(i)$ by definition.
		We also have $Q \tupling{i}{x} \le (\dot{T}_{0} Q \tupling{i}{{-}})(\eta^T \comp x)$ because $\dot{T}$ is a graded monad lifting of $T$, and we have $\eta^T : Q \tupling{i}{{-}} \le \dot{T}_{0} Q \tupling{i}{{-}}$.
		\item (Multiplication): We show
		\[ (\identity{}, \mu^T \comp \pi_2) : \ddot{T}_{f} \ddot{T}_{g} (I, X, P, Q) \to \ddot{T}_{f + g} (I, X, P, Q) \]
		is a morphism.
		It suffices to prove
		\[ P(i) \land (\dot{T}_{f(i)} (\lambda y. P(i) \land \dot{T}_{g(i)} Q \tupling{i}{y}))(x) \le P(i) \land (\dot{T}_{f(i) + g(i)} Q \tupling{i}{{-}})(\mu^T \comp x) \]
		for any $\tupling{i}{x} \in \category{C}(1, I \times T^2 X)$.
		If $P(i) = \bot_{\Omega}$, then both sides are $\bot_{\Omega}$.
		Suppose $P(i) \neq \bot_{\Omega}$.
		Since we have
		\[ (\dot{T}_{f(i)} (\lambda y. P(i) \land \dot{T}_{g(i)} Q \tupling{i}{y}))(x) \quad\le\quad (\dot{T}_{f(i)} (\dot{T}_{g(i)} Q \tupling{i}{{-}}))(x), \]
		it suffices to prove
		\[ (\dot{T}_{f(i)} (\dot{T}_{g(i)} Q \tupling{i}{{-}}))(x) \le (\dot{T}_{f(i) + g(i)} Q \tupling{i}{{-}})(\mu^T \comp x), \]
		which follows from $\mu^T : \dot{T}_{f(i)} (\dot{T}_{g(i)} Q \tupling{i}{{-}}) \to \dot{T}_{f(i) + g(i)} Q \tupling{i}{{-}}$.
		\item (Reindexing): Let $u : (J, Q) \to (I, P)$.
		We show
		\[ u^{*} \ddot{T}_f (I, X, P, R) = \ddot{T}_{u^{*} f} u^{*} (I, X, P, R). \]
		By unfolding the definition, we obtain the following.
		\begin{align}
			u^{*} \ddot{T}_f (I, X, P, R) &= (J, T X, Q, \lambda \tupling{j}{x}. Q(j) \land P(u \comp j) \land (\dot{T}_{f(u \comp j)} R \tupling{u \comp j}{{-}})(x)) \\
			&= (J, T X, Q, \lambda \tupling{j}{x}. Q(j) \land (\dot{T}_{f(u \comp j)} R \tupling{u \comp j}{{-}})(x)) \\
			\ddot{T}_{u^{*} f} u^{*} (I, X, P, R) &= \ddot{T}_{f(u \comp {-})} (J, X, Q, \lambda \tupling{j}{x}. Q(j) \land R \tupling{u \comp j}{x}) \\
			&= (J, T X, Q, \lambda \tupling{j}{x}. Q(j) \land (\dot{T}_{f(u \comp j)}(\lambda y. Q(j) \land R \tupling{u \comp j}{y}))(x))
		\end{align}
		To show that these two are equal, it suffices to show
		\[ Q' \land (\dot{T}_m R')(x) \le \dot{T}_m (\lambda y. Q' \land R'(y))(x) \]
		for any $Q' \in (\mathbf{Pred}_{\Omega}(\category{C}))_1$ and $R' \in (\mathbf{Pred}_{\Omega}(\category{C}))_X$ and $x \in \category{C}(1, X)$.
		This is proved as follows.
		\begin{align}
			Q' \land (\dot{T}_m R')(x) &= (Q' \dotTimes \dot{T}_m R')(\tupling{{!}}{\identity{}} \comp x) \\
			&\le \dot{T}_m (Q' \dotTimes R')(\strength^T \comp \tupling{{!}}{\identity{}} \comp x) && \text{because $\dot{T}$ is a strong lifting} \\
			&= \dot{T}_m (Q' \dotTimes R')(T \tupling{{!}}{\identity{}} \comp x) \\
			&= \dot{T}_m (\lambda y. Q' \land R'(y))(x) && \text{because $T \tupling{{!}}{\identity{}}$ is an isomorphism}
		\end{align}
		Here, $\tupling{{!}}{\identity{}} : (\lambda y. Q' \land R'(y)) \to Q' \dotTimes R'$ is an isomorphism, and thus, $T \tupling{{!}}{\identity{}} : \dot{T}_m (\lambda y. Q' \land R'(y)) \to \dot{T}_m (Q' \dotTimes R')$ is also an isomorphism.
	\end{itemize}
\end{appendixproof}

Proposition~\ref{prop:indexed-graded-monad-omega-valued} can be viewed as a construction of a model of a \emph{dependent} effect system from that of a \emph{simple} effect system, since a simple effect system is typically modelled by a strong graded monad lifting along a fibration of (some kind of) predicates (e.g., \cite{AguirreICFP2021}).
The models of all instances in Section~\ref{sec:instances} are obtained by applying this construction to suitable strong graded monad liftings.

\subsection{Soundness of the Dependent Effect System}\label{sec:dependent-effect-system-soundness}

We define models of our dependent effect system defined in Section~\ref{sec:dependent-effect-system} and prove its soundness.
The models defined below are specialized for the constructions in Definition~\ref{def:indexed-graded-monoid-over-omega-valued-predicates} and Proposition~\ref{prop:indexed-graded-monad-omega-valued}, but it is also possible to define models in a more general setting, as we remark in Remark~\ref{rem:generalizations-of-soundness}.
\begin{definition}\label{def:dependent-effect-system-model}
	A \emph{model of the dependent effect system} consists of the following.
	\begin{itemize}
		\item A simple FGCBV model $(\category{C}, T, \interpret{{-}}_s)$ (Definition~\ref{def:simple-fgcbv-model}). The subscript $s$ indicates that $\interpret{{-}}_s$ is the interpretation for the underlying simple type system.
		\item A complete Heyting algebra $\Omega$ and an interpretation $\interpret{a} \in \mathbf{Pred}_{\Omega}(\category{C})_{\interpret{A}}$ for each predicate symbol $\mathtt{a} : A \to \mathbf{Fml}$, which make the $\Omega$-valued predicate fibration $\mathrm{pred}^{\Omega}_{\category{C}} : \mathbf{Pred}_{\Omega}(\category{C}) \to \category{C}$ (Example~\ref{ex:omega-valued-predicates-fibration}) a model of formulas.
		\item A preordered monoid $\GradingMonoid$ and an interpretation $\interpret{\mathtt{be}} \in (\category{C} \sslash_{\Omega} \GradingMonoid)_{\top \interpret{A}}$ for each basic effect symbol $\mathtt{be} : A \to \mathbf{Effect}$, which make the indexed preordered monoid $(\category{C} \sslash_{\Omega} \GradingMonoid)_{({-})}$ (Definition~\ref{def:indexed-graded-monoid-over-omega-valued-predicates}) a model of dependent effect terms.
		Here, $\top : \category{C} \to \mathbf{Pred}_{\Omega}(\category{C})$ is the terminal object functor, which maps $X \in \category{C}$ to the $\Omega$-valued predicate $\top X$ defined by $\top X (x) = \top_{\Omega}$ for any $x \in \category{C}(1, X)$.
		In this situation, the semantic subeffecting relation is defined as follows.
		\begin{equation}
			\dot{\Gamma} \vDash \mathcal{E}_1 \le \mathcal{E}_2 \quad\coloneqq\quad \text{$\interpret{\mathcal{E}_1} \le \interpret{\mathcal{E}_2}$ in $(\category{C} \sslash_{\Omega} \GradingMonoid)_{\interpret{\dot{\Gamma}}}$}
			\label{eq:semantic-subeffecting}
		\end{equation}
		\item A strong $\GradingMonoid$-graded monad lifting $\ddot{T}$ of $T$ along $\mathrm{pred}^{\Omega}_{\category{C}} : \mathbf{Pred}_{\Omega}(\category{C}) \to \category{C}$ (cf.\ Proposition~\ref{prop:indexed-graded-monad-omega-valued}).
	\end{itemize}
	These data are required to satisfy the following properties.
	\begin{enumerate}[align=left,labelwidth=1ex,label=Ax~\arabic*]
		\item For each effect-free operation $\mathtt{op} : (x : \dot{A}) \rightarrowtriangle \dot{B}$ and generic effect $\mathtt{gef} : (x : \dot{A}) \stackrel{\mathcal{E}}{\rightarrowtriangle} \dot{B}$, the interpretations $\interpret{\mathtt{op}}_s$ and $\interpret{\mathtt{gef}}_s$ have the following liftings along $\mathrm{pred}^{\Omega}_{\category{C}} : \mathbf{Pred}_{\Omega}(\category{C}) \to \category{C}$.
		\[ \tupling{\identity{}}{\interpret{\mathtt{op}}_s} : \interpret{x : \dot{A}} \dotTo \interpret{x : \dot{A}, y : \dot{B}},\ \tupling{\identity{}}{\interpret{\mathtt{gef}}_s} : \interpret{x : \dot{A}} \dotTo \interpret{x : \dot{A}, y : T_{\mathcal{E}} \dot{B}} \]
		Here, for any $\Omega$-valued predicate $X, Y$ in $\mathbf{Pred}_{\Omega}(\category{C})$ and a morphism $f$ in $\category{C}$, we write $f : X \dotTo Y$ if there exists a morphism (a lifting) $g : X \to Y$ in $\mathbf{Pred}_{\Omega}(\category{C})$ such that $\mathrm{pred}^{\Omega}_{\category{C}}(g) = f$.
		\label{item:model-axiom-lifting}
		\item The functor $\category{C}(1, {-})$ preserves binary coproducts (cf.\ Prop.~\ref{prop:refinement-fibration-omega-valued}).
		\label{item:model-axiom-coproduct}
		\qed
	\end{enumerate}
\end{definition}

\begin{theorem}[soundness]\label{thm:soundness}
	Given a model as in Definition~\ref{def:dependent-effect-system-model}, if $\dot{\Gamma} \vdash M : \dot{C}$ is well-typed and recursion-free, then there exists a lifting $\tupling{\identity{}}{\interpret{M}_s} : \interpret{\dot{\Gamma}} \dotTo \interpret{\dot{\Gamma}, x : \dot{C}}$ along $\mathrm{pred}^{\Omega}_{\category{C}} : \mathbf{Pred}_{\Omega}(\category{C}) \to \category{C}$ where $x$ is a fresh variable, and similarly for well-typed value terms $\dot{\Gamma} \vdash V : \dot{A}$.
	\ifthenelse{\boolean{longversion}}{%
	\begin{equation}
		\begin{tikzcd}[row sep=small,ampersand replacement=\&]
			\mathbf{Pred}_{\Omega}(\category{C}) \ar[d, swap, "{\mathrm{pred}^{\Omega}_{\category{C}}}"] \& \interpret{\dot{\Gamma}} \ar[r, "\exists"] \& \interpret{\dot{\Gamma}, x : \dot{C}} \\
			\category{C} \&  \interpret{\underlying{\dot{\Gamma}}}_s \ar[r, "{\tupling{\identity{}}{\interpret{M}_s}}"] \& \interpret{\underlying{\dot{\Gamma}}}_s \times \interpret{\underlying{\dot{C}}}_s
		\end{tikzcd}
	\end{equation}%
	}{}
\end{theorem}
\begin{proof}
	By induction on type derivation.
	See \referappendix{Theorem}{51}{thm:soundness-wo-recursion-terms} for details.
	\qed
\end{proof}
The above theorem states that if $\dot{\Gamma} \vdash M : \dot{C}$ is well-typed, then for any element $\gamma \in \category{C}(1, \interpret{\dot{\Gamma}})$, we have $\interpret{\dot{\Gamma}}(\gamma) \le \interpret{\dot{\Gamma}, x : \dot{C}}(\gamma, \interpret{M}_s(\gamma))$ with respect to the order relation in $\Omega$.
In particular, when $\Omega = \mathbf{2}$, for any $\gamma$ satisfying the precondition $\interpret{\dot{\Gamma}}$, the soundness theorem guarantees that the result of the computation $\interpret{M}_s(\gamma)$ satisfies the postcondition $\interpret{\dot{\Gamma}, x : \dot{C}}(\gamma, {-})$.

In Theorem~\ref{thm:soundness}, we restrict attention to terms without recursion, purely for simplicity of presentation.
Extending the soundness theorem to terms with recursion is relatively straightforward, and we present such an extension in \referappendix{Theorem}{79}{thm:soundness-w-recursion}.
The main idea is as follows.
We interpret the underlying simple type system with recursion using $\omega$CPO-enriched models (cf.\ \referappendix{Definition}{70}{def:omegaCPO-enriched-fgcbv-model}).
By imposing suitable admissibility conditions on graded monads, the construction in Proposition~\ref{prop:indexed-graded-monad-omega-valued} gives a model of the dependent effect system with recursion.
In Section~\ref{sec:instances}, we use the soundness theorem with recursion to reason about programs with recursive functions.

\begin{remark}\label{rem:generalizations-of-soundness}
	Theorem~\ref{thm:soundness} is specialized to models constructed from Proposition~\ref{prop:indexed-graded-monad-omega-valued}, since this construction already covers all the instances in Section~\ref{sec:instances}.
	Nevertheless, the soundness theorem can be stated in a more general setting.
	We present such a formulation in \referappendix{Theorem}{51}{thm:soundness-wo-recursion-terms}, using more general predicate fibrations, indexed preordered monoids, and indexed graded monads.
	Moreover, our semantic framework based on indexed graded monads can be also applied to dependent effect systems \emph{without} refinement types by replacing refinement fibrations with more general SCCompCs.
	However, we focused on the type system with refinement types in this paper because we aim at refinement-type-based automated verification tools in the future.
\end{remark}

\begin{toappendix}

\subsection{Without Recursion}

We divide Definition~\ref{def:dependent-effect-system-model} into three parts.
Using the first part, we define the semantics of contexts and types.
Then, we extend the semantics to terms without recursion by adding the second part.
Finally, we extend the semantics to terms with recursion by adding the third part.
We also slightly generalize models by allowing arbitrary models of formulas instead of restricting to $\Omega$-valued predicate fibrations.

\begin{definition}\label{def:dependent-effect-system-model-part1}
	We consider the following data that are a part of a model of the dependent effect system (Definition~\ref{def:dependent-effect-system-model}).
	We call these data a \emph{pre-model of the dependent effect system without recursion}.
	\begin{itemize}
		\item A simple FGCBV model $(\category{C}, T, \interpret{{-}}_s)$ (Definition~\ref{def:simple-fgcbv-model}).
		\item A model of formulas (Definition~\ref{def:formula-model}) $p : \category{P} \to \category{C}$.
		\item A model of dependent effect terms (Definition~\ref{def:dependent-effect-system-model}) $m : \category{M} \to \category{P}$ together with the interpretation of basic effects $\interpret{\mathtt{be}} \in \category{M}_{\top \interpret{A}}$ for each basic effect $\mathtt{be} : A \to \mathbf{Effect}$.
		\item A fibred $m$-graded monad $\dot{T}$ on the $p$-refinement fibration $\{ s(\category{C}) \mid \category{P} \} \to \category{P}$.
	\end{itemize}
	These data satisfy the following properties.
	\begin{itemize}
		\item The $p$-refinement fibration $\{ s(\category{C}) \mid \category{P} \} \to \category{P}$ is a SCCompC with strong fibred binary coproducts and these structures are preserved by the functor~\eqref{eq:refinement-fibration-morphism}.
		\item The fibred graded monad $\dot{T}$ is a lifting of the fibred monad $\SimpleFibredMonad{T}$ along the functor~\eqref{eq:refinement-fibration-morphism} where $\SimpleFibredMonad{T}$ is the fibred monad induced by the strong monad $T$ (Lemma~\ref{lem:simple-fibration-strong-monad}).
	\end{itemize}
\end{definition}

Now, we can define the interpretation of contexts and types.
\begin{theorem}\label{thm:soundness-wo-recursion-types}
	Given a pre-model in Definition~\ref{def:dependent-effect-system-model-part1}, we have the following interpretation.
	\begin{itemize}
		\item If $\dot{\Gamma} \vdash$ is well-formed, then we have $\interpret{\dot{\Gamma}} \in \category{P}$ such that $p \interpret{\dot{\Gamma}} = \interpret{\underlying{\dot{\Gamma}}}$.
		\item If $\dot{\Gamma} \vdash \dot{A}$ is well-formed, then we have $\interpret{\dot{\Gamma} \vdash \dot{A}} \in \{ s(\category{C}) \mid \category{P} \}_{\interpret{\dot{\Gamma}}}$ such that $u \interpret{\dot{\Gamma} \vdash \dot{A}} = (\interpret{\underlying{\dot{\Gamma}}}, \interpret{\underlying{\dot{A}}})$.
		\item If $\dot{\Gamma} \vdash \dot{C}$ is well-formed, then we have $\interpret{\dot{\Gamma} \vdash \dot{C}} \in \{ s(\category{C}) \mid \category{P} \}_{\interpret{\dot{\Gamma}}}$ such that $u \interpret{\dot{\Gamma} \vdash \dot{C}} = (\interpret{\underlying{\dot{\Gamma}}}, \interpret{\underlying{\dot{C}}})$.
		\item If $\dot{\Gamma} \vdash \mathcal{E} : \mathbf{Effect}$ is well-formed, then we have $\interpret{\dot{\Gamma} \vdash \mathcal{E} : \mathbf{Effect}} \in \category{M}_{\interpret{\dot{\Gamma}}}$.
	\end{itemize}
\end{theorem}
\begin{proof}[Proof of Theorem~\ref{thm:soundness-wo-recursion-types}]
	By simultaneous induction.
	The interpretation is given as follows.

	\paragraph{Contexts}
	$\interpret{\dot{\Gamma}} \in \category{P}$ such that $p \interpret{\dot{\Gamma}} = \interpret{\underlying{\dot{\Gamma}}}$.

	\begin{mathpar}
		\inferrule{ }{
			\interpret{\diamond} \coloneqq \top 1 \in \category{P}_{\interpret{\diamond}}
		}
		\and
		\inferrule{
			\interpret{\dot{\Gamma}} \in \category{P} \\
			\interpret{\dot{\Gamma} \vdash \dot{A}} \in \{ s(\category{C}) \mid \category{P} \}_{\interpret{\dot{\Gamma}}}
		}{
			\interpret{\dot{\Gamma}, x : \dot{A}} \coloneqq \{ \interpret{\dot{\Gamma} \vdash \dot{A}} \} \in \category{P}
		}
	\end{mathpar}
	where $\top : \category{C} \to \category{P}$ is the fibred terminal object functor.
	Note the following equalities.
	\[ p \{ \interpret{\dot{\Gamma} \vdash \dot{A}} \}
	= \{ u \interpret{\dot{\Gamma} \vdash \dot{A}} \} 
	= \{ \interpret{\underlying{\dot{\Gamma}} \vdash \underlying{\dot{A}}} \}
	= \interpret{\underlying{\dot{\Gamma}}, x : \underlying{\dot{A}}} \]

	\paragraph{Value types}
	$\interpret{\dot{\Gamma} \vdash \dot{A}} \in \{ s(\category{C}) \mid \category{P} \}_{\interpret{\dot{\Gamma}}}$ such that $u \interpret{\dot{\Gamma} \vdash \dot{A}} = (\interpret{\underlying{\dot{\Gamma}}}, \interpret{\underlying{\dot{A}}})$.

	\begin{mathpar}
		\inferrule{
			\interpret{\dot{\Gamma}} \in \category{P}_{\interpret{\underlying{\dot{\Gamma}}}} \\
			\interpret{\phi} \in \category{P}_{\interpret{\underlying{\dot{\Gamma}}, x : b}}
		}{
			\interpret{\dot{\Gamma} \vdash \{ x : b \mid \phi \}} \coloneqq (\interpret{\underlying{\Gamma}}, \interpret{b}, \interpret{\dot{\Gamma}}, \pi^{*} \interpret{\dot{\Gamma}} \land \interpret{\phi})
		}
		\and
		\inferrule{
			\interpret{\dot{\Gamma}} \in \category{P}_{\interpret{\underlying{\dot{\Gamma}}}} \\
			\interpret{\phi} \in \category{P}_{\interpret{\underlying{\dot{\Gamma}}, x : \mathbf{unit}}}
		}{
			\interpret{\dot{\Gamma} \vdash \{ x : \mathbf{unit} \mid \phi \}} \coloneqq (\interpret{\underlying{\Gamma}}, 1, \interpret{\dot{\Gamma}}, \pi^{*} \interpret{\dot{\Gamma}} \land \interpret{\phi})
		}
		\and
		\inferrule{
			\interpret{\dot{\Gamma}, x : \dot{A} \vdash \dot{C}} \in \{ s(\category{C}) \mid \category{P} \}_{\interpret{\dot{\Gamma}, x : \dot{A}}}
		}{
			\interpret{\dot{\Gamma} \vdash (x : \dot{A}) \to \dot{C}} \coloneqq \prod_{\interpret{\dot{\Gamma} \vdash \dot{A}}} \interpret{\dot{\Gamma}, x : \dot{A} \vdash \dot{C}}
		}
		\and
		\inferrule{
			\interpret{\dot{\Gamma}, x : \dot{A} \vdash \dot{B}} \in \{ s(\category{C}) \mid \category{P} \}_{\interpret{\dot{\Gamma}, x : \dot{A}}}
		}{
			\interpret{\dot{\Gamma} \vdash (x : \dot{A}) \times \dot{B}} \coloneqq \coprod_{\interpret{\dot{\Gamma} \vdash \dot{A}}} \interpret{\dot{\Gamma}, x : \dot{A} \vdash \dot{B}}
		}
		\and
		\inferrule{
			\interpret{\dot{\Gamma} \vdash \dot{A}} \in \{ s(\category{C}) \mid \category{P} \}_{\interpret{\dot{\Gamma}}} \\
			\interpret{\dot{\Gamma} \vdash \dot{B}} \in \{ s(\category{C}) \mid \category{P} \}_{\interpret{\dot{\Gamma}}}
		}{
			\interpret{\dot{\Gamma} \vdash \dot{A} + \dot{B}} \coloneqq \interpret{\dot{\Gamma} \vdash \dot{A}} + \interpret{\dot{\Gamma} \vdash \dot{B}}
		}
	\end{mathpar}

	\paragraph{Computation type}
	$\interpret{\dot{\Gamma} \vdash \dot{C}} \in \{ s(\category{C}) \mid \category{P} \}_{\interpret{\dot{\Gamma}}}$ such that $u \interpret{\dot{\Gamma} \vdash \dot{C}} = (\interpret{\underlying{\dot{\Gamma}}}, \interpret{\underlying{\dot{C}}})$
	\begin{mathpar}
		\inferrule{
			\interpret{\dot{\Gamma} \vdash \mathcal{E} : \mathbf{Effect}} \in \category{M}_{\interpret{\dot{\Gamma}}} \\
			\interpret{\dot{\Gamma} \vdash \dot{A}} \in \{ s(\category{C}) \mid \category{P} \}_{\interpret{\dot{\Gamma}}}
		}{
			\interpret{\dot{\Gamma} \vdash T_{\mathcal{E}} \dot{A}} \coloneqq \dot{T}_{\interpret{\dot{\Gamma} \vdash \mathcal{E} : \mathbf{Effect}}} \interpret{\dot{\Gamma} \vdash \dot{A}}
		}
	\end{mathpar}
	\[ u \dot{T}_{\interpret{\mathcal{E}}} \interpret{\dot{\Gamma} \vdash \dot{A}} = T u \interpret{\dot{\Gamma} \vdash \dot{A}} = \hat{T} (\interpret{\underlying{\dot{\Gamma}}}, \interpret{\underlying{\dot{A}}}) = (\interpret{\underlying{\dot{\Gamma}}}, \interpret{\underlying{T_{\mathcal{E}} \dot{A}}}) \]

	\paragraph{Effect term}
	$\interpret{\dot{\Gamma} \vdash \mathcal{E} : \mathbf{Effect}} \in \category{M}_{\interpret{\dot{\Gamma}}}$

	\begin{mathpar}
		\inferrule{
			\interpret{\dot{\Gamma}} \in \category{P}
		}{
			\interpret{\dot{\Gamma} \vdash \GradeUnit : \mathbf{Effect}} \coloneqq I \interpret{\dot{\Gamma}}
		}
		\and
		\inferrule{
			\interpret{\dot{\Gamma} \vdash \mathcal{E}_1 : \mathbf{Effect}} \in \category{M}_{\interpret{\dot{\Gamma}}} \\
			\interpret{\dot{\Gamma} \vdash \mathcal{E}_2 : \mathbf{Effect}} \in \category{M}_{\interpret{\dot{\Gamma}}}
		}{
			\interpret{\dot{\Gamma} \vdash \mathcal{E}_1 \GradeMult \mathcal{E}_2 : \mathbf{Effect}} \coloneqq \interpret{\dot{\Gamma} \vdash \mathcal{E}_1 : \mathbf{Effect}} \otimes \interpret{\dot{\Gamma} \vdash \mathcal{E}_2 : \mathbf{Effect}}
		}
		\and
		\inferrule{
			\interpret{\dot{\Gamma}} \in \category{P} \\
			\interpret{V} : \interpret{\underlying{\dot{\Gamma}}} \to \interpret{A} \\
			\interpret{e} \in \category{M}_{\top \interpret{A}}
		}{
			\interpret{\dot{\Gamma} \vdash \mathtt{be}(V) : \mathbf{Effect}} \coloneqq (\top \interpret{V} \comp (\interpret{\dot{\Gamma}} \le \top \interpret{\underlying{\dot{\Gamma}}}))^{*} \interpret{e} \in \category{M}_{\interpret{\dot{\Gamma}}}
		}
	\end{mathpar}
	Here, we have $\interpret{\dot{\Gamma}} \le \top \interpret{\underlying{\dot{\Gamma}}}$ and $\interpret{V} : \top \interpret{\underlying{\dot{\Gamma}}} \dotTo \top \interpret{A}$.
	\begin{equation}
		\begin{tikzcd}
			\category{M} \ar[d] & \interpret{\dot{\Gamma} \vdash \mathtt{be}(V) : \mathbf{Effect}} \ar[rr, "\text{(cartesian)}"] & & \interpret{e} \\
			\category{P} & \interpret{\dot{\Gamma}} \ar[r, "\le"] & \top \interpret{\underlying{\dot{\Gamma}}} \ar[r, "\top \interpret{V}"] & \top \interpret{A}
		\end{tikzcd}
	\end{equation}
\end{proof}

As the next step, we show that the interpretation of terms in $\category{C}$ can be lifted along $p : \category{P} \to \category{C}$.
\begin{definition}\label{def:dependent-effect-system-model-part2}
	A \emph{model of the dependent effect system without recursion} is a pre-model of the dependent effect system without recursion (Definition~\ref{def:dependent-effect-system-model-part1}) that satisfies the following additional properties.
	\begin{itemize}
		\item For each $\mathtt{op} : (x : \dot{A}) \rightarrowtriangle \dot{B}$, we have $\tupling{\identity{}}{\interpret{\mathtt{op}} \comp \pi_2} : \interpret{x : \dot{A}} \dotTo \{ \interpret{x : \dot{A} \vdash \dot{B}} \}$ in $\category{P}$.
		\item For each $\mathtt{gef} : (x : \dot{A}) \stackrel{\mathcal{E}}{\rightarrowtriangle} \dot{B}$, we have $\tupling{\identity{}}{\interpret{\mathtt{gef}} \comp \pi_2} : \interpret{x : \dot{A}} \to \{\interpret{x : \dot{A} \vdash T_{\mathcal{E}} \dot{B}}\}$.
	\end{itemize}
	Here, we have $u \interpret{x : \dot{A}} = 1 \times \interpret{\underlying{\dot{A}}}$, which is why we insert $\pi_2 : 1 \times \interpret{\underlying{\dot{A}}} \to \interpret{\underlying{\dot{A}}}$ before $\interpret{c}$ and $\interpret{\mathtt{gef}}$.
\end{definition}

\begin{theorem}[soundness]\label{thm:soundness-wo-recursion-terms}
	Consider the recursion-free fragment.
	Given a model in Definition~\ref{def:dependent-effect-system-model-part2}, we have the following interpretation.
	\begin{itemize}
		\item If $\dot{\Gamma} \vdash V : \dot{A}$ is well-typed, then we have $\tupling{\identity{}}{\interpret{V}} : \interpret{\dot{\Gamma}} \dotTo \{ \interpret{\dot{\Gamma} \vdash \dot{A}} \}$ in $\category{P}$.
		\item If $\dot{\Gamma} \vdash M : \dot{C}$ is well-typed, then we have $\tupling{\identity{}}{\interpret{M}} : \interpret{\dot{\Gamma}} \dotTo \{ \interpret{\dot{\Gamma} \vdash \dot{C}} \}$ in $\category{P}$.
		\item If $\dot{\Gamma} \vdash \dot{A} <: \dot{B}$, then $\interpret{\dot{\Gamma} \vdash \dot{A}} <: \interpret{\dot{\Gamma} \vdash \dot{B}}$ holds.
		\item If $\dot{\Gamma} \vdash \dot{C} <: \dot{D}$, then $\interpret{\dot{\Gamma} \vdash \dot{C}} <: \interpret{\dot{\Gamma} \vdash \dot{D}}$ holds.
	\end{itemize}
\end{theorem}

\begin{definition}
	Let $\Gamma_1$ and $\Gamma_2$ be contexts and $A$ be a type in the underlying type system.
	We define the semantic weakening morphism $\mathrm{proj}_{\Gamma_1; x : \dot{A}; \Gamma_2} : \interpret{\Gamma_1, x : A, \Gamma_2} \to \interpret{\Gamma_1, \Gamma_2}$ as follows.
	\begin{align}
		\mathrm{proj}_{\Gamma_1; x : A; \diamond} &\coloneqq \pi_1 \\
		\mathrm{proj}_{\Gamma_1; x : A; \Gamma_2, y : B} &\coloneqq \mathrm{proj}_{\Gamma_1; x : A; \Gamma_2} \times \identity{}
	\end{align}
\end{definition}

\begin{lemma}[weakening]\label{lem:weakening}
	Assume that the interpretation of value terms of ground types and formulas satisfies the semantic weakening property.
	\begin{itemize}
		\item If $B$ is a ground type and $\Gamma_1, \Gamma_2 \vdash V : B$ is a well-typed value term, then we have $\interpret{\Gamma_1, x : A, \Gamma_2 \vdash V} = \interpret{\Gamma_1, \Gamma_2 \vdash V} \comp \mathrm{proj}_{\Gamma_1; x : A; \Gamma_2}$.
		\item If $\Gamma_1, \Gamma_2 \vdash \phi : \mathbf{Fml}$ is well formed, then $\interpret{\Gamma_1, x : A, \Gamma_2 \vdash \phi : \mathbf{Fml}} = \mathrm{proj}_{\Gamma_1; x : \dot{A}; \Gamma_2}^{*} \interpret{\Gamma_1, \Gamma_2 \vdash \phi : \mathbf{Fml}}$.
	\end{itemize}
	If $\dot{\Gamma}_1 \vdash \dot{A}$ is well-formed, then we have the following.
	\begin{itemize}
		\item If $\dot{\Gamma}_1, \dot{\Gamma}_2 \vdash$ is well-formed, then we have the following.
		\[ \mathrm{proj}_{\underlying{\dot{\Gamma}_1}; x : \underlying{\dot{A}}; \underlying{\dot{\Gamma}_2}} : \interpret{\dot{\Gamma}_1, x : \dot{A}, \dot{\Gamma}_2} \dotTo \interpret{\dot{\Gamma}_1, \dot{\Gamma}_2} \]
		We write $\mathrm{proj}_{\dot{\Gamma}_1; x : \dot{A}; \dot{\Gamma}_2}: \interpret{\dot{\Gamma}_1, x : \dot{A}, \dot{\Gamma}_2} \to \interpret{\dot{\Gamma}_1, \dot{\Gamma}_2}$ for the morphism in $\category{P}$ above $\mathrm{proj}_{\underlying{\dot{\Gamma}_1}; x : \underlying{\dot{A}}; \underlying{\dot{\Gamma}_2}}$.
		\item If $\dot{\Gamma}_1, \dot{\Gamma}_2 \vdash \dot{B}$ is well-formed, then we have the following.
		\begin{equation}
			\interpret{\dot{\Gamma}_1, x : \dot{A}, \dot{\Gamma}_2 \vdash \dot{B}} = \mathrm{proj}_{\dot{\Gamma}_1; x : \dot{A}; \dot{\Gamma}_2}^{*} \interpret{\dot{\Gamma}_1, \dot{\Gamma}_2 \vdash \dot{B}}
		\end{equation}
		\item If $\dot{\Gamma}_1, \dot{\Gamma}_2 \vdash \dot{C}$ is well-formed, then we have the following.
		\[ \interpret{\dot{\Gamma}_1, x : \dot{A}, \dot{\Gamma}_2 \vdash \dot{C}} = \mathrm{proj}_{\dot{\Gamma}_1; x : \dot{A}; \dot{\Gamma}_2}^{*} \interpret{\dot{\Gamma}_1, \dot{\Gamma}_2 \vdash \dot{C}} \]
		\item If $\dot{\Gamma}_1, \dot{\Gamma}_2 \vdash \mathcal{E} : \mathbf{Effect}$ is well-formed, then we have the following.
		\[ \interpret{\dot{\Gamma}_1, x : \dot{A}, \dot{\Gamma}_2 \vdash \mathcal{E} : \mathbf{Effect}} = \mathrm{proj}_{\dot{\Gamma}_1; x : \dot{A}; \dot{\Gamma}_2}^{*} \interpret{\dot{\Gamma}_1, \dot{\Gamma}_2 \vdash \mathcal{E} : \mathbf{Effect}} \]
	\end{itemize}
\end{lemma}
\begin{proof}
	By simultaneous induction on the well-formedness of contexts and types.
	\begin{itemize}
		\item If $\dot{\Gamma}_2 = \diamond$, then we have
		\[ \pi_{\interpret{\dot{\Gamma}_1 \vdash \dot{A}}} : \interpret{\dot{\Gamma}_1, x : \dot{A}} \to \interpret{\dot{\Gamma}_1} \qquad p \pi_{\interpret{\dot{\Gamma}_1 \vdash \dot{A}}} = \pi_1 = \mathrm{proj}_{\underlying{\dot{\Gamma}_1}; x : \underlying{\dot{A}}; \diamond} \]
		\item If $\dot{\Gamma}_2 = \dot{\Gamma}'_2, y : \dot{B}$, then we have $\interpret{\dot{\Gamma}_1, x : \dot{A}, \dot{\Gamma}_2 \vdash \dot{B}} = \mathrm{proj}_{\dot{\Gamma}_1; x : \dot{A}; \dot{\Gamma}_2}^{*} \interpret{\dot{\Gamma}_1, \dot{\Gamma}_2 \vdash \dot{B}}$ and $\mathrm{proj}_{\dot{\Gamma}_1; x : \dot{A}; \dot{\Gamma}_2}: \interpret{\dot{\Gamma}_1, x : \dot{A}, \dot{\Gamma}_2} \to \interpret{\dot{\Gamma}_1, \dot{\Gamma}_2}$ by IH.
		Therefore, we have
		\[ \{ \overline{\mathrm{proj}_{\dot{\Gamma}_1; x : \dot{A}; \dot{\Gamma}_2}} (\interpret{\dot{\Gamma}_1, \dot{\Gamma}_2 \vdash \dot{B}}) \} : \interpret{\dot{\Gamma}_1, x : \dot{A}, \dot{\Gamma}_2, y : \dot{B}} \to \interpret{\dot{\Gamma}_1, \dot{\Gamma}_2, y : \dot{B}} \]
		which is above $\mathrm{proj}_{\underlying{\dot{\Gamma}_1}; x : \underlying{\dot{A}}; \underlying{\dot{\Gamma}_2}, y : \underlying{\dot{B}}}$.
		\begin{align}
			p \{ \overline{\mathrm{proj}_{\dot{\Gamma}_1; x : \dot{A}; \dot{\Gamma}_2}} (\interpret{\dot{\Gamma}_1, \dot{\Gamma}_2 \vdash \dot{B}}) \} &= \{ \overline{p \mathrm{proj}_{\dot{\Gamma}_1; x : \dot{A}; \dot{\Gamma}_2}} (u \interpret{\dot{\Gamma}_1, \dot{\Gamma}_2 \vdash \dot{B}}) \} \\
			&= \mathrm{proj}_{\underlying{\dot{\Gamma}_1}; x : \underlying{\dot{A}}; \underlying{\dot{\Gamma}_2}} \times \identity{} \\
			&= \mathrm{proj}_{\underlying{\dot{\Gamma}_1}; x : \underlying{\dot{A}}; \underlying{\dot{\Gamma}_2}, y : \underlying{\dot{B}}}
		\end{align}
		\item We omit the cases for well-formed value types $\dot{\Gamma}_1, \dot{\Gamma}_2 \vdash \dot{B}$, as the proof is almost the same as that for pure refinement type systems.
		For base types and the unit type, we use the weakening property for predicates.
		For dependent function types and dependent pair types, we use the Beck--Chevalley condition.
		For sum types, we use the fact that fibred coproducts are preserved by reindexing.
		\item If $\dot{\Gamma}_1, \dot{\Gamma}_2 \vdash T_{\mathcal{E}} \dot{B}$ is well-formed, then we use the fibred-ness of the graded monad $\dot{T}$.
		\begin{align}
			\mathrm{proj}_{\dot{\Gamma}_1; x : \dot{A}; \dot{\Gamma}_2}^{*} \interpret{\dot{\Gamma}_1, \dot{\Gamma}_2 \vdash T_{\mathcal{E}} \dot{B}} &= \mathrm{proj}_{\dot{\Gamma}_1; x : \dot{A}; \dot{\Gamma}_2}^{*} \dot{T}_{\interpret{\dot{\Gamma}_1, \dot{\Gamma}_2 \vdash \mathcal{E} : \mathbf{Effect}}} \interpret{\dot{\Gamma}_1, \dot{\Gamma}_2 \vdash \dot{B}} \\
			&= \dot{T}_{\mathrm{proj}_{\dot{\Gamma}_1; x : \dot{A}; \dot{\Gamma}_2}^{*} \interpret{\dot{\Gamma}_1, \dot{\Gamma}_2 \vdash \mathcal{E} : \mathbf{Effect}}} \mathrm{proj}_{\dot{\Gamma}_1; x : \dot{A}; \dot{\Gamma}_2}^{*} \interpret{\dot{\Gamma}_1, \dot{\Gamma}_2 \vdash \dot{B}} \\
			&= \dot{T}_{\interpret{\dot{\Gamma}_1, x : \dot{A}, \dot{\Gamma}_2 \vdash \mathcal{E} : \mathbf{Effect}}} \interpret{\dot{\Gamma}_1, x : \dot{A}, \dot{\Gamma}_2 \vdash \dot{B}} \\
			&= \interpret{\dot{\Gamma}_1, x : \dot{A}, \dot{\Gamma}_2 \vdash T_{\mathcal{E}} \dot{B}}
		\end{align}
		\item If $\dot{\Gamma}_1, \dot{\Gamma}_2 \vdash \GradeUnit : \mathbf{Effect}$ is well-formed, then we have
		\begin{align}
			\mathrm{proj}_{\dot{\Gamma}_1; x : \dot{A}; \dot{\Gamma}_2}^{*} \interpret{\dot{\Gamma}_1, \dot{\Gamma}_2 \vdash \GradeUnit : \mathbf{Effect}} &= \mathrm{proj}_{\dot{\Gamma}_1; x : \dot{A}; \dot{\Gamma}_2}^{*} I \interpret{\dot{\Gamma}_1, \dot{\Gamma}_2} \\
			&= I \interpret{\dot{\Gamma}_1, x : \dot{A}, \dot{\Gamma}_2} \\
			&= \interpret{\dot{\Gamma}_1, x : \dot{A}, \dot{\Gamma}_2 \vdash \GradeUnit : \mathbf{Effect}}
		\end{align}
		Note that the structure of the indexed monoidal category is preserved by reindexing.
		\item If $\dot{\Gamma}_1, \dot{\Gamma}_2 \vdash \mathcal{E}_1 \GradeMult \mathcal{E}_2 : \mathbf{Effect}$ is well-formed, then we have
		\begin{align}
			&\mathrm{proj}_{\dot{\Gamma}_1; x : \dot{A}; \dot{\Gamma}_2}^{*} \interpret{\dot{\Gamma}_1, \dot{\Gamma}_2 \vdash \mathcal{E}_1 \GradeMult \mathcal{E}_2 : \mathbf{Effect}} \\
			&= \mathrm{proj}_{\dot{\Gamma}_1; x : \dot{A}; \dot{\Gamma}_2}^{*} (\interpret{\dot{\Gamma}_1, \dot{\Gamma}_2 \vdash \mathcal{E}_1 : \mathbf{Effect}} \otimes \interpret{\dot{\Gamma}_1, \dot{\Gamma}_2 \vdash \mathcal{E}_2 : \mathbf{Effect}}) \\
			&= \mathrm{proj}_{\dot{\Gamma}_1; x : \dot{A}; \dot{\Gamma}_2}^{*} \interpret{\dot{\Gamma}_1, \dot{\Gamma}_2 \vdash \mathcal{E}_1 : \mathbf{Effect}} \otimes \mathrm{proj}_{\dot{\Gamma}_1; x : \dot{A}; \dot{\Gamma}_2}^{*} \interpret{\dot{\Gamma}_1, \dot{\Gamma}_2 \vdash \mathcal{E}_2 : \mathbf{Effect}} \\
			&= \interpret{\dot{\Gamma}_1, x : \dot{A}, \dot{\Gamma}_2 \vdash \mathcal{E}_1 : \mathbf{Effect}} \otimes \interpret{\dot{\Gamma}_1, x : \dot{A}, \dot{\Gamma}_2 \vdash \mathcal{E}_2 : \mathbf{Effect}} \\
			&= \interpret{\dot{\Gamma}_1, x : \dot{A}, \dot{\Gamma}_2 \vdash \mathcal{E}_1 \GradeMult \mathcal{E}_2 : \mathbf{Effect}}
		\end{align}
		\item If $\dot{\Gamma}_1, \dot{\Gamma}_2 \vdash \mathtt{be}(V) : \mathbf{Effect}$ is well-formed and $\mathtt{be} : B \to \mathbf{Effect}$, then we have
		\begin{align}
			&\mathrm{proj}_{\dot{\Gamma}_1; x : \dot{A}; \dot{\Gamma}_2}^{*} \interpret{\dot{\Gamma}_1, \dot{\Gamma}_2 \vdash \mathtt{be}(V) : \mathbf{Effect}} \\
			&= \mathrm{proj}_{\dot{\Gamma}_1; x : \dot{A}; \dot{\Gamma}_2}^{*} (\top \interpret{V} \comp ({\le}))^{*} \interpret{e} \\
			&= \interpret{\dot{\Gamma}_1, x : \dot{A}, \dot{\Gamma}_2 \vdash \mathtt{be}(V) : \mathbf{Effect}}
		\end{align}
		\begin{equation}
			\begin{tikzcd}
				\interpret{\dot{\Gamma}_1, x : \dot{A}, \dot{\Gamma}_2} \ar[r, "\le"] \ar[d, "\mathrm{proj}_{\dot{\Gamma}_1; x : \dot{A}; \dot{\Gamma}_2}"] & \top \interpret{\underlying{\dot{\Gamma}_1}, x : \underlying{\dot{A}}, \underlying{\dot{\Gamma}_2}} \ar[d, swap, "\top \mathrm{proj}_{\underlying{\dot{\Gamma}_1}; x : \underlying{\dot{A}}; \underlying{\dot{\Gamma}_2}}"] \ar[rd, "\top \interpret{\underlying{\dot{\Gamma}_1}, x : \underlying{\dot{A}}, \underlying{\dot{\Gamma}_2} \vdash V : B}"] \\
				\interpret{\dot{\Gamma}_1, \dot{\Gamma}_2} \ar[r, "\le"] & \top \interpret{\underlying{\dot{\Gamma}_1}, \underlying{\dot{\Gamma}_2}} \ar[r, swap, "\top \interpret{\underlying{\dot{\Gamma}_1}, \underlying{\dot{\Gamma}_2} \vdash V : B}"] & \top \interpret{B}
			\end{tikzcd}
		\end{equation}
	\end{itemize}
\end{proof}

\begin{definition}
	Let $\Gamma_1 \vdash V : A$ be a well-typed term in the underlying type system.
	We define the semantic substitution morphism as follows.
	\[ \mathrm{subst}_{\Gamma_1; x : A; \Gamma_2; V} : \interpret{\Gamma_1, \Gamma_2} \to \interpret{\Gamma_1, x : A, \Gamma_2} \]
	\begin{align}
		\mathrm{subst}_{\Gamma_1; x : A; \diamond; V} &\coloneqq \tupling{\identity{}}{\interpret{V}} \\
		\mathrm{subst}_{\Gamma_1; x : A; \Gamma_2, y : B; V} &\coloneqq \mathrm{subst}_{\Gamma_1; x : A; \Gamma_2; V} \times \identity{}
	\end{align}
\end{definition}

\begin{lemma}[substitution]\label{lem:subst}
	Assume that the interpretation of terms of ground types and formulas satisfies the semantic substitution property.
	\begin{itemize}
		\item If $\Gamma_1 \vdash V : A$ and $\Gamma_1, x : A, \Gamma_2 \vdash W : B$ are well-typed terms and $B$ is a ground type, then we have $\interpret{W[V/x]} = \interpret{W} \comp \mathrm{subst}_{\Gamma_1; x : A; \Gamma_2; V}$.
		\item If $\Gamma_1 \vdash V : A$ is a well-typed term and $\Gamma_1, x : A, \Gamma_2 \vdash \phi : \mathbf{Fml}$ is a well-formed formula, then we have $\interpret{\phi[V/x]} = \mathrm{subst}_{\Gamma_1; x : A; \Gamma_2; V}^{*} \interpret{\phi}$.
	\end{itemize}
	If $\dot{\Gamma}_1 \vdash V : \dot{A}$ is well-typed and a lifting $\tupling{\identity{}}{\interpret{V}} : \interpret{\dot{\Gamma}_1} \dotTo \{\interpret{\dot{\Gamma}_1 \vdash \dot{A}}\}$ exists, then we have the following.
	\begin{itemize}
		\item If $\dot{\Gamma}_1, \dot{\Gamma}_2 \vdash$ is well-formed, then we have
		\[ \mathrm{subst}_{\underlying{\dot{\Gamma}_1}; x : \underlying{\dot{A}}; \underlying{\dot{\Gamma}_2}; V} : \interpret{\dot{\Gamma}_1, \dot{\Gamma}_2[V/x]} \dotTo \interpret{\dot{\Gamma}_1, x : \dot{A}, \dot{\Gamma}_2} \]
		We write $\mathrm{subst}_{\dot{\Gamma}_1; x : \dot{A}; \dot{\Gamma}_2; V} : \interpret{\dot{\Gamma}_1, \dot{\Gamma}_2[V/x]} \to \interpret{\dot{\Gamma}_1, x : \dot{A}, \dot{\Gamma}_2}$ for the morphism in $\category{P}$ above $\mathrm{subst}_{\underlying{\dot{\Gamma}_1}; x : \underlying{\dot{A}}; \underlying{\dot{\Gamma}_2}; V}$.
		\item If $\dot{\Gamma}, x : \dot{A}, \dot{\Gamma}_2 \vdash \dot{B}$ is well-formed, then we have
		\[ \interpret{\dot{\Gamma}_1, \dot{\Gamma}_2[V/x] \vdash \dot{B}[V/x]} = \mathrm{subst}_{\dot{\Gamma}_1; x : \dot{A}; \dot{\Gamma}_2; V}^{*} \interpret{\dot{\Gamma}_1, x : \dot{A}, \dot{\Gamma}_2 \vdash \dot{B}} \]
		\item $\interpret{\dot{\Gamma}_1, \dot{\Gamma}_2[V/x] \vdash \dot{C}[V/x]} = \mathrm{subst}_{\dot{\Gamma}_1; x : \dot{A}; \dot{\Gamma}_2; V}^{*} \interpret{\dot{\Gamma}_1, x : \dot{A}, \dot{\Gamma}_2 \vdash \dot{C}}$
		\item $\interpret{\dot{\Gamma}_1, \dot{\Gamma}_2[V/x] \vdash \mathcal{E}[V/x] : \mathbf{Effect}} = \mathrm{subst}_{\dot{\Gamma}_1; x : \dot{A}; \dot{\Gamma}_2; V}^{*} \interpret{\dot{\Gamma}_1, x : \dot{A}, \dot{\Gamma}_2 \vdash \mathcal{E} : \mathbf{Effect}}$
	\end{itemize}
\end{lemma}
\begin{proof}
	By simultaneous induction on the well-formedness of contexts and types.
	The proof follows almost the same pattern as that for the semantic weakening property.
	\begin{itemize}
		\item If $\dot{\Gamma}_2 = \diamond$, then we have $\mathrm{subst}_{\underlying{\dot{\Gamma}_1}; x : \underlying{\dot{A}}; \diamond; V} : \interpret{\dot{\Gamma}_1} \dotTo \interpret{\dot{\Gamma}_1, x : \dot{A}}$ by assumption.
		\item If $\dot{\Gamma}_1, x : \dot{A}, \dot{\Gamma}_2 \vdash \mathtt{be}(W) : \mathbf{Effect}$ is well-formed and $\mathtt{be} : B \to \mathbf{Effect}$, then we have
		\begin{align}
			&\mathrm{subst}_{\dot{\Gamma}_1; x : \dot{A}; \dot{\Gamma}_2; V}^{*} \interpret{\dot{\Gamma}_1, x : \dot{A}, \dot{\Gamma}_2 \vdash \mathtt{be}(W) : \mathbf{Effect}} \\
			&= \mathrm{subst}_{\dot{\Gamma}_1; x : \dot{A}; \dot{\Gamma}_2; V}^{*} (\top \interpret{W} \comp ({\le}))^{*} \interpret{e} \\
			&= \interpret{\dot{\Gamma}_1, \dot{\Gamma}_2[V/x] \vdash \mathtt{be}(W[V/x]) : \mathbf{Effect}}
		\end{align}
		\begin{equation}
			\begin{tikzcd}[column sep=huge]
				\interpret{\dot{\Gamma}_1, \dot{\Gamma}_2[V/x]} \ar[r, "\le"] \ar[d, swap, "\mathrm{subst}_{\dot{\Gamma}_1; x : \dot{A}; \dot{\Gamma}_2; V}"] & \top \interpret{\underlying{\dot{\Gamma}_1}, \underlying{\dot{\Gamma}_2}} \ar[d, swap, "\top \mathrm{subst}_{\underlying{\dot{\Gamma}_1}; x : \underlying{\dot{A}}; \underlying{\dot{\Gamma}_2}; V}"] \ar[rd, "\top \interpret{\underlying{\dot{\Gamma}_1}, \underlying{\dot{\Gamma}_2} \vdash W[V/x] : B}"] \\
				\interpret{\dot{\Gamma}_1, x : \dot{A}, \dot{\Gamma}_2} \ar[r, "\le"] & \top \interpret{\underlying{\dot{\Gamma}_1}, x : \underlying{\dot{A}}, \underlying{\dot{\Gamma}_2}} \ar[r, swap, "\top \interpret{\underlying{\dot{\Gamma}_1}, x : \underlying{\dot{A}}, \underlying{\dot{\Gamma}_2} \vdash W : B}"] & \top \interpret{B}
			\end{tikzcd}
		\end{equation}
	\end{itemize}
\end{proof}

\begin{proof}[Proof of Theorem~\ref{thm:soundness-wo-recursion-terms}]
By simultaneous induction.

\paragraph{Value term}
$\interpret{\dot{\Gamma} \vdash V : \dot{A}} = (\identity{}, \interpret{\underlying{\dot{\Gamma}} \vdash V : \underlying{\dot{A}}} \comp \pi_1) \in \{ s(\category{C}) \mid \category{P} \}_{\interpret{\dot{\Gamma}}} (1 \interpret{\dot{\Gamma}}, \interpret{\dot{\Gamma} \vdash \dot{A}})$

Note that $(\identity{}, f \comp \pi_1) \in \{ s(\category{C}) \mid \category{P} \}_{\interpret{\dot{\Gamma}}} (1 \interpret{\dot{\Gamma}}, \interpret{\dot{\Gamma} \vdash \dot{A}})$ bijectively corresponds to $\tupling{\identity{}}{f} : \interpret{\dot{\Gamma}} \dotTo \{\interpret{\dot{\Gamma} \vdash \dot{A}}\}$ in $\category{P}$.

\begin{lemma}
	\hyperlink{rule:VT-Var}{\textsc{VT-Var}} is sound.
\end{lemma}
\begin{nestedproof}
	By induction on the length of the context $\dot{\Gamma}$.
	In \hyperlink{rule:VT-Var}{\textsc{VT-Var}}, $\dot{\Gamma}$ must be non-empty.
	Thus, we consider the conclusion of the form $\dot{\Gamma}, x : \dot{A} \vdash y : \dot{B}$.
	If $x = y$:
	\begin{mathpar}
		\inferrule{
			\interpret{\dot{\Gamma} \vdash \dot{A}} \in \{ s(\category{C}) \mid \category{P} \}_{\interpret{\dot{\Gamma}}}
		}{
			\interpret{\dot{\Gamma}, x : \dot{A} \vdash x : \dot{A}} \qquad\coloneqq\qquad
			{\begin{tikzcd}
				1 \interpret{\dot{\Gamma}, x : \dot{A}} \ar[d, equal] \\
				\pi_{\interpret{\dot{\Gamma} \vdash \dot{A}}}^{*} 1 \interpret{\dot{\Gamma}} \ar[d, "\eta"] \\
				\pi_{\interpret{\dot{\Gamma} \vdash \dot{A}}}^{*} \coprod_{\interpret{\dot{\Gamma} \vdash \dot{A}}} \pi_{\interpret{\dot{\Gamma} \vdash \dot{A}}}^{*} 1 \interpret{\dot{\Gamma}} \ar[d, "\pi_{\interpret{\dot{\Gamma} \vdash \dot{A}}}^{*} \mathbf{fst}"] \\
				\pi_{\interpret{\dot{\Gamma} \vdash \dot{A}}}^{*} \interpret{\dot{\Gamma} \vdash \dot{A}} \ar[d, equal, "\text{Lemma~\ref{lem:weakening}}"] \\
				\interpret{\dot{\Gamma}, x : \dot{A} \vdash \dot{A}}
			\end{tikzcd}}
		}
	\end{mathpar}
	\begin{align}
		u \interpret{\dot{\Gamma}, x : \dot{A} \vdash x : \dot{A}} &= u \pi_{\interpret{\dot{\Gamma} \vdash \dot{A}}}^{*} \mathbf{fst} \comp u \eta^{\coprod \dashv \pi^{*}} \\
		&= (\identity{}, \pi_1 \comp \pi_2 \comp (\pi_1 \times \identity{})) \comp (\identity{}, \pi_2 \times \identity{}) \\
		&= (\identity{}, \pi_1 \comp \pi_2 \comp (\pi_1 \times \identity{}) \comp \tupling{\pi_1}{\pi_2 \times \identity{}}) \\
		&= (\identity{}, \pi_2 \comp \pi_1) \\
		&= (\identity{}, \interpret{\underlying{\dot{\Gamma}}, x : \underlying{\dot{A}} \vdash x : \underlying{\dot{A}}} \comp \pi_1)
	\end{align}
	If $x \neq y$:
	\begin{mathpar}
		\inferrule{
			\interpret{\dot{\Gamma} \vdash \dot{A}} \in \{ s(\category{C}) \mid \category{P} \}_{\interpret{\dot{\Gamma}}} \\
			\interpret{\dot{\Gamma} \vdash \dot{B}} \in \{ s(\category{C}) \mid \category{P} \}_{\interpret{\dot{\Gamma}}}
		}{
			\interpret{\dot{\Gamma}, y : \dot{B} \vdash x : \dot{A}} \qquad\coloneqq\qquad
			{\begin{tikzcd}
				1 \interpret{\dot{\Gamma}, y : \dot{B}} \ar[d, equal] \\
				\pi_{\interpret{\dot{\Gamma} \vdash \dot{B}}}^{*} 1 \interpret{\dot{\Gamma}} \ar[d, "\pi^{*} \interpret{\dot{\Gamma} \vdash x : \dot{A}}"] \\
				\pi_{\interpret{\dot{\Gamma} \vdash \dot{B}}}^{*} \interpret{\dot{\Gamma} \vdash \dot{A}} \ar[d, equal, "\text{Lemma~\ref{lem:weakening}}"] \\
				\interpret{\dot{\Gamma}, y : \dot{B} \vdash \dot{A}}
			\end{tikzcd}}
		}
	\end{mathpar}
	\begin{align}
		u \interpret{\dot{\Gamma}, y : \dot{B} \vdash x : \dot{A}} &= u \pi^{*} \interpret{\dot{\Gamma} \vdash x : \dot{A}} \\
		&= \pi^{*}(\identity{}, \interpret{\underlying{\dot{\Gamma}} \vdash x : \underlying{\dot{A}}} \comp \pi_1) \\
		&= (\identity{}, \interpret{\underlying{\dot{\Gamma}} \vdash x : \underlying{\dot{A}}} \comp \pi_1 \comp (\pi_1 \times \identity{})) \\
		&= (\identity{}, \interpret{\underlying{\dot{\Gamma}} \vdash x : \underlying{\dot{A}}} \comp \pi_1 \comp \pi_1) \\
		&= (\identity{}, \interpret{\underlying{\dot{\Gamma}}, y : \underlying{\dot{B}} \vdash x : \underlying{\dot{A}}} \comp \pi_1)
	\end{align}	
\end{nestedproof}

\begin{lemma}
	\hyperlink{rule:VT-EffFreeOp}{\textsc{VT-EffFreeOp}} is sound.
\end{lemma}
\begin{proof}
	\begin{mathpar}
		\inferrule{
			\interpret{\dot{\Gamma} \vdash V : \dot{A}} \in \{ s(\category{C}) \mid \category{P} \}_{\interpret{\dot{\Gamma}}}(1 \interpret{\dot{\Gamma}}, \interpret{\dot{\Gamma} \vdash \dot{A}}) \\
			\tupling{\identity{}}{\interpret{\mathtt{op}} \comp \pi_2} : \interpret{x : \dot{A}} \dotTo \{ \interpret{x : \dot{A} \vdash \dot{B}} \}
		}{
			\interpret{\dot{\Gamma} \vdash \mathtt{op}\ V : \dot{B}[V/x]} \qquad\coloneqq\qquad
			{\begin{tikzcd}
				1 \interpret{\dot{\Gamma}} \ar[d, equal] \\
				(s\interpret{\dot{\Gamma} \vdash V : \dot{A}})^{*} \pi_{\interpret{\dot{\Gamma} \vdash \dot{A}}}^{*} 1 \interpret{\dot{\Gamma}} \ar[d, equal] \\
				(s\interpret{\dot{\Gamma} \vdash V : \dot{A}})^{*} 1 \interpret{\dot{\Gamma}, x : \dot{A}} \ar[d, equal, "\text{Lemma~\ref{lem:weakening}}"] \\
				(s\interpret{\dot{\Gamma} \vdash V : \dot{A}})^{*} \{ \overline{!}(\interpret{\vdash \dot{A}}) \}^{*} 1 \interpret{x : \dot{A}} \ar[d, "(s\interpret{\dot{\Gamma} \vdash V : \dot{A}})^{*} \{ \overline{!}(\interpret{\vdash \dot{A}}) \}^{*} \dot{\interpret{\mathtt{op}}}"] \\
				(s\interpret{\dot{\Gamma} \vdash V : \dot{A}})^{*} \{ \overline{!}(\interpret{\vdash \dot{A}}) \}^{*} \interpret{x : \dot{A} \vdash \dot{B}} \ar[d, equal, "\text{Lemma~\ref{lem:weakening}}"] \\
				(s\interpret{\dot{\Gamma} \vdash V : \dot{A}})^{*} \interpret{\dot{\Gamma}, x : \dot{A} \vdash \dot{B}} \ar[d, equal, "\text{Lemma~\ref{lem:subst}}"] \\
				\interpret{\dot{\Gamma} \vdash \dot{B}[V/x]}
			\end{tikzcd}}
		}
	\end{mathpar}
	
	Here, note that for any context $x_1 : \dot{A}_1, \dots, x_n : \dot{A}_n$ we have
	\[ \interpret{x_1 : \dot{A}_1, \dots, x_n : \dot{A}_n} \xrightarrow{!} 1 \quad=\quad \interpret{x_1 : \dot{A}_1, \dots, x_n : \dot{A}_n} \xrightarrow{\mathrm{proj}} \interpret{x_1 : \dot{A}_1, \dots, x_{n - 1} : \dot{A}_{n - 1}} \xrightarrow{\dots} \interpret{\diamond} = 1 \]
	
	\begin{align}
		u \interpret{\dot{\Gamma} \vdash \mathtt{op}\ V : \dot{B}[V/x]} &= u (s\interpret{\dot{\Gamma} \vdash V : \dot{A}})^{*} \{ \overline{!}(\interpret{\vdash \dot{A}}) \}^{*} \dot{\interpret{c}} \\
		&= (\identity{}, \interpret{c} \comp \pi_2 \comp \pi_1 \comp (({!} \times \identity{}) \times \identity{}) \comp (\tupling{\identity{}}{\interpret{V}} \times \identity{})) \\
		&= (\identity{}, \interpret{c} \comp \interpret{V} \comp \pi_1) \\
	\end{align}	
\end{proof}

\begin{lemma}
	\hyperlink{rule:VT-Lam}{\textsc{VT-Lam}} is sound.
\end{lemma}
\begin{proof}
	\begin{mathpar}
		\inferrule{
			\interpret{\dot{\Gamma}, x : \dot{A} \vdash M : \dot{C}} \in \{ s(\category{C}) \mid \category{P} \}_{\interpret{\dot{\Gamma}, x : \dot{A}}}(1 \interpret{\dot{\Gamma}, x : \dot{A}}, \interpret{\dot{\Gamma}, x : \dot{A} \vdash \dot{C}})
		}{
			\interpret{\dot{\Gamma} \vdash \lambda x. M : (x : \dot{A}) \to \dot{C}} \qquad\coloneqq\qquad
			{\begin{tikzcd}
				1 \interpret{\dot{\Gamma}} \ar[d, "\eta"] \\
				\prod_{\interpret{\dot{\Gamma} \vdash \dot{A}}} \pi_{\interpret{\dot{\Gamma} \vdash \dot{A}}}^{*} 1 \interpret{\dot{\Gamma}} \ar[d, equal] \\
				\prod_{\interpret{\dot{\Gamma} \vdash \dot{A}}} 1 \interpret{\dot{\Gamma}, x : \dot{A}} \ar[d, "\prod \interpret{\dot{\Gamma}, x : \dot{A} \vdash M : \dot{C}}"] \\
				\prod_{\interpret{\dot{\Gamma} \vdash \dot{A}}} \interpret{\dot{\Gamma}, x : \dot{A} \vdash \dot{C}} \ar[d, equal] \\
				\interpret{\dot{\Gamma} \vdash (x : \dot{A}) \to \dot{C}}
			\end{tikzcd}}
		}
	\end{mathpar}
	\begin{align}
		&u \interpret{\dot{\Gamma} \vdash \lambda x. M : (x : \dot{A}) \to \dot{C}} \\
		&= u \prod \interpret{\dot{\Gamma}, x : \dot{A} \vdash M : \dot{C}} \comp u \eta\\
		&= (\identity{}, \Lambda(\interpret{\underlying{\dot{\Gamma}}, x : \underlying{\dot{A}} \vdash M : \underlying{\dot{C}}} \comp \pi_1 \comp (\identity{} \times \eval) \comp \tupling{\pi_1 \times \identity{}}{\pi_2 \times \identity{}})) \comp (\identity{}, \Lambda (\pi_2 \comp \pi_1)) \\
		&= (\identity{}, \Lambda(\interpret{\underlying{\dot{\Gamma}}, x : \underlying{\dot{A}} \vdash M : \underlying{\dot{C}}} \comp (\pi_1 \times \identity{})) \comp \tupling{\pi_1}{\Lambda (\pi_2 \comp \pi_1)}) \\
		&= (\identity{}, \Lambda(\interpret{\underlying{\dot{\Gamma}}, x : \underlying{\dot{A}} \vdash M : \underlying{\dot{C}}} \comp (\pi_1 \times \identity{}) \comp (\tupling{\pi_1}{\Lambda (\pi_2 \comp \pi_1)} \times \identity{}))) \\
		&= (\identity{}, \Lambda(\interpret{\underlying{\dot{\Gamma}}, x : \underlying{\dot{A}} \vdash M : \underlying{\dot{C}}} \comp (\pi_1 \times \identity{}))) \\
		&= (\identity{}, \Lambda(\interpret{\underlying{\dot{\Gamma}}, x : \underlying{\dot{A}} \vdash M : \underlying{\dot{C}}}) \comp \pi_1)
	\end{align}	
\end{proof}

\begin{lemma}
	\hyperlink{rule:VT-Unit}{\textsc{VT-Unit}} is sound.
\end{lemma}
\begin{proof}
	\begin{mathpar}
		\inferrule{
			\interpret{\dot{\Gamma}} \in \category{P}
		}{
			\interpret{\dot{\Gamma} \vdash () : \mathbf{unit}} \qquad\coloneqq\qquad
			{\begin{tikzcd}
				1 \interpret{\dot{\Gamma}} \ar[d, equal] \\
				\interpret{\dot{\Gamma} \vdash \mathbf{unit}}
			\end{tikzcd}}
		}
	\end{mathpar}
	\[ u \interpret{\dot{\Gamma} \vdash () : \mathbf{unit}} = (\identity{}, {!} \comp \pi_1) : (\interpret{\underlying{\dot{\Gamma}}}, 1) \to (\interpret{\underlying{\dot{\Gamma}}}, 1) \]	
\end{proof}

\begin{lemma}
	\hyperlink{rule:VT-Pair}{\textsc{VT-Pair}} is sound.
\end{lemma}
\begin{proof}
\begin{mathpar}
	\inferrule{
		\interpret{\dot{\Gamma} \vdash V : \dot{A}} \in \{ s(\category{C}) \mid \category{P} \}_{\interpret{\dot{\Gamma}}}(1 \interpret{\dot{\Gamma}}, \interpret{\dot{\Gamma} \vdash \dot{A}}) \\
		\interpret{\dot{\Gamma} \vdash W : \dot{B}[V/x]} \in \{ s(\category{C}) \mid \category{P} \}_{\interpret{\dot{\Gamma}}}(1 \interpret{\dot{\Gamma}}, \interpret{\dot{\Gamma} \vdash \dot{B}[V/x]})
	}{
		\interpret{\dot{\Gamma} \vdash (V, W) : (x : \dot{A}) \times \dot{B}} \qquad\coloneqq\qquad
		{\begin{tikzcd}
			1 \interpret{\dot{\Gamma}} \ar[d, "\interpret{\dot{\Gamma} \vdash W : \dot{B}[V/x]}"] \\
			\interpret{\dot{\Gamma} \vdash \dot{B}[V/x]} \ar[d, equal, "\text{Lemma~\ref{lem:subst}}"] \\
			(s\interpret{\dot{\Gamma} \vdash V : \dot{A}})^{*} \interpret{\dot{\Gamma}, x : \dot{A} \vdash \dot{B}} \ar[d, "(s \interpret{\dot{\Gamma} \vdash V : \dot{A}})^{*} \eta"] \\
			(s\interpret{\dot{\Gamma} \vdash V : \dot{A}})^{*} \pi_{\interpret{\dot{\Gamma} \vdash \dot{A}}}^{*} \coprod_{\interpret{\dot{\Gamma} \vdash \dot{A}}} \interpret{\dot{\Gamma}, x : \dot{A} \vdash \dot{B}} \ar[d, equal] \\
			\interpret{\dot{\Gamma} \vdash (x : \dot{A}) \times \dot{B}}
		\end{tikzcd}}
	}
\end{mathpar}
\begin{align}
	u \interpret{\dot{\Gamma} \vdash (V, W) : (x : \dot{A}) \times \dot{B}} &= u (s\interpret{\dot{\Gamma} \vdash V : \dot{A}})^{*} \eta \comp u \interpret{\dot{\Gamma} \vdash W : \dot{B}[V/x]} \\
	&= (\identity{}, (\pi_2 \times \identity{}) \comp (\tupling{\identity{}}{\interpret{V}} \times \identity{})) \comp (\identity{}, \interpret{W} \comp \pi_1) \\
	&= (\identity{}, (\interpret{V} \times \identity{}) \comp \tupling{\pi_1}{\interpret{W} \comp \pi_1}) \\
	&= (\identity{}, \tupling{\interpret{V}}{\interpret{W}} \comp \pi_1)
\end{align}
\end{proof}

\begin{lemma}
	\hyperlink{rule:VT-Inl}{\textsc{VT-Inl}} and \hyperlink{rule:VT-Inr}{\textsc{VT-Inr}} are sound.
\end{lemma}
\begin{proof}
	We show only \hyperlink{rule:VT-Inl}{\textsc{VT-Inl}} as \hyperlink{rule:VT-Inr}{\textsc{VT-Inr}} is similar.
	\begin{mathpar}
		\inferrule{
			\interpret{\dot{\Gamma} \vdash V : \dot{A}_i} \in \{ s(\category{C}) \mid \category{P} \}_{\interpret{\dot{\Gamma}}}(1 \interpret{\dot{\Gamma}}, \interpret{\dot{\Gamma} \vdash \dot{A}_i})
		}{
			\interpret{\dot{\Gamma} \vdash \leftinj{V} : \dot{A}_1 + \dot{A}_2} \qquad\coloneqq\qquad
			{\begin{tikzcd}
				1 \interpret{\dot{\Gamma}} \ar[d, "\interpret{\dot{\Gamma} \vdash V : \dot{A}_1}"] \\
				\interpret{\dot{\Gamma} \vdash \dot{A}_1} \ar[d, "\iota_1"] \\
				\interpret{\dot{\Gamma} \vdash \dot{A}_1 + \dot{A}_2}
			\end{tikzcd}}
		}
	\end{mathpar}
	\begin{align}
		u \interpret{\dot{\Gamma} \vdash \leftinj{V} : \dot{A}_1 + \dot{A}_2} &= (\identity{}, \iota_1 \comp \pi_2) \comp (\identity{}, \interpret{V} \comp \pi_1) \\
		&= (\identity{}, \iota_1 \comp \pi_2 \comp \tupling{\pi_1}{\interpret{V} \comp \pi_1}) \\
		&= (\identity{}, \iota_1 \comp \interpret{V} \comp \pi_1)
		\qedhere
	\end{align}
\end{proof}

\begin{lemma}
	\hyperlink{rule:VT-Sub}{\textsc{VT-Sub}} is sound.
\end{lemma}
\begin{proof}
	\begin{mathpar}
		\inferrule{
			\interpret{\dot{\Gamma} \vdash V : \dot{A}} \in \{ s(\category{C}) \mid \category{P} \}_{\interpret{\dot{\Gamma}}}(1 \interpret{\dot{\Gamma}}, \interpret{\dot{\Gamma} \vdash \dot{A}}) \\
			\interpret{\dot{\Gamma} \vdash \dot{A}} <: \interpret{\dot{\Gamma} \vdash \dot{B}}
		}{
			\interpret{\dot{\Gamma} \vdash V : \dot{B}} \qquad\coloneqq\qquad
			{\begin{tikzcd}
				1 \interpret{\dot{\Gamma}} \ar[d, "\interpret{\dot{\Gamma} \vdash V : \dot{A}}"] \\
				\interpret{\dot{\Gamma} \vdash \dot{A}} \ar[d, "<:"] \\
				\interpret{\dot{\Gamma} \vdash \dot{B}}
			\end{tikzcd}}
		}
	\end{mathpar}
	\begin{align}
		u \interpret{\dot{\Gamma} \vdash V : \dot{B}} &= u ({<:}) \comp u \interpret{\dot{\Gamma} \vdash V : \dot{A}} \\
		&= u \interpret{\dot{\Gamma} \vdash V : \dot{A}}
	\end{align}	
\end{proof}

\paragraph{Computation term}
$\interpret{\dot{\Gamma} \vdash M : \dot{C}} = (\identity{}, \interpret{\underlying{\dot{\Gamma}} \vdash M : \underlying{\dot{C}}} \comp \pi_1) \in \{ s(\category{C}) \mid \category{P} \}_{\interpret{\dot{\Gamma}}} (1 \interpret{\dot{\Gamma}}, \interpret{\dot{\Gamma} \vdash \dot{C}})$

\begin{lemma}
	\hyperlink{rule:CT-App}{\textsc{CT-App}} is sound.
\end{lemma}
\begin{proof}
	\begin{mathpar}
		\inferrule{
			\interpret{\dot{\Gamma} \vdash V : (x : \dot{A}) \to \dot{C}} \in \{ s(\category{C}) \mid \category{P} \}_{\interpret{\dot{\Gamma}}} (1 \interpret{\dot{\Gamma}}, \interpret{\dot{\Gamma} \vdash (x : \dot{A}) \to \dot{C}}) \\
			\interpret{\dot{\Gamma} \vdash W : \dot{A}} \in \{ s(\category{C}) \mid \category{P} \}_{\interpret{\dot{\Gamma}}} (1 \interpret{\dot{\Gamma}}, \interpret{\dot{\Gamma} \vdash \dot{A}})
		}{
			\interpret{\dot{\Gamma} \vdash V \ W : \dot{C}[V/x]} \qquad\coloneqq\qquad
			{\begin{tikzcd}
				1 \interpret{\dot{\Gamma}} \ar[d, equal] \\
				(s\interpret{\dot{\Gamma} \vdash W : \dot{A}})^{*} \pi_{\interpret{\dot{\Gamma} \vdash \dot{A}}}^{*} 1 \interpret{\dot{\Gamma}} \ar[d, "(s\interpret{\dot{\Gamma} \vdash W : \dot{A}})^{*} \pi_{\interpret{\dot{\Gamma} \vdash \dot{A}}}^{*} \interpret{\dot{\Gamma} \vdash V : (x : \dot{A}) \to \dot{C}}"] \\
				(s\interpret{\dot{\Gamma} \vdash W : \dot{A}})^{*} \pi_{\interpret{\dot{\Gamma} \vdash \dot{A}}}^{*} \prod_{\interpret{\dot{\Gamma} \vdash \dot{A}}} \interpret{\dot{\Gamma}, x : \dot{A} \vdash \dot{C}} \ar[d, "(s\interpret{\dot{\Gamma} \vdash W : \dot{A}})^{*} \epsilon"] \\
				(s\interpret{\dot{\Gamma} \vdash W : \dot{A}})^{*} \interpret{\dot{\Gamma}, x : \dot{A} \vdash \dot{C}} \ar[d, equal, "\text{Lemma~\ref{lem:subst}}"] \\
				\interpret{\dot{\Gamma} \vdash \dot{C}[V/x]}
			\end{tikzcd}}
		}
	\end{mathpar}
	\begin{align}
		u \interpret{\dot{\Gamma} \vdash V \ W : \dot{C}[V/x]} &= (\identity{}, \eval \comp \braiding \comp (\pi_2 \times \identity{}) \comp (\tupling{\identity{}}{\interpret{W}} \times \identity{})) \comp (\identity{}, \interpret{V} \comp \pi_1) \\
		&= (\identity{}, \eval \comp \braiding \comp (\interpret{W} \times \identity{}) \comp \tupling{\pi_1}{\interpret{V} \comp \pi_1}) \\
		&= (\identity{}, \eval \comp \tupling{\interpret{V}}{\interpret{W}} \comp \pi_1)
	\end{align}
\end{proof}

\begin{lemma}
	\hyperlink{rule:CT-Ret}{\textsc{CT-Ret}} is sound.
\end{lemma}
\begin{proof}
\begin{mathpar}
	\inferrule{
		\interpret{\dot{\Gamma} \vdash V : \dot{A}} \in \{ s(\category{C}) \mid \category{P} \}_{\interpret{\dot{\Gamma}}} (1 \interpret{\dot{\Gamma}}, \interpret{\dot{\Gamma} \vdash \dot{A}})
	}{
		\interpret{\dot{\Gamma} \vdash \return{V} : \dot{A}} \qquad\coloneqq\qquad
		{\begin{tikzcd}
			1 \interpret{\dot{\Gamma}} \ar[d, "\interpret{\dot{\Gamma} \vdash V : \dot{A}}"] \\
			\interpret{\dot{\Gamma} \vdash \dot{A}} \ar[d, "\eta^{\dot{T}}"] \\
			\interpret{\dot{\Gamma} \vdash T_{\GradeUnit} \dot{A}}
		\end{tikzcd}}
	}
\end{mathpar}
\begin{align}
	u \interpret{\dot{\Gamma} \vdash \return{V} : \dot{A}} &= (\identity{}, \eta^T \comp \pi_2) \comp (\identity{}, \interpret{V} \comp \pi_1) \\
	&= (\identity{}, \eta^T \comp \interpret{V} \comp \pi_1)
\end{align}
\end{proof}

\begin{lemma}
	\hyperlink{rule:CT-Let}{\textsc{CT-Let}} is sound.
\end{lemma}
\begin{proof}
\begin{mathpar}
	\inferrule{
		\interpret{\dot{\Gamma} \vdash M : T_{\mathcal{E}_1} \dot{A}} \in \{ s(\category{C}) \mid \category{P} \}_{\interpret{\dot{\Gamma}}} (1 \interpret{\dot{\Gamma}}, \interpret{\dot{\Gamma} \vdash T_{\mathcal{E}_1} \dot{A}}) \\
		\interpret{\dot{\Gamma} \vdash T_{\mathcal{E}_2} \dot{B}} \in \{ s(\category{C}) \mid \category{P} \}_{\interpret{\dot{\Gamma}}} \\
		\interpret{\dot{\Gamma}, x : \dot{A} \vdash N : T_{\mathcal{E}_2} \dot{B}} \in \{ s(\category{C}) \mid \category{P} \}_{\interpret{\dot{\Gamma}}} (1 \interpret{\dot{\Gamma}, x : \dot{A}}, \interpret{\dot{\Gamma}, x : \dot{A} \vdash T_{\mathcal{E}_2} \dot{B}})
	}{
		\interpret{\dot{\Gamma} \vdash \letin{x}{M}{N} : T_{\mathcal{E}_1 \GradeMult \mathcal{E}_2} \dot{B}} \qquad\coloneqq\qquad
		{\begin{tikzcd}
			1 \interpret{\dot{\Gamma}} \ar[d, "\interpret{\dot{\Gamma} \vdash M : T_{\mathcal{E}_1} \dot{A}}"] \\
			\dot{T}_{\interpret{\dot{\Gamma} \vdash \mathcal{E}_1 : \mathbf{Effect}}} \interpret{\dot{\Gamma} \vdash \dot{A}} \ar[d, "\dot{T}_{\interpret{\dot{\Gamma} \vdash \mathcal{E}_1 : \mathbf{Effect}}} \tupling{\identity{}}{!}"] \\
			\dot{T}_{\interpret{\dot{\Gamma} \vdash \mathcal{E}_1 : \mathbf{Effect}}} \coprod_{\interpret{\dot{\Gamma} \vdash \dot{A}}} \pi_{\interpret{\dot{\Gamma} \vdash \dot{A}}}^{*} 1 \interpret{\dot{\Gamma}} \ar[d, equal] \\
			\dot{T}_{\interpret{\dot{\Gamma} \vdash \mathcal{E}_1 : \mathbf{Effect}}} \coprod_{\interpret{\dot{\Gamma} \vdash \dot{A}}} 1 \interpret{\dot{\Gamma}, x : \dot{A}} \ar[d, "\dot{T}_{\interpret{\mathcal{E}_1}} \coprod_{\interpret{\dot{\Gamma} \vdash \dot{A}}} \interpret{N}"] \\
			\dot{T}_{\interpret{\dot{\Gamma} \vdash \mathcal{E}_1 : \mathbf{Effect}}} \coprod_{\interpret{\dot{\Gamma} \vdash \dot{A}}} \interpret{\dot{\Gamma}, x : \dot{A} \vdash T_{\mathcal{E}_2} \dot{B}} \ar[d, equal, "\text{Lemma~\ref{lem:weakening}}"] \\
			\dot{T}_{\interpret{\dot{\Gamma} \vdash \mathcal{E}_1 : \mathbf{Effect}}} \coprod_{\interpret{\dot{\Gamma} \vdash \dot{A}}} \pi_{\interpret{\dot{\Gamma} \vdash \dot{A}}}^{*} \dot{T}_{\interpret{\dot{\Gamma} \vdash \mathcal{E}_2 : \mathbf{Effect}}} \interpret{\dot{\Gamma} \vdash \dot{B}} \ar[d, "\dot{T}_{\interpret{\mathcal{E}_1}} \epsilon"] \\
			\dot{T}_{\interpret{\dot{\Gamma} \vdash \mathcal{E}_1 : \mathbf{Effect}}} \dot{T}_{\interpret{\dot{\Gamma} \vdash \mathcal{E}_2 : \mathbf{Effect}}} \interpret{\dot{\Gamma} \vdash \dot{B}} \ar[d, "\mu^{\dot{T}}"] \\
			\interpret{\dot{\Gamma} \vdash T_{\mathcal{E}_1 \GradeMult \mathcal{E}_2} \dot{B}}
		\end{tikzcd}}
	}
\end{mathpar}
Here, recall that a CCompC has fibred products, which are given by $X \times Y = \coprod_X \pi_X^{*} Y$.
\begin{align}
	&u \interpret{\dot{\Gamma} \vdash \letin{x}{M}{N} : T_{\mathcal{E}_1 \GradeMult \mathcal{E}_2} \dot{B}} \\
	&= (\identity{}, \mu^T \comp \pi_2) \comp (\identity{}, T (\pi_2 \comp \pi_2) \comp \strength^T) \comp (\identity{}, T \tupling{\pi_1 \comp \pi_2}{\interpret{N} \comp \pi_1 \comp \associator^{-1}} \comp \strength^T) \\
	&\qquad\qquad \comp (\identity{}, T (\tupling{\identity{}}{!} \comp \pi_2) \comp \strength^T) \comp (\identity{}, \interpret{M} \comp \pi_1) \\
	&= (\identity{}, \mu^T \comp \pi_2) \comp (\identity{}, T (\pi_2 \comp \pi_2 \comp \tupling{\pi_1}{\tupling{\pi_1 \comp \pi_2}{\interpret{N} \comp \pi_1 \comp \associator^{-1}}} \comp \tupling{\pi_1}{\tupling{\identity{}}{!} \comp \pi_2}) \comp \strength^T) \\
	&\qquad\qquad \comp (\identity{}, \interpret{M} \comp \pi_1) \\
	&= (\identity{}, \mu^T \comp \pi_2) \comp (\identity{}, T \interpret{N} \comp \strength^T) \comp (\identity{}, \interpret{M} \comp \pi_1) \\
	&= (\identity{}, \mu^T \comp \pi_2 \comp \tupling{\pi_1}{T \interpret{N} \comp \strength^T} \comp \tupling{\pi_1}{\interpret{M} \comp \pi_1}) \\
	&= (\identity{}, \mu^T \comp T \interpret{N} \comp \strength^T \comp \tupling{\identity{}}{\interpret{M}} \comp \pi_1) \qedhere
\end{align}
\end{proof}

\begin{lemma}
	\hyperlink{rule:CT-GenEff}{\textsc{CT-GenEff}} is sound.
\end{lemma}
\begin{proof}
\begin{mathpar}
	\inferrule{
		\interpret{\dot{\Gamma} \vdash V : \dot{A}} \in \{ s(\category{C}) \mid \category{P} \}_{\interpret{\dot{\Gamma}}} (1 \interpret{\dot{\Gamma}}, \interpret{\dot{\Gamma} \vdash \dot{A}}) \\
		\tupling{\identity{}}{\interpret{\mathtt{gef}} \comp \pi_2} : \interpret{x : \dot{A}} \dotTo \{ \interpret{x : \dot{A} \vdash T_{\mathcal{E}} \dot{B}} \}
	}{
		\interpret{\dot{\Gamma} \vdash \mathtt{gef}\ V : T_{\mathcal{E}[V/x]} \dot{B}[V/x]} \qquad\coloneqq\qquad
		{\begin{tikzcd}
			\text{(similar to $\mathtt{op}\ V$)}
		\end{tikzcd}}
	}
\end{mathpar}
\end{proof}

\begin{lemma}
	\hyperlink{rule:CT-Sub}{\textsc{CT-Sub}} is sound.
\end{lemma}
\begin{proof}
\begin{mathpar}
	\inferrule{
		\interpret{\dot{\Gamma} \vdash M : \dot{C}} \\
		\interpret{\dot{\Gamma} \vdash \dot{C}} <: \interpret{\dot{\Gamma} \vdash \dot{D}}
	}{
		\interpret{\dot{\Gamma} \vdash M : \dot{D}} \qquad\coloneqq\qquad
		{\begin{tikzcd}
			1 \interpret{\dot{\Gamma}} \ar[d, "\interpret{\dot{\Gamma} \vdash M : \dot{C}}"] \\
			\interpret{\dot{\Gamma} \vdash \dot{C}} \ar[d, "<:"] \\
			\interpret{\dot{\Gamma} \vdash \dot{D}}
		\end{tikzcd}}
	}
\end{mathpar}
\end{proof}

\begin{lemma}
	\hyperlink{rule:CT-PatternMatch}{\textsc{CT-PatternMatch}} is sound.
\end{lemma}
\begin{proof}
\begin{mathpar}
	\footnotesize
	\inferrule{
		\interpret{\dot{\Gamma} \vdash V : (x : \dot{A}) \times \dot{B}} \in \{ s(\category{C}) \mid \category{P} \}_{\interpret{\dot{\Gamma}}} (1 \interpret{\dot{\Gamma}}, \interpret{\dot{\Gamma} \vdash (x : \dot{A}) \times \dot{B}}) \\
		\interpret{\dot{\Gamma}, z : (x : \dot{A}) \times \dot{B} \vdash \dot{C}} \in \{ s(\category{C}) \mid \category{P} \}_{\interpret{\dot{\Gamma}, z : (x : \dot{A}) \times \dot{B}}} \\
		\interpret{\dot{\Gamma}, x : \dot{A}, y : \dot{B} \vdash M : \dot{C}[(x, y)/ z]} \in \{ s(\category{C}) \mid \category{P} \}_{\interpret{\dot{\Gamma}, x : \dot{A}, y : \dot{B}}}(1 \interpret{\dot{\Gamma}, x : \dot{A}, y : \dot{B}}, \interpret{\dot{\Gamma}, x : \dot{A}, y : \dot{B} \vdash \dot{C}[(x, y)/ z]})
	}{
		\interpret{\dot{\Gamma} \vdash \patternmatch{V}{x}{y}{M} : \dot{C}[V/z]} \qquad\coloneqq\qquad
		{\begin{tikzcd}
			1 \interpret{\dot{\Gamma}} \ar[d, equal] \\
			(s \interpret{\dot{\Gamma} \vdash V : (x : \dot{A}) \times \dot{B}})^{*} (\kappa^{-1})^{*} 1 \interpret{\dot{\Gamma}, x : \dot{A}, y : \dot{B}} \ar[d, "(s \interpret{\dot{\Gamma} \vdash V : (x : \dot{A}) \times \dot{B}})^{*} (\kappa^{-1})^{*} \interpret{\dot{\Gamma}, x : \dot{A}, y : \dot{B} \vdash M : \dot{C}}"] \\
			(s \interpret{\dot{\Gamma} \vdash V : (x : \dot{A}) \times \dot{B}})^{*} (\kappa^{-1})^{*} \kappa^{*} \interpret{\dot{\Gamma}, z : (x : \dot{A}) \times \dot{B} \vdash \dot{C}} \ar[d, equal] \\
			(s \interpret{\dot{\Gamma} \vdash V : (x : \dot{A}) \times \dot{B}})^{*} \interpret{\dot{\Gamma}, z : (x : \dot{A}) \times \dot{B} \vdash \dot{C}} \ar[d, equal, "\text{Lemma~\ref{lem:subst}}"] \\
			\interpret{\dot{\Gamma} \vdash \dot{C}[V/z]}
		\end{tikzcd}}
	}
\end{mathpar}

Here, note that we have
\[ \kappa^{*} \interpret{\dot{\Gamma}, z : (x : \dot{A}) \times \dot{B} \vdash \dot{C}} = \interpret{\dot{\Gamma}, x : \dot{A}, y : \dot{B} \vdash \dot{C}[(x, y)/z]} \]
because $\kappa$ is obtained as the following composite.
\begin{equation}
	\begin{tikzcd}
		\interpret{\dot{\Gamma}, x : \dot{A}, y : \dot{B}} \ar[d, "\mathrm{subst}_{\dot{\Gamma}, x : \dot{A}, y : \dot{B}; z : (x : \dot{A}) \times \dot{B}; \diamond; (x, y)}"] \\
		\interpret{\dot{\Gamma}, x : \dot{A}, y : \dot{B}, z : (x : \dot{A}) \times \dot{B}} \ar[d, "\mathrm{proj}_{\dot{\Gamma}, x : \dot{A}; y : \dot{B}; z : (x : \dot{A}) \times \dot{B}}"] \\
		\interpret{\dot{\Gamma}, x : \dot{A}, z : (x : \dot{A}) \times \dot{B}} \ar[d, "\mathrm{proj}_{\dot{\Gamma}; x : \dot{A}; z : (x : \dot{A}) \times \dot{B}}"] \\
		\interpret{\dot{\Gamma}, z : (x : \dot{A}) \times \dot{B}}
	\end{tikzcd}
\end{equation}
where
\begin{align}
	\mathrm{subst}_{\dot{\Gamma}, x : \dot{A}, y : \dot{B}; z : (x : \dot{A}) \times \dot{B}; \diamond; (x, y)} &= s \interpret{\dot{\Gamma}, x : \dot{A}, y : \dot{B} \vdash (x, y) : (x : \dot{A}) \times \dot{B}} \\
	\mathrm{proj}_{\dot{\Gamma}, x : \dot{A}; y : \dot{B}; z : (x : \dot{A}) \times \dot{B}} &= \{ \overline{\pi_{\interpret{\dot{\Gamma}, x : \dot{A} \vdash \dot{B}}}}(\interpret{\dot{\Gamma}, x : \dot{A} \vdash (x : \dot{A}) \times \dot{B}}) \} \\
	\mathrm{proj}_{\dot{\Gamma}; x : \dot{A}; z : (x : \dot{A}) \times \dot{B}} &= \{ \overline{\pi_{\interpret{\dot{\Gamma} \vdash \dot{A}}}}(\interpret{\dot{\Gamma} \vdash (x : \dot{A}) \times \dot{B}}) \} \\
\end{align}

\begin{align}
	u \interpret{\dot{\Gamma} \vdash \patternmatch{V}{x}{y}{M} : \dot{C}[V/z]} &= (\identity{}, \interpret{M} \comp \pi_1 \comp (\associator^{-1} \times \identity{}) \comp (\tupling{\identity{}}{\interpret{V}} \times \identity{})) \\
	&= (\identity{}, \interpret{M} \comp \associator^{-1} \comp \tupling{\identity{}}{\interpret{V}} \comp \pi_1)
\end{align}
\end{proof}

\begin{lemma}
	\hyperlink{rule:CT-Case}{\textsc{CT-Case}} is sound.
\end{lemma}
\begin{proof}
\begin{mathpar}
	\footnotesize
	\inferrule{
		\interpret{\dot{\Gamma} \vdash V : \dot{A} + \dot{B}} \in \{ s(\category{C}) \mid \category{P} \}_{\interpret{\dot{\Gamma}}}(1 \interpret{\dot{\Gamma}}, \interpret{\dot{\Gamma} \vdash \dot{A} + \dot{B}}) \\
		\interpret{\dot{\Gamma}, z : \dot{A} + \dot{B} \vdash \dot{C}} \in \{ s(\category{C}) \mid \category{P} \}_{\interpret{\dot{\Gamma}, z : \dot{A} + \dot{B}}} \\
		\interpret{\dot{\Gamma}, x : \dot{A} \vdash M : \dot{C}[\iota_1\ x/z]} \in \{ s(\category{C}) \mid \category{P} \}_{\interpret{\dot{\Gamma}, x : \dot{A}}}(1 \interpret{\dot{\Gamma}, x : \dot{A}}, \interpret{\dot{\Gamma}, x : \dot{A} \vdash \dot{C}[\iota_1\ x/z]}) \\
		\interpret{\dot{\Gamma}, y : \dot{B} \vdash N : \dot{C}[\iota_2\ y/z]} \in \{ s(\category{C}) \mid \category{P} \}_{\interpret{\dot{\Gamma}, y : \dot{B}}}(1 \interpret{\dot{\Gamma}, y : \dot{B}}, \interpret{\dot{\Gamma}, y : \dot{B} \vdash \dot{C}[\iota_2\ x/z]})
	}{
		\interpret{\dot{\Gamma} \vdash \caseof{V}{\casepattern{\leftinj{x}}{M}, \casepattern{\rightinj{y}}{N}} : \dot{C}[V/z]} \qquad\coloneqq\qquad
		{\begin{tikzcd}
			1 \interpret{\dot{\Gamma}} \ar[d, equal] \\
			(s \interpret{\dot{\Gamma} \vdash V : \dot{A} + \dot{B}})^{*} 1 \{ \interpret{\dot{\Gamma} \vdash \dot{A}} + \interpret{\dot{\Gamma} \vdash \dot{B}} \} \ar[d] \\
			(s \interpret{\dot{\Gamma} \vdash V : \dot{A} + \dot{B}})^{*} \interpret{\dot{\Gamma}, z : \dot{A} + \dot{B} \vdash \dot{C}} \ar[d, equal, "\text{Lemma~\ref{lem:subst}}"] \\
			\interpret{\dot{\Gamma} \vdash \dot{C}[V/z]}
		\end{tikzcd}}
	}
\end{mathpar}
Note the following.
\begin{itemize}
	\item If a SCCompC $\category{E} \to \category{B}$ has strong fibred coproducts, then we have the isomorphism $\tupling{\{ \iota_1 \}^{*}}{\{ \iota_2 \}^{*}} : \category{E}_{\{ X + Y \}}(A, B) \to \category{E}_{\{ X \}}(\{ \iota_1 \}^{*} A, \{ \iota_1 \}^{*} B) \times \category{E}_{\{ Y \}}(\{ \iota_2 \}^{*} A, \{ \iota_2 \}^{*} B)$.
	In the simple fibration, this isomorphism is given as follows.
	For each $(\identity{}, f) : (I \times (X + Y), A) \to (I \times (X + Y), B)$, we have $(\identity{}, f \comp ((\identity{} \times \iota_1) \times \identity{})) : (I \times X, A) \to (I \times X, B)$ and $(\identity{}, f \comp ((\identity{} \times \iota_2) \times \identity{})) : (I \times Y, A) \to (I \times Y, B)$.
	Conversely, for each $(\identity{}, f_1) : (I \times X, A) \to (I \times X, B)$ and $(\identity{}, f_2) : (I \times Y, A) \to (I \times Y, B)$, we have $(\identity{}, [f_1, f_2] \comp d) : (I \times (X + Y), A) \to (I \times (X + Y), B)$ where $d : (I \times (X + Y)) \times A \to (I \times X) \times A + (I \times Y) \times A$ is the inverse of $[(\identity{} \times \iota_1) \times \identity{}, (\identity{} \times \iota_2) \times \identity{}]$, which exists in any distributive category.
	\item We have $\{ \iota_1 \}^{*} 1 \interpret{\dot{\Gamma}, z : \dot{A} + \dot{B}} = 1 \interpret{\dot{\Gamma}, x : \dot{A}}$ and $\{ \iota_2 \}^{*} 1 \interpret{\dot{\Gamma}, z : \dot{A} + \dot{B}} = 1 \interpret{\dot{\Gamma}, y : \dot{B}}$.
	\item By Lemma~\ref{lem:weakening} and~\ref{lem:subst}, we have
	\begin{align}
		&\interpret{\dot{\Gamma}, x : \dot{A} \vdash \dot{C}[\iota_1\ x/z]} \\
		&= (s \interpret{\dot{\Gamma}, x : \dot{A} \vdash \iota_1\ x : \dot{A} + \dot{B}})^{*} \interpret{\dot{\Gamma}, x : \dot{A}, z : \dot{A} + \dot{B} \vdash \dot{C}} \\
		&= (s \interpret{\dot{\Gamma}, x : \dot{A} \vdash \iota_1\ x : \dot{A} + \dot{B}})^{*} \{ \overline{\pi_{\interpret{\dot{\Gamma} \vdash \dot{A}}}}(\interpret{\dot{\Gamma} \vdash \dot{A} + \dot{B}}) \}^{*} \interpret{\dot{\Gamma}, z : \dot{A} + \dot{B} \vdash \dot{C}} \\
		&= \{ \iota_1 \}^{*} \interpret{\dot{\Gamma}, z : \dot{A} + \dot{B} \vdash \dot{C}}
	\end{align}
\end{itemize}

\begin{align}
	&u \interpret{\dot{\Gamma} \vdash \caseof{V}{\casepattern{\leftinj{x}}{M}, \casepattern{\rightinj{y}}{N}} : \dot{C}[V/z]} \\
	&= (\identity{}, [\interpret{M} \comp \pi_1, \interpret{N} \comp \pi_1] \comp d \comp (\tupling{\identity{}}{\interpret{V}} \times \identity{})) \\
	&= (\identity{}, [\interpret{M}, \interpret{N}] \comp d' \comp \pi_1 \comp (\tupling{\identity{}}{\interpret{V}} \times \identity{})) \\
	&= (\identity{}, [\interpret{M}, \interpret{N}] \comp d' \comp \tupling{\identity{}}{\interpret{V}} \comp \pi_1)
\end{align}
where $d' : \interpret{\underlying{\dot{\Gamma}}} \times (\interpret{\underlying{\dot{A}}} + \interpret{\underlying{\dot{A}}}) \to \interpret{\underlying{\dot{\Gamma}}} \times \interpret{\underlying{\dot{A}}} + \interpret{\underlying{\dot{\Gamma}}} \times \interpret{\underlying{\dot{B}}}$ is the canonical isomorphism.
\end{proof}

\paragraph{Subtyping}
Most of the cases are standard.

\begin{mathpar}
	\inferrule{
		\interpret{\mathcal{E}_1} \le \interpret{\mathcal{E}_2} \in \category{M}_{\interpret{\dot{\Gamma}}} \\
		\interpret{\dot{\Gamma} \vdash \dot{A}} <: \interpret{\dot{\Gamma} \vdash \dot{B}}
	}{
		\interpret{\dot{\Gamma} \vdash T_{\mathcal{E}_1} \dot{A}} \xrightarrow{\dot{T}_{\interpret{\mathcal{E}_1} \le \interpret{\mathcal{E}_2}} (\interpret{\dot{\Gamma} \vdash \dot{A}} <: \interpret{\dot{\Gamma} \vdash \dot{B}})} \interpret{\dot{\Gamma} \vdash T_{\mathcal{E}_2} \dot{B}} 
	}
\end{mathpar}
\begin{align}
	u \dot{T}_{\interpret{\mathcal{E}_1} \le \interpret{\mathcal{E}_2}} (\interpret{\dot{\Gamma} \vdash \dot{A}} <: \interpret{\dot{\Gamma} \vdash \dot{B}}) &= \hat{T} u (\interpret{\dot{\Gamma} \vdash \dot{A}} <: \interpret{\dot{\Gamma} \vdash \dot{B}}) \\
	&= \identity{} \\
	r \dot{T}_{\interpret{\mathcal{E}_1} \le \interpret{\mathcal{E}_2}} (\interpret{\dot{\Gamma} \vdash \dot{A}} <: \interpret{\dot{\Gamma} \vdash \dot{B}}) &= r (\interpret{\dot{\Gamma} \vdash \dot{A}} <: \interpret{\dot{\Gamma} \vdash \dot{B}}) \\
	&= \identity{}
\end{align}
\end{proof}

\subsection{With Recursion}\label{sec:fgcbv-model-with-recursion}

To interpret recursion, we require the category $\category{C}$ to be an $\omegaCPO$-enriched category and the monad $T$ to have a bottom element so that we can interpret recursive functions as least fixed points.
We follow the definition of models in \cite{KatsumataInfComput2013}.

\begin{definition}[$\omegaCPO$-enriched FGCBV model]\label{def:omegaCPO-enriched-fgcbv-model}
	A bicartesian closed category $\category{C}$ is an \emph{$\omegaCPO$-enriched bicartesian closed category} if $\category{C}$ is an $\omegaCPO$-enriched category; and tupling $\tupling{{-}}{{-}} : \category{C}(X, Y) \times \category{C}(X, Z) \to \category{C}(X, Y \times Z)$, currying $\Lambda : \category{C}(X \times Y, Z) \to \category{C}(X, \exponential{Y}{Z})$, and cotupling $[{-}. {-}] : \category{C}(X, Z) \times \category{C}(Y, Z) \to \category{C}(X + Y, Z)$ are Scott continuous.
	A strong monad $T$ on an $\omegaCPO$-enriched bicartesian closed category $\category{C}$ is a \emph{pseudo-lifting strong monad} if it has a morphism $\bot_{T 0} \in \category{C}(1, T 0)$ such that $T {?} \comp \bot_{T 0} \in \category{C}(1, T X)$ is the bottom element for any $X$ where ${?} : 0 \to X$ is the unique morphism from the initial object $0$.
	An \emph{$\omegaCPO$-enriched FGCBV model} is a FGCBV model $(\category{C}, T, \interpret{{-}})$ such that $\category{C}$ is an $\omegaCPO$-enriched bicartesian closed category, and $T$ is a pseudo-lifting strong monad \cite{KatsumataInfComput2013}.
\end{definition}

Given an $\omegaCPO$-enriched FGCBV model, the interpretation $\interpret{\recfun{f}{x}{M}} : \interpret{\Gamma} \to \interpret{A \to T B}$ of a recursive function $\Gamma \vdash \recfun{f}{x}{M} : A \to T B$ is defined as the (parameterized) least fixed point of $\interpret{\lambda x. M} = \Lambda \interpret{M} : \interpret{\Gamma} \times \interpret{A \to T B} \to \interpret{A \to T B}$.

\begin{example}[cost analysis]\label{ex:cost-analysis-model}
	We consider an $\omegaCPO$-enriched version of Example~\ref{ex:cost-analysis-model-wo-recursion}.
	An $\omegaCPO$-enriched FGCBV model for cost analysis is given as $(\omegaCPO, ({-})_{\bot} \times \mathbb{N}_{\infty}, \interpret{{-}})$ where $\mathbb{N}_{\infty} = (\mathbb{N} \cup \{ \infty \}, {\le})$ is the $\omega$cpo of extended natural numbers with the usual order, and $({-}) \times \mathbb{N}_{\infty}$ is the cost monad induced by the additive monoid $(\mathbb{N}_{\infty}, 0, {+})$.
	The lifting monad $({-})_{\bot}$ and the cost monad is composed by a distributive law, which gives us the pseudo-lifting strong monad $({-})_{\bot} \times \mathbb{N}_{\infty}$ with the bottom element $\bot_{T 0} = (\bot, 0) \in 0_{\bot} \times \mathbb{N}_{\infty}$.
	Note that this combination of monads keeps track of the cost even if the computation is non-terminating.
\end{example}

Let $(\category{C}, T, \interpret{{-}})$ be an $\omegaCPO$-enriched FGCBV model.
Then, it satisfies the following properties.

\begin{lemma}
	For any Eilenberg--Moore $T$-algebra $\alpha : T A \to A$, $\bot_{A} \coloneqq \alpha \comp \bot_{T A}$ is the least element in $\category{C}(1, A)$.
	\qed
\end{lemma}

\begin{lemma}
	$\bot_A \comp {!}_X \in \category{C}(X, A)$ is the least element for any $X$ and any Eilenberg--Moore algebra $A$.
\end{lemma}
\begin{proof}
	Since $\category{C}$ is $\omegaCPO$-enriched bi-cartesian closed, we have the canonical isomorphism $\category{C}(X, A) \cong \category{C}(1, \exponential{X}{A})$ between $\omega$cpos.
	The least element in $\category{C}(X, A)$ is given by $\Lambda^{-1}(\bot_{\exponential{X}{A}}) \comp \tupling{!}{\identity{}}$, which is equal to $\bot_A \comp {!}_X$ as follows.
	\begin{align}
		\bot_{\exponential{X}{A}} &= \alpha_{\exponential{X}{A}} \comp \bot_{T \exponential{X}{A}} \\
		&= \Lambda(\alpha_A \comp T \eval \comp \strength') \comp T {?}_{\exponential{X}{A}} \comp \bot_{T 0} \\
		\Lambda^{-1}(\bot_{\exponential{X}{A}}) \comp \tupling{!}{\identity{}} &= \alpha_A \comp T \eval \comp \strength' \comp \tupling{T {?}_{\exponential{X}{A}} \comp \bot_{T 0} \comp {!}}{\identity{}} \\
		&= \alpha_A \comp T \eval \comp T ({?}_{\exponential{X}{A}} \times \identity{}) \comp \strength' \comp \tupling{\bot_{T 0} \comp {!}}{\identity{}} \\
		&= \alpha_A \comp T {?}_A \comp T \pi_1 \comp \strength' \comp \tupling{\bot_{T 0} \comp {!}}{\identity{}} \\
		&= \alpha_A \comp T {?}_A \comp \pi_1 \comp \tupling{\bot_{T 0} \comp {!}}{\identity{}} \\
		&= \alpha_A \comp T {?}_A \comp \bot_{T 0} \comp {!} \\
		&= \bot_A \comp {!}_X
	\end{align}
\end{proof}

\begin{proposition}
	The parameterized uniform $T$-fixed-point operator is given by
	\[ f^{\dagger} = \mathrm{lfp} (f \comp \tupling{\identity{}}{{-}}) = \sup_n (f \comp \tupling{\identity{}}{{-}})^n(\bot \comp {!}) \]
	for each $f : X \times A \to A$.
\end{proposition}

To prove the soundness of the typing rule for recursion, we need to impose an additional condition on predicates so that the fixed-point operator does not escape from the predicate.
We use the classical notion of \emph{admissibility} of predicates to achieve this.

Let $X$ be an $\omega$cpo with a least element $\bot_X$.
A subset $P \subseteq X$ is \emph{admissible} if $P$ contains the bottom element and closed under $\sup$ of $\omega$-chains, that is, $P$ is admissible if we have $\bot_X \in P$ and $\sup_n x_n \in P$ for any $\omega$-chain $\{ x_n \}_{n \in \mathbb{N}} \subseteq P$.
For any Scott-continuous function $f : X \to X$, if $P \subseteq X$ is an admissible subset that is preserved by $f$ (i.e.\ $f(x) \in P$ for any $x \in P$), then the least fixed point $\mu f$ of $f$ is in $P$.
Admissibility of predicates can be adapted to the refinement fibrations \cite{KuraICFP2024}.

\begin{definition}\label{def:admissible-em-algebra}
	Let $(X, A, P, Q) \in \{ s(\category{C}) \mid \category{P} \}$ be an object and $\nu : T A \to A$ be an Eilenberg--Moore $T$-algebra.
	The object $(X, A, P, Q)$ is \emph{admissible} if the following conditions are satisfied.
	\begin{itemize}
		\item $\tupling{\identity{}}{\bot_{A} \comp {!}} : P \dotTo Q$.
		\item For each $\omega$-chain $\{ f_n : X \to A \}_{n=0}^{\infty}$ such that $\tupling{\identity{}}{f_n} : P \dotTo Q$ for each $n$, we have $\tupling{\identity{}}{\sup_n f_n} : P \dotTo Q$.
	\end{itemize}
\end{definition}

\begin{definition}\label{def:admissible-graded-monad}
	Let $\dot{T}$ be an indexed graded monad over $\{ s(\category{C}) \mid \category{P} \}$, $(X, Y, P, Q) \in \{ s(\category{C}) \mid \category{P} \}$, and $f \in \category{M}_{P}$.
	We say $\dot{T}_f (X, Y, P, Q)$ is \emph{admissible} if the following conditions are satisfied.
	\begin{itemize}
		\item $\tupling{\identity{}}{\bot_{T X} \comp {!}} : P \dotTo Q'$ where $Q'$ is the last component of $\dot{T}_f (X, Y, P, Q) = (X, T Y, P, Q')$.
		\item For each $\omega$-chain $\{ f_n : X \to T Y \}_{n=0}^{\infty}$ such that $\tupling{\identity{}}{f_n} : P \dotTo Q'$ for each $n$, we have $\tupling{\identity{}}{\sup_n f_n} : P \dotTo Q'$.
	\end{itemize}
	The indexed graded monad $\dot{T}$ is \emph{admissible} on $(X, Y, P, Q) \in \{ s(\category{C}) \mid \category{P} \}$ if for each $f \in \category{M}_{P}$, $\dot{T}_f (X, Y, P, Q)$ is admissible.
\end{definition}

Now, the model of the dependent effect system with recursion is defined as follows.
\begin{definition}\label{def:dependent-effect-system-model-part3}
	An \emph{$\omegaCPO$-enriched model of the dependent effect system} is a model of the dependent effect system without recursion (Definition~\ref{def:dependent-effect-system-model-part2}) such that the following conditions are satisfied.
	\begin{itemize}
		\item $\category{C}$ is $\omegaCPO$-enriched bi-cartesian closed.
		\item $T$ is a pseudo-lifting strong monad. The least element in $\category{C}(1, T X)$ is given by $\bot_{T X} = T {?}_X \comp \bot_{T 0}$ where ${?}_X : 0 \to X$.
		\item For any (well-formed) computation type $\dot{\Gamma} \vdash T_{\mathcal{E}} \dot{A}$, the interpretation $\interpret{\dot{\Gamma} \vdash T_{\mathcal{E}} \dot{A}}$ is admissible (Definition~\ref{def:admissible-graded-monad}).
	\end{itemize}
\end{definition}

\begin{lemma}\label{lem:product-admissible}
	If $(X \times Y, A, Q, R)  \in \{ s(\category{C}) \mid \category{P} \}_{\{ (X, Y, P, Q) \}}$ is admissible, then so is $\prod_{(X, Y, P, Q)} (X \times Y, A, Q, R)$.
\end{lemma}
\begin{proof}
	In any SCCompC $\category{E} \to \category{B}$, we have the following isomorphisms for any $X \in \category{E}$ and $Y \in \category{E}_{\{ X \}}$.
	\begin{align}
		\{ f \in \category{B}(p X, \{ \prod_X Y \}) \mid \pi_X \comp f = \identity{} \} &\cong \category{E}_{p X}(1 p X, \prod_X Y) \\
		&\cong \category{E}_{\{X\}}(\pi_X^{*} 1 p X, Y) \\
		&= \category{E}_{\{X\}}(1 \{X\}, Y) \\
		&\cong \{ f \in \category{B}(\{X\}, \{ Y \}) \mid \pi_Y \comp f = \identity{} \}
	\end{align}
	By instantiating this isomorphism to the current situation, we have the following: for any $f : X \to \exponential{Y}{A}$, we have $\tupling{\identity{}}{f} : P \dotTo \{ \prod_{(X, Y, P, Q)} (X \times Y, A, Q, R) \}$ if and only if $\tupling{\identity{}}{\Lambda^{-1} f} : Q \dotTo R$.
	\begin{itemize}
		\item We have $\tupling{\identity{}}{\bot_{\exponential{Y}{A}} \comp {!}} : P \dotTo \{ \prod_{(X, Y, P, Q)} (X \times Y, A, Q, R) \}$ because $\Lambda^{-1} (\bot_{\exponential{Y}{A}} \comp {!}) = \Lambda^{-1} (\bot_{\exponential{Y}{A}}) \comp ({!} \times \identity{}) = \bot_A \comp {!}$ and $\tupling{\identity{}}{\bot_A \comp {!}} : Q \dotTo R$.
		\item Let $\{ f_n : X \to \exponential{Y}{A} \}_{n=0}^{\infty}$ be an $\omega$-chain such that $\tupling{\identity{}}{f_n} : P \dotTo \{ \prod_{(X, Y, P, Q)} (X \times Y, A, Q, R) \}$ for each $n$.
		Then, $\{ \tupling{\identity{}}{\Lambda^{-1} f_n} \}_{n = 0}^{\infty}$ is an $\omega$-chain such that $\tupling{\identity{}}{\Lambda^{-1} f_n} : Q \dotTo R$ for each $n$.
		Since $(X \times Y, A, Q, R)$ is admissible, we have $\tupling{\identity{}}{\sup_n \Lambda^{-1} f_n} = \tupling{\identity{}}{\Lambda^{-1} \sup_n f_n} : Q \dotTo R$.
		Therefore, we have $\tupling{\identity{}}{\sup_n f_n} : P \dotTo \{ \prod_{(X, Y, P, Q)} (X \times Y, A, Q, R) \}$.
	\end{itemize}
\end{proof}

\begin{theorem}[soundness]\label{thm:soundness-w-recursion}
	Given a model in Definition~\ref{def:dependent-effect-system-model-part3}, the soundness theorem (Theorem~\ref{thm:soundness-wo-recursion-terms}) is true including recursion.
\end{theorem}

\begin{lemma}
	Given a model in Definition~\ref{def:dependent-effect-system-model-part3}, \hyperlink{rule:CT-Rec}{\textsc{CT-Rec}} is sound.
\end{lemma}
\begin{proof}
	It suffices to prove $\tupling{\identity{}}{\interpret{\recfun{f}{x}{M}}} : \interpret{\dot{\Gamma}} \dotTo \{ \interpret{\dot{\Gamma} \vdash (x : \dot{A}) \to T_{\mathcal{E}} \dot{B}} \}$.
	By IH, we have $\interpret{\dot{\Gamma}, f : (x : \dot{A}) \to T_{\mathcal{E}} \dot{B}, x : \dot{A} \vdash M : T_{\mathcal{E}} \dot{B}} \in \{ s(\category{C}) \mid \category{P} \}_{\interpret{\dot{\Gamma}, f : (x : \dot{A}) \to T_{\mathcal{E}} \dot{B}, x : \dot{A}}} (1 \interpret{\dot{\Gamma}, f : (x : \dot{A}) \to T_{\mathcal{E}} \dot{B}, x : \dot{A}}, \interpret{\dot{\Gamma}, f : (x : \dot{A}) \to T_{\mathcal{E}} \dot{B}, x : \dot{A} \vdash T_{\mathcal{E}} \dot{B}})$.
	By the properties of SCCompCs, we have
	\begin{equation}
		\tupling{\pi_1}{\Lambda \interpret{M}} : \interpret{\dot{\Gamma}, f : (x : \dot{A}) \to T_{\mathcal{E}} \dot{B}} \dotTo \interpret{\dot{\Gamma}, f : (x : \dot{A}) \to T_{\mathcal{E}} \dot{B}} \label{eq:recursion}
	\end{equation}

	By Lemma~\ref{lem:product-admissible}, $\interpret{\dot{\Gamma}, f : (x : \dot{A}) \to T_{\mathcal{E}} \dot{B}}$ is admissible.
	By definition of admissibility, we have
	\[ \tupling{\identity{}}{\bot \comp {!}} : \interpret{\dot{\Gamma}} \dotTo \interpret{\dot{\Gamma}, f : (x : \dot{A}) \to T_{\mathcal{E}} \dot{B}}. \]
	By \eqref{eq:recursion}, we have
	\[ \tupling{\pi_1}{\Lambda \interpret{M}}^n \comp \tupling{\identity{}}{\bot \comp {!}} = \tupling{\identity{}}{(\Lambda \interpret{M} \comp \tupling{\identity{}}{{-}})^n (\bot \comp {!})} : \interpret{\dot{\Gamma}} \dotTo \interpret{\dot{\Gamma}, f : (x : \dot{A}) \to T_{\mathcal{E}} \dot{B}}. \]
	By admissibility, we have
	\[ \tupling{\identity{}}{\sup_n (\Lambda \interpret{M} \comp \tupling{\identity{}}{{-}})^n (\bot \comp {!})} = \tupling{\identity{}}{(\Lambda \interpret{M})^{\dagger}} : \interpret{\dot{\Gamma}} \dotTo \interpret{\dot{\Gamma}, f : (x : \dot{A}) \to T_{\mathcal{E}} \dot{B}}. \]
\end{proof}	
\end{toappendix}

\begin{toappendix}
\subsection{Fibrational Presentation of Indexed Graded Monads}\label{sec:fibrational-presentation}
By the Grothendieck construction, indexed preordered monoids and indexed graded monads can be presented in terms of fibrations.
We briefly describe this fibrational presentation here.
A $\category{B}$-indexed preordered monoid can be presented as preordered fibrations with monoid structures.

\begin{definition}[indexed preordered monoid]
	Let $\CategoryOfPreorderedMonoids$ be the category of preordered monoids.
	A \emph{$\category{B}$-indexed preordered monoid} is a split preordered fibration $m : \category{M} \to \category{B}$ together with a split fibred functor $\GradeUnit : \mathbf{Id}_{\category{B}} \to m$ and $({\GradeMult}) : m \times m \to m$ that satisfy the monoid laws.
	Here, we use the product $m \times m : \category{M} \times_{\category{B}} \category{M} \to \category{B}$ in the 2-category $\mathbf{Fib}(\category{B})$ of fibrations over $\category{B}$, which is given by pullback.
\end{definition}

\begin{example}\label{ex:indexed-preordered-monoid-as-fibration}
	Let $\GradingMonoid$ be a preordered monoid.
	By the Grothendieck construction, the indexed preordered monoid in Example~\ref{ex:preordered-monoid-valued-functions} corresponds to the domain fibration $\dom : \Set / \GradingMonoid \to \Set$.
	The indexed preordered monoid in Example~\ref{ex:simple-effect-indexed-preordered-monoid} corresponds to the projection fibration $\mathrm{pr}_{\category{B}, \GradingMonoid} : \category{B} \times \GradingMonoid \to \category{B}$.
\end{example}

Indexed graded monads can also be presented as graded monads internal to the 2-category of fibrations over a fixed base category.

\begin{definition}
	Let $\mathcal{K}$ be a 2-category with finite products and $\GradingMonoid = (M, I, {\otimes})$ be a monoidal object in $\mathcal{K}$.
	An \emph{$\GradingMonoid$-graded monad} in $\mathcal{K}$ is a tuple $(T, \eta, \mu)$ where $T : M \times C \to C$ is a 1-cell, $\eta : \identity{C} \to T \comp (I \times C) : C \to C$ is a 2-cell called a \emph{unit}, and $\mu : T \comp (M \times T) \to T \comp (({\otimes}) \times C) : M \times M \times C \to C$ is a 2-cell called a \emph{multiplication}.
	These components are required to satisfy the graded monad laws.
	Here, note that we omitted the associativity and the unit laws for the cartesian products in $\mathcal{K}$.
\end{definition}

\begin{definition}[indexed graded monad]
	Let $m : \category{M} \to \category{B}$ be a $\category{B}$-indexed preordered monoid and $p : \category{E} \to \category{B}$ be a fibration.
	An \emph{indexed $m$-graded monad} on $p$ is a graded monad $(T, \eta, \mu)$ on $p$ in the 2-category $\mathbf{Fib}(\category{B})$ of fibrations over base category $\category{B}$.
	Concretely, this consists of a tuple $(T, \eta, \mu)$ where $T : m \times p \to p$ is a fibred functor and $\eta, \mu$ are vertical natural transformations that satisfy the monad laws.
\end{definition}
\end{toappendix}

\section{Instances}\label{sec:instances}
\begin{figure}[tbp]
	\centering
	\setlength{\tabcolsep}{3pt}
	\begin{tabular}{ll|cc}
		\textbf{Instance} & & \textbf{Grading monoid} $\GradingMonoid$ & \textbf{Heyting algebra} $\Omega$ \\
		\hline
		Cost analysis & Section~\ref{sec:cost-analysis} & $(\mathbb{N}_{\infty},\ 0,\ {+},\ {\le})$ & $(2, {\le})$ \\ %
		Expectation bound & Section~\ref{sec:expectation-logic} & $([0, \infty],\ 0,\ {+},\ {\le})$ & $([0, \infty], {\ge})$ \\ %
		Temporal safety & \referappendix{Section}{6.3}{sec:temporal-safety} & $(\PowerSet{\Sigma^{*}},\ \{ \varepsilon \},\ ({\cdot}),\ {\subseteq})$ & $(2, {\le})$ \\ %
		Union bound & \referappendix{Section}{6.4}{sec:union-bound-logic} & $([0, \infty),\ 0,\ {+},\ {\le})$ & $(2, {\le})$ %
	\end{tabular}
	\caption{Instances of graded Hoare logics for simple effects.}
	\label{fig:instances}
\end{figure}

We extend several existing simple effect systems (Fig.~\ref{fig:instances}) to dependent effect systems using the results in Section~\ref{sec:indexed-graded-monad}.
\ifthenelse{\boolean{longversion}}{}{%
Due to space constraints, we present only two instances: cost analysis and expectation logic.
Other instances are presented in the long version~\cite{arxiv}.}

\subsection{Cost Analysis}\label{sec:cost-analysis}

\begin{toappendix}
	\subsection{Cost Analysis}
\end{toappendix}

We consider a dependent effect system for cost analysis.
Following Fig.~\ref{fig:instances}, we use the complete Heyting algebra $\Omega = \mathbf{2}$ of two-valued boolean algebra to interpret formulas.
The grading monoid is given by $\GradingMonoid = \mathbb{N}_{\infty} = (\mathbb{N}_{\infty}, 0, {+}, {\le})$.

\subsubsection{Generic Effect}
We use a generic effect $\mathtt{Tick} : (n : \NatType) \stackrel{n}{\rightarrowtriangle} \UnitType$ that takes a natural number $n$ and incurs $n$ units of cost.

\subsubsection{Model}
We define a model of the dependent effect system for cost analysis following Definition~\ref{def:dependent-effect-system-model}.
We use the simple FGCBV model $(\omegaCPO, ({-})_{\bot} \times \mathbb{N}_{\infty}, \interpret{{-}}_s)$ where the strong monad is given by the composite of the cost monad $({-}) \times \mathbb{N}_{\infty}$ and the lifting monad $({-})_{\bot}$.
We impose a mild additional assumption that the interpretation $\interpret{b}_s \in \omegaCPO$ of any base type $b \in \mathbf{Base}$ is \emph{discrete}, i.e., the order relation is given by the identity relation $({=})$.
Although this assumption is not necessary to interpret simple FGCBV, we need this to soundly interpret recursion in the dependent effect system (see \referappendix{Appendix}{D.2}{sec:discrete-types} for details).
As a model of formulas, we use the $\mathbf{2}$-valued predicate fibration $\mathrm{pred}_{\omegaCPO} : \mathbf{Pred}(\omegaCPO) \to \omegaCPO$ over $\omegaCPO$.

Now, we define a strong $\mathbb{N}_{\infty}$-graded monad lifting along the predicate fibration $\mathrm{pred}_{\omegaCPO}$.
Since the strong monad $({-})_{\bot} \times \mathbb{N}_{\infty}$ in the base category is defined as a composite, we define the graded monad lifting as the composite of two graded monad liftings.
As a lifting of the cost monad $({-}) \times \mathbb{N}_{\infty}$, we define a strong $\mathbb{N}_{\infty}$-graded monad lifting $\CostGradedMonad$ as follows: for $(X, P) \in \mathbf{Pred}(\omegaCPO)$,
\[ \CostGradedMonad_m (X, P) \quad\coloneqq\quad (X \times \mathbb{N}_{\infty}, \{ (x, n) \in X \times \mathbb{N}_{\infty} \mid x \in P \land n \le m \}). \]
Here, we implicitly identify $P \subseteq \omegaCPO(1, X)$ with a subset $P \subseteq X$.
\ifthenelse{\boolean{longversion}}{
It is easy to see that the unit and the multiplication for the cost monad have liftings, which give the unit and the multiplication for the graded monad lifting $\CostGradedMonad$.
As for a lifting of the lifting monad $({-})_{\bot}$, we define a strong ($1$-graded) monad lifting $\PartialCorrectnessMonad$ as follows, which represents partial correctness, since diverging computation $\bot \in X_{\bot}$ satisfies the predicate $\PartialCorrectnessMonad (X, P)$.
\begin{equation}
	\PartialCorrectnessMonad (X, P) \quad\coloneqq\quad (X_{\bot}, P \cup \{ \bot \}) \qquad \text{for $(X, P) \in \mathbf{Pred}(\omegaCPO)$}
	\label{eq:partial-correctness-monad}
\end{equation}
\vspace{-\baselineskip}
}{As for a lifting of the lifting monad $({-})_{\bot}$, we define a strong ($1$-graded) monad lifting $\PartialCorrectnessMonad$ as $\PartialCorrectnessMonad (X, P) \coloneqq (X_{\bot}, P \cup \{ \bot \})$ for each $(X, P) \in \mathbf{Pred}(\omegaCPO)$, which represents partial correctness, since diverging computation $\bot \in X_{\bot}$ satisfies the predicate $\PartialCorrectnessMonad (X, P)$.}
\begin{lemma}
	The composite $(m, (X, P)) \mapsto \CostGradedMonad_m \PartialCorrectnessMonad (X, P)$ is a strong $\mathbb{N}_{\infty}$-graded monad lifting of $({-})_{\bot} \times \mathbb{N}_{\infty}$ along $\mathrm{pred}_{\omegaCPO} : \mathbf{Pred}(\omegaCPO) \to \omegaCPO$.
	\qed
\end{lemma}

As for the interpretation of basic effects, we define the interpretation of $\mathtt{nat2eff} : \NatType \to \mathbf{Effect}$ as the embedding $i : \mathbb{N} \to \mathbb{N}_{\infty}$.

Now, we show that the model satisfies the conditions in Definition~\ref{def:dependent-effect-system-model}.
It is straightforward to show \ref{item:model-axiom-lifting}.
For example, consider the generic effect $\mathtt{Tick} : (n : \NatType) \stackrel{n}{\rightarrowtriangle} \UnitType$.
The interpretation $\interpret{\mathtt{Tick}} : \mathbb{N} \to 1_{\bot} \times \mathbb{N}_{\infty}$ in the simple FGCBV model has a lifting along the predicate fibration $\mathrm{pred}_{\omegaCPO}$ when paired with the identity: $\tupling{\identity{}}{\interpret{\mathtt{Tick}}} : \{ n \in \mathbb{N} \mid \top \} \dotTo \{ (n, (r, c)) \in \mathbb{N} \times (1_{\bot} \times \mathbb{N}_{\infty}) \mid c \le n \}$.
This validates the refinement type signature $\mathtt{Tick} : (n : \NatType) \stackrel{n}{\rightarrowtriangle} \UnitType$.
\ref{item:model-axiom-coproduct} is satisfied by $\omegaCPO$ since we have $\omegaCPO(1, X + Y) \cong \omegaCPO(1, X) + \omegaCPO(1, Y)$.

\begin{toappendix}
\begin{definition}[monad lifting for partial correctness]
	Assume $\category{C} = \Set, \omegaCPO, \omegaQBS$.
	We define $\PartialCorrectnessMonad$ as a monad lifting of $({-})_{\bot}$ along $\mathbf{Pred}_{\Omega}(\category{C}) \to \category{C}$ as follows.
	\[ \PartialCorrectnessMonad (X, P) \coloneqq (X_{\bot}, [P, \top_{\Omega}]) \]
	where $[P, \top_{\Omega}] : \category{C}(1, X_{\bot}) \cong \category{C}(1, X) + 1 \to \Omega$ is the co-tupling of $P$ and $\top_{\Omega}$.
\end{definition}

\begin{lemma}\label{lem:partial-correctness-distributive-law}
	Let $\category{C} = \omegaCPO, \omegaQBS$.
	If
	\begin{itemize}
		\item $\dot{T}$ is a strong $\GradingMonoid$-graded monad lifting of a strong monad $T$,
		\item $\eta^T$ is strict (i.e.\ $\eta^T$ maps bottom element to bottom element), and
		\item $\eta^T$ has a lifting $\eta^T : \PartialCorrectnessMonad (0, {?}) \dotTo T_m \PartialCorrectnessMonad (0, {?})$ for each $m \in \GradingMonoid$ (i.e.\ $\top_{\Omega} \le (T_m \PartialCorrectnessMonad P)(\eta^T \comp \bot)$ for any $(X, P) \in \mathbf{Pred}_{\Omega}(\category{C})$ and $m \in \GradingMonoid$),
	\end{itemize}
	then $\dot{T} \PartialCorrectnessMonad$ is a strong $\GradingMonoid$-graded monad lifting of $T({-})_{\bot}$ along $\mathbf{Pred}_{\Omega}(\category{C}) \to \category{C}$.
	Note that $\eta^T : \PartialCorrectnessMonad (0, {?}) \dotTo T_m \PartialCorrectnessMonad (0, {?})$ always exists if the unit $\GradeUnit \in \GradingMonoid$ is the least element in $\GradingMonoid$.
\end{lemma}
\begin{proof}
	We have a distributive law $d_X : (T X)_{\bot} \to T (X_{\bot})$ between strong monads if $\eta^T$ is strict.
	\[ d = [T \eta^{({-})_{\bot}}, \eta^T \comp \bot] \]
	We show $d$ lifts to a distributive law $d_X : \PartialCorrectnessMonad \dot{T}_m (X, P) \to \dot{T}_m \PartialCorrectnessMonad (X, P)$.
	Let $x \in \category{C}(1, (T X)_{\bot})$.
	\begin{itemize}
		\item If $x = \bot$, then $d \comp x = \eta^T \comp \bot$.
		We have $(\PartialCorrectnessMonad \dot{T}_m P)(x) \le (\dot{T}_m \PartialCorrectnessMonad P)(d \comp x)$ because $\top_{\Omega} \le (T_m \PartialCorrectnessMonad P)(\eta^T \comp \bot)$
		\item If $x \in \category{C}(1, T X)$, then $d \comp x = T \eta^{({-})_{\bot}} \comp x$ and $(\PartialCorrectnessMonad \dot{T}_m P)(x) = (\dot{T}_m P)(x)$.
		Since $T \eta^{({-})_{\bot}} : \dot{T}_m P \to \dot{T}_m \PartialCorrectnessMonad P$, we conclude $(\PartialCorrectnessMonad \dot{T}_m P)(x) \le (\dot{T}_m \PartialCorrectnessMonad P)(d \comp x)$.
		\qedhere
	\end{itemize}
\end{proof}
\end{toappendix}

\begin{toappendix}
\subsection{Discreteness of Ground Types}\label{sec:discrete-types}

\begin{lemma}
	If $\Gamma \vdash V : A$ is well-typed and $A$ is a ground type, then $V$ is in the following fragment of value terms.
	We call them \emph{ground value terms}.
	\[ V, W \coloneqq x \mid c\ V \mid () \mid (V, W) \mid \leftinj{V} \mid \rightinj{V} \mid \leftproj{V} \mid \rightproj{V} \]
	Moreover, we have $\mathrm{Gnd}(\Gamma) \vdash V : A$ where $\mathrm{Gnd}(\Gamma)$ is the restriction of $\Gamma$ to ground types.
	\[ \mathrm{Gnd}(\diamond) \coloneqq \diamond \qquad \mathrm{Gnd}(\Gamma, x : A) = \begin{cases}
		\mathrm{Gnd}(\Gamma), x : A & \text{if $A$ is a ground type} \\
		\mathrm{Gnd}(\Gamma) & \text{otherwise}
	\end{cases} \]
\end{lemma}
\begin{proof}
	By induction on the typing derivation $\Gamma \vdash V : A$.
\end{proof}

\begin{lemma}
	Consider the interpretation of simply typed FGCBV.
	Let $(\category{C}, T, \interpret{{-}})$ be a simple FGCBV model.
	If $\Gamma \vdash V : A$ is well-typed and $A$ is a ground type, then we have the following.
	\[ \interpret{\Gamma \vdash V : A} = \interpret{\mathrm{Gnd}(\Gamma) \vdash V : A} \comp \pi_{\mathrm{Gnd}} \]
	where $\pi_{\mathrm{Gnd}} : \interpret{\Gamma} \to \interpret{\mathrm{Gnd}(\Gamma)}$ is the projection.
\end{lemma}
\begin{proof}
	By induction.
\end{proof}

In the definition of formulas, we assume that atomic formulas take only ground value terms.
As a consequence, we have the following.
\begin{lemma}
	If $\Gamma \vdash \phi : \mathbf{Fml}$ is well-formed, then $\mathrm{Gnd}(\Gamma) \vdash \phi : \mathbf{Fml}$ is well-formed.
\end{lemma}
\begin{proof}
	By induction.
\end{proof}

\begin{lemma}\label{lem:ground-type-projection-formula}
	Let $p : \category{P} \to \category{C}$ be a model of formulas.
	Assume that all logical connectives are preserved by reindexing functors of $p$.
	Note that this assumption is already included in the definition of models of formulas as long as we consider only $\top, \bot, {\land}, {\lor}, {\implies}$, but we assume this in case we add other logical connectives in the syntax of formulas.
	Then, for each well-formed formula $\Gamma \vdash \phi : \mathbf{Fml}$, we have the following.
	\[ \interpret{\Gamma \vdash \phi : \mathbf{Fml}} = \pi_{\mathrm{Gnd}}^{*} \interpret{\mathrm{Gnd}(\Gamma) \vdash \phi : \mathbf{Fml}} \]
\end{lemma}
\begin{proof}
	By induction.
\end{proof}

We define $\mathrm{Gnd}(\dot{\Gamma})$ for refinement contexts $\dot{\Gamma}$ in the same way as for simple contexts $\Gamma$.
\[ \mathrm{Gnd}(\diamond) \coloneqq \diamond \qquad \mathrm{Gnd}(\dot{\Gamma}, x : \dot{A}) = \begin{cases}
	\mathrm{Gnd}(\dot{\Gamma}), x : \dot{A} & \text{if $\underlying{\dot{A}}$ is a ground type} \\
	\mathrm{Gnd}(\dot{\Gamma}) & \text{otherwise}
\end{cases} \]
\begin{lemma}
	\begin{itemize}
		\item If $\dot{\Gamma} \vdash$ is well-formed, then $\mathrm{Gnd}(\dot{\Gamma}) \vdash$ is well-formed.
		\item If $\dot{\Gamma} \vdash \dot{A}$ is well-formed, then $\mathrm{Gnd}(\dot{\Gamma}) \vdash \dot{A}$ is well-formed.
		\item If $\dot{\Gamma} \vdash \dot{C}$ is well-formed, then $\mathrm{Gnd}(\dot{\Gamma}) \vdash \dot{C}$ is well-formed.
		\item If $\dot{\Gamma} \vdash \mathcal{E} : \mathbf{Effect}$ is well-formed, then $\mathrm{Gnd}(\dot{\Gamma}) \vdash \mathcal{E} : \mathbf{Effect}$ is well-formed.
	\end{itemize}
\end{lemma}
\begin{proof}
	By induction.
	We need weakening and $\underlying{\mathrm{Gnd}(\dot{\Gamma})} = \mathrm{Gnd}(\underlying{\dot{\Gamma}})$ in the proof.
\end{proof}

\begin{lemma}\label{lem:ground-type-projection-refinement-type}
	Consider a $\mathbf{2}$-valued model of the dependent effect system where $\category{C} = \omegaCPO$.
	Let $(X, Y, P, Q)$ be the interpretation $\interpret{\dot{\Gamma} \vdash \dot{A}}$ or $\interpret{\dot{\Gamma} \vdash \dot{C}}$ of a type.
	Then, for any $x, x' \in P = \interpret{\dot{\Gamma}}$ and $y \in Y$, if $\pi_{\mathrm{Gnd}}(x) = \pi_{\mathrm{Gnd}}(x')$, then $(x, y) \in Q$ if and only if $(x', y) \in Q$.
\end{lemma}
\begin{proof}
	By induction.
	\begin{itemize}
		\item If $(X, Y, P, Q) = \interpret{\dot{\Gamma} \vdash \{ x : b \mid \phi \}}$, then
		\begin{align}
			(x, y) \in Q &\iff (x, y) \in \interpret{\underlying{\dot{\Gamma}}, x : b \vdash \phi : \mathbf{Fml}} \qquad \text{($x \in P$ by assumption)} \\
			&\iff (\pi_{\mathrm{Gnd}}(x), y) \in \interpret{\mathrm{Gnd}(\underlying{\dot{\Gamma}}), x : b \vdash \phi : \mathbf{Fml}}
		\end{align}
		\item For $\dot{\Gamma} \vdash (x : \dot{A}) \to \dot{C}$, let $(X, Y, P, Q) = \interpret{\dot{\Gamma} \vdash \dot{A}}$ and $(X \times Y, Z, Q, R) = \interpret{\dot{\Gamma}, x : \dot{A} \vdash \dot{C}}$.
		Assume $x, x' \in P$ are such that $\pi_{\mathrm{Gnd}}(x) = \pi_{\mathrm{Gnd}}(x')$, and $f$ satisfies $\forall y, (x, y) \in Q \implies (x, y, f(y)) \in R$.
		Then, by IH, for any $y$ such that $(x', y) \in Q$, we have $(x, y) \in Q$, and thus $(x, y, f(y)) \in R$.
		Again by IH, we have $(x', y, f(y)) \in R$.
		\item For $\dot{\Gamma} \vdash (x : \dot{A}) \times \dot{B}$, let $(X, Y, P, Q) = \interpret{\dot{\Gamma} \vdash \dot{A}}$ and $(X \times Y, Z, Q, R) = \interpret{\dot{\Gamma}, x : \dot{A} \vdash \dot{B}}$.
		Assume $x, x' \in P$ are such that $\pi_{\mathrm{Gnd}}(x) = \pi_{\mathrm{Gnd}}(x')$, and $(y, z)$ is such that $((x, y), z) \in R$.
		Then, we have $(x, y), (x', y) \in Q$ and $((x', y), z) \in R$.
		\item For $\dot{\Gamma} \vdash \dot{A} + \dot{B}$, let $(X, Y, P, Q) = \interpret{\dot{\Gamma} \vdash \dot{A}}$ and $(X, Z, P, R) = \interpret{\dot{\Gamma} \vdash \dot{B}}$.
		Assume $x, x' \in P$ are such that $\pi_{\mathrm{Gnd}}(x) = \pi_{\mathrm{Gnd}}(x')$, and $y$ is such that $(x, y) \in Q \cup R$.
		Then by IH, we have $(x', y) \in Q \cup R$.
		\item For $\dot{\Gamma} \vdash T_{\mathcal{E}} \dot{A}$, let $(X, Y, P, Q) = \interpret{\dot{\Gamma} \vdash \dot{A}}$.
		\[ \interpret{\dot{\Gamma} \vdash T_{\mathcal{E}} \dot{A}} = (X, T Y, P, \{ (x, y) \mid x \in P \land y \in \ddot{T}_{\interpret{\mathcal{E}}(x)} Q(x) \}) \]
		By IH, we have $Q(x) = Q(x')$.
		Moreover, we can easily prove $\interpret{\mathcal{E}}(x) = \interpret{\mathcal{E}}(x')$ by induction on the structure of $\mathcal{E}$. %
		\qedhere
	\end{itemize}
\end{proof}

We conjecture that a more general version of the above lemma holds, but we do not use it in this paper.
\begin{conjecture}
	Suppose that we are given a model of the dependent effect system (Definition~\ref{def:dependent-effect-system-model}).
	\begin{itemize}
		\item For any well-formed refinement context $\dot{\Gamma}$, the projection $\pi_{\mathrm{Gnd}} : \interpret{\underlying{\dot{\Gamma}}} \to \interpret{\mathrm{Gnd}(\underlying{\dot{\Gamma}})}$ has a lifting $\pi_{\mathrm{Gnd}} : \interpret{\dot{\Gamma}} \to \interpret{\mathrm{Gnd}(\dot{\Gamma})}$ along $p : \category{P} \to \category{C}$.
		\begin{equation}
			\begin{tikzcd}
				\category{P} \ar[d, "p"] & \interpret{\dot{\Gamma}} \ar[r] & \interpret{\mathrm{Gnd}(\dot{\Gamma})} \\
				\category{C} & \interpret{\underlying{\dot{\Gamma}}} \ar[r, "\pi_{\mathrm{Gnd}}"] & \interpret{\mathrm{Gnd}(\underlying{\dot{\Gamma}})}
			\end{tikzcd}
		\end{equation}
		\item For any well-formed type $\dot{\Gamma} \vdash \dot{A}$ and $\dot{\Gamma} \vdash \dot{C}$, we have the following cartesian liftings.
		\begin{equation}
			\begin{tikzcd}
				\{ s(\category{C}) \mid \category{P} \} \ar[d] & \interpret{\dot{\Gamma} \vdash \dot{A}} \ar[r] & \interpret{\mathrm{Gnd}(\dot{\Gamma}) \vdash \dot{A}} & \interpret{\dot{\Gamma} \vdash \dot{C}} \ar[r] & \interpret{\mathrm{Gnd}(\dot{\Gamma}) \vdash \dot{C}} \\
				\category{P} & \interpret{\dot{\Gamma}} \ar[r, "\pi_{\mathrm{Gnd}}"] & \interpret{\mathrm{Gnd}(\dot{\Gamma})} & \interpret{\dot{\Gamma}} \ar[r, "\pi_{\mathrm{Gnd}}"] & \interpret{\mathrm{Gnd}(\dot{\Gamma})}
			\end{tikzcd}
		\end{equation}
		\item For any well-formed effect term $\dot{\Gamma} \vdash \mathcal{E} : \mathbf{Effect}$, we have the following cartesian lifting.
		\begin{equation}
			\begin{tikzcd}
				\category{M} \ar[d] & \interpret{\dot{\Gamma} \vdash \mathcal{E} : \mathbf{Effect}} \ar[r] & \interpret{\mathrm{Gnd}(\dot{\Gamma}) \vdash \mathcal{E} : \mathbf{Effect}} \\
				\category{P} & \interpret{\dot{\Gamma}} \ar[r, "\pi_{\mathrm{Gnd}}"] & \interpret{\mathrm{Gnd}(\dot{\Gamma})}
			\end{tikzcd}
		\end{equation}
	\end{itemize}
	\qed
\end{conjecture}

\begin{lemma}\label{lem:omega-cpo-type-admissible}
	Consider a $\mathbf{2}$-valued model of the dependent effect system where $\category{C} = \omegaCPO$.
	Assume that all base types are interpreted by $\omega$-cpo's with the trivial order ${=}$.
	Assume also that the given graded monad lifting $\ddot{T}$ maps predicates closed under $\sup$ of $\omega$-chains to predicates closed under $\sup$ of $\omega$-chains.
	\begin{itemize}
		\item For the interpretation $(X, P) = \interpret{\dot{\Gamma}}$ of any context $\dot{\Gamma}$, $P$ is closed under the supremum of $\omega$-chains.
		\item For the interpretation $(X, Y, P, Q) = \interpret{\dot{\Gamma} \vdash \dot{A}}$ or $(X, Y, P, Q) = \interpret{\dot{\Gamma} \vdash \dot{C}}$ of any type $\dot{\Gamma} \vdash \dot{A}$ or $\dot{\Gamma} \vdash \dot{C}$, $Q$ is closed under the supremum of $\omega$-chains.
	\end{itemize}
\end{lemma}
\begin{proof}
	By simultaneous induction on contexts and types.
	\begin{itemize}
		\item $\interpret{\diamond} = (1, 1 \subseteq 1)$. This is trivial.
		\item $\interpret{\dot{\Gamma}, x : \dot{A}} = (X \times Y, Q)$ where $(X, Y, P, Q) = \interpret{\dot{\Gamma} \vdash \dot{A}}$.
		By IH, $Q$ is closed under the supremum of $\omega$-chains.
		\item For $\dot{\Gamma} \vdash \{ x : b \mid \phi \}$ where $b$ is a base type or the unit type, we have
		\[ \interpret{\dot{\Gamma} \vdash \{ x : b \mid \phi \}} = (\dots, \pi^{*} \interpret{\dot{\Gamma}} \land \interpret{\phi}) \]
		Since the interpretation of ground types has the trivial order too, and the interpretation of formulas is (trivially) closed under the supremum of $\omega$-chains.
		By IH, $\pi^{*} \interpret{\dot{\Gamma}}$ is closed under the supremum of $\omega$-chains.
		\item For $\dot{\Gamma} \vdash (x : \dot{A}) \times \dot{B}$,
		\begin{mathpar}
			\inferrule{
				\interpret{\dot{\Gamma} \vdash \dot{A}} = (X, Y, P, Q) \\
				\interpret{\dot{\Gamma}, x : \dot{A} \vdash \dot{B}} = (X \times Y, Z, Q, R)
			}{
				\interpret{\dot{\Gamma} \vdash (x : \dot{A}) \times \dot{B}} = (X, Y \times Z, P, (\associator^{-1})^{*} R)
			}
		\end{mathpar}
		Since $R$ is closed under the supremum of $\omega$-chains, so is $(\associator^{-1})^{*} R = \{ (x, (y, z)) \mid ((x, y), z) \in R \}$.
		\item For $\dot{\Gamma} \vdash (x : \dot{A}) \to \dot{C}$
		\begin{mathpar}
			\inferrule{
				\interpret{\dot{\Gamma} \vdash \dot{A}} = (X, Y, P, Q) \\
				\interpret{\dot{\Gamma}, x : \dot{A} \vdash \dot{C}} = (X \times Y, Z, Q, R)
			}{
				\interpret{\dot{\Gamma} \vdash (x : \dot{A}) \to \dot{C}} = (X, \exponential{Y}{Z}, P, \{ (x, f) \mid x \in P \land \forall y, (x, y) \in Q \implies (x, y, f(y)) \in R \})
			}
		\end{mathpar}
		We use Lemma~\ref{lem:ground-type-projection-refinement-type}.
		Suppose we have an $\omega$-chain $\{ (x_n, f_n) \}$ that satisfies $x_n \in P \land \forall y, (x_n, y) \in Q \implies (x_n, y, f_n(y)) \in R$.
		Since $P$ is closed under the supremum of $\omega$-chains, we have $\sup_n x_n \in P$.
		Let $y$ be such that $(\sup_n x_n, y) \in Q$.
		Then, we have $(x_n, y) \in Q$ for any $n$ because $\pi_{\mathrm{Gnd}}(\sup_n x_n) = \pi_{\mathrm{Gnd}}(x_n)$ holds by the trivial order structure on $\mathrm{Gnd}(\underlying{\dot{\Gamma}})$.
		By assumption, we have $(x_n, y, f_n(y)) \in R$ for any $n$.
		Since $R$ is closed under the supremum of $\omega$-chains, we have $(\sup_n x_n, y, \sup_n f_n(y)) \in R$.
		\item For $\dot{\Gamma} \vdash \dot{A} + \dot{B}$
		\begin{mathpar}
			\inferrule{
				\interpret{\dot{\Gamma} \vdash \dot{A}} = (X, Y, P, Q) \\
				\interpret{\dot{\Gamma} \vdash \dot{B}} = (X, Z, P, R)
			}{
				\interpret{\dot{\Gamma} \vdash \dot{A} + \dot{B}} = (X, Y + Z, P, Q \cup R)
			}
		\end{mathpar}
		Let $(x_n, w_n) \in X \times (Y + Z)$ be an $\omega$-chain such that $(x_n, w_n) \in Q \cup R$.
		Since the order relation on $Y + Z$ is defined by
		\[ w_m \le_{Y+Z} w_n \quad\iff\quad w_m \le_Y w_n \lor w_m \le_Z w_n, \]
		we have either $w_n \in Y$ for all $n$ or $w_n \in Z$ for all $n$.
		That is, $(x_n, w_n) \in Q$ for all $n$ or $(x_n, w_n) \in R$ for all $n$.
		Since $Q$ and $R$ are closed under the supremum of $\omega$-chains, we have $\sup_n (x_n, w_n) \in Q$ or $\sup_n (x_n, w_n) \in R$.
		\item For $\dot{\Gamma} \vdash T_{\mathcal{E}} \dot{A}$, we have the following.
		\[ \interpret{\dot{\Gamma} \vdash T_{\mathcal{E}} \dot{A}} = (X, T Y, P, \{ (x, y) \mid x \in P \land y \in \ddot{T}_{\interpret{\mathcal{E}}(x)} Q(x) \}) \]
		where $(X, Y, P, Q) = \interpret{\dot{\Gamma} \vdash \dot{A}}$.
		Let $\{ (x_n, y_n) \in X \times T Y \}_n$ be an $\omega$-chain such that $x_n \in P \land y_n \in \ddot{T}_{\interpret{\mathcal{E}}(x_n)} Q(x_n)$ for each $n$.
		By Lemma~\ref{lem:ground-type-projection-refinement-type}, $Q(x_n)$ and $\interpret{\mathcal{E}}(x_n)$ are constant for all $n$.
		By assumption, $\ddot{T}_{\interpret{\mathcal{E}}(x_n)} Q(x_n)$ is closed under the supremum of $\omega$-chains.
		Thus, we have $\sup_n (x_n, y_n) \in \{ (x, y) \mid x \in P \land y \in \ddot{T}_{\interpret{\mathcal{E}}(x)} Q(x) \}$.
	\end{itemize}
\end{proof}

\end{toappendix}

By Theorem~\ref{thm:soundness}, we obtain a model of the dependent effect system.
By the general soundness theorem, the cost of a well-typed computation term is upper bounded by the dependent effect.

\begin{corollary}[soundness of cost analysis]\label{cor:cost-soundness}
	Suppose that $\dot{\Gamma} \vdash M : T_{\mathcal{E}} \dot{A}$ is well-typed.
	Let $\mathsf{cost}(M) \coloneqq \pi_2 \comp \interpret{M}_s : \interpret{\underlying{\dot{\Gamma}}}_s \to \mathbb{N}_{\infty}$ be the cost of $M$.
	Then, $\mathsf{cost}(M)$ is upper bounded by $\mathcal{E}$ whenever the predicates in $\dot{\Gamma}$ are satisfied.
	\[ \forall \gamma \in \interpret{\underlying{\dot{\Gamma}}}_s,\quad \gamma \in \interpret{\dot{\Gamma}} \quad\implies\quad \mathsf{cost}(M)(\gamma) \le \interpret{\mathcal{E}}(\gamma) \]
	Here, we identify $\interpret{\mathcal{E}} \in (\omegaCPO \sslash \mathbb{N}_{\infty})_{\interpret{\dot{\Gamma}}}$ with a function $\interpret{\underlying{\dot{\Gamma}}}_s \to \mathbb{N}_{\infty}$.
	\qed
\end{corollary}

\subsubsection{Example Program}

We consider an example program that computes the insertion sort of a list of lists, which is taken from benchmarks for Resource aware ML \cite{HoffmannPOPL2017}.
The insertion function is defined in Fig.~\ref{fig:insertion-sort}.
Here, we use OCaml-style comments to describe the effect (i.e., the cost in this case).
We write recursive function definitions using the \texttt{let rec} syntax, which is defined by appropriate syntactic sugar.
Since the current type system does not explicitly support inductive data types, we include a type of lists of real numbers as a base type denoted as $\mathtt{list}$.
We assume that basic operations for lists such as $\mathtt{len} : \mathtt{list} \rightarrowtriangle \NatType$, $\mathtt{head} : \mathtt{list} \rightarrowtriangle \RealType$, $\mathtt{tail} : \mathtt{list} \rightarrowtriangle \mathtt{list}$, $[] : \mathtt{list}$, and $(::) : \RealType \times \mathtt{list} \rightarrowtriangle \mathtt{list}$ are available as effect-free operations with appropriate refinement type signatures.
We also include a base type of lists of lists of real numbers denoted as $\mathtt{list\_list}$ together with basic operations for $\mathtt{list\_list}$.
Then, the functions $\mathtt{leq}$ and $\mathtt{insert}$ are typed as $\mathtt{leq} : (l_1 : \mathtt{list}) \times (l_2 : \mathtt{list}) \to T_{\min \{ \mathtt{len}\ l_1, \mathtt{len}\ l_2 \}} \mathtt{bool}$ and $\mathtt{insert} : (x : \mathtt{list}) \times (l : \mathtt{list\_list}) \to T_{(\mathtt{len}\ x + 1) \times \mathtt{len}\ l} \mathtt{list\_list}$, in which the effects depend on the input lists.
By the soundness (Corollary~\ref{cor:cost-soundness}), it follows that the cost of $\mathtt{insert}\ (x, l)$ is upper bounded by $(\mathtt{len}\ x + 1) \times \mathtt{len}\ l$.

\begin{toappendix}
	\subsection{Refinement Type Signatures for List Operations}
We assume that refinement type signatures for list operations are given as follows.
\begin{verbatim}
len : list -> nat
head : (l : {l : list | len l > 0}) -> { x : real | x = l[0] }
tail : (l : {l : list | len l > 0}) ->
  { l' : list | len l' = len l - 1 /\ forall i. l'[i] = l[i + 1] }
[] : {l : list | len l = 0}
(::) : (x : real) -> (l : list) ->
  { l' : list | len l' = len l + 1 /\ l'[0] = x /\ forall i. l'[i + 1] = l[i] }
\end{verbatim}
\end{toappendix}

\begin{figure}[tb]
	\scriptsize
\begin{verbatim}
(* leq : (l1 : list) -> (l2 : list) -> T (min (len l1) (len l2)) bool *)
let rec leq l1 l2 =
  if len l1 = 0 then return true
  else if len l2 = 0 then return false
  else
    let _ = Tick 1 in                   (* cost: 1 *)
    let b = leq (tail l1) (tail l2) in  (* cost: min (len (tail l1)) (len (tail l2)) *)
    return ((head l1) <= (head l2)) && b

(* insert : (x : list) -> (l : list_list) -> T ((len x + 1) * len l) list_list *)
let rec insert x l =
  if len l = 0 then return x :: []
  else
    let _ = Tick 1 in                (* cost: 1 *)
    let b = leq x (head l) in        (* cost: min (len x) (len (head l)) *)
    if b then return x :: l
    else
      let tl = insert x (tail l) in  (* cost: ((len x + 1) * len (tail l)) *)
      return (head l) :: tl
\end{verbatim}
\caption{An example of cost analysis (insertion sort).}
\label{fig:insertion-sort}
\end{figure}

\subsection{Expectation Logic}\label{sec:expectation-logic}

\begin{toappendix}
	\subsection{Expectation Logic}\label{sec:expectation-logic-details}
\end{toappendix}

The expectation logic \cite[Section~4.2]{AguirreICFP2021} is a logic for reasoning about expectations of probabilistic programs.
This logic is graded by the difference between a given pre-expectation, which is a $[0, \infty]$-valued function, and the weakest pre-expectation, which is the expected value of a given post-expectation.
We can instantiate our framework to the expectation logic.
One notable difference between the expectation logic and other instances is that the expectation logic interprets predicates as \emph{real-valued predicates} rather than boolean-valued predicates.
Formally, we use $\Omega = [0, \infty]^{\op} = ([0, \infty], {\ge})$ as a complete Heyting algebra.
Note that the order is reversed:
the meet operation $x \land^{\op} y$ is $\max \{ x, y \}$, and the join operation $x \lor^{\op} y$ is $\min \{ x, y \}$.
Since the expectation logic bounds the difference between $[0, \infty]^{\op}$-valued predicates, we use $\GradingMonoid = ([0, \infty], 0, {+}, {\le})$ as a grading monoid.

\subsubsection{Generic Effects}
For simplicity, we consider only one generic effect $\mathtt{Bern}$ that samples a value of type $\mathtt{bool} = \UnitType + \UnitType$ from the Bernoulli distribution.
However, adding other distributions, including continuous distributions, is straightforward.
Let $A$ and $B$ be $[0, \infty]^{\op}$-valued predicates and $e \ge 0$.
The refinement type signature of $\mathtt{Bern}$ is given as follows.
\begin{align}
	\mathtt{Bern} :\ &(p : \{ p {:} \RealType \mid \langle 0 \le p \le 1 \rangle \land^{\op} C \}) \stackrel{e}{\rightarrowtriangle} \{ t {:} \UnitType \mid A \} + \{ f {:} \UnitType \mid B \} \\
	&\text{where} \quad C = p \cdot A + (1 - p) \cdot B - e
\end{align}
Here, $p$ is a parameter for the Bernoulli distribution, and $\langle \phi \rangle$ embeds a boolean predicate $\phi$ into a $[0, \infty]^{\op}$-valued predicate: $\langle \phi \rangle = 0$ if $\phi$ holds, and $\langle \phi \rangle = \infty$ if $\phi$ does not hold.
Note that $\langle \phi \rangle \land^{\op} A$ is equal to $A$ if $\phi$ holds, and equal to $\infty$ otherwise.
The above type signature means if $0 \le p \le 1$, then the difference between the expected value $p \cdot A + (1 - p) \cdot B$ and the precondition $C = p \cdot A + (1 - p) \cdot B - e$ is upper bounded by $e$.
In the above type signature, $A$, $B$, and $e$ are constants, but we may also consider them as \emph{ghost parameters}, which are parameters of a generic effect that are not used in the actual computation but used in the type signature.
Using ghost parameters allows us more flexibility in typing, since we can instantiate them with value terms rather than constants.

\subsubsection{Model}
We use the simple FGCBV model $(\omegaQBS, \ProbMonad ({-})_{\bot}, \interpret{-}_s)$ where $\omegaQBS$ is the category of $\omega$-quasi-Borel spaces \cite{VakarPOPL2019}, $\ProbMonad$ is the probabilistic powerdomain monad on $\omegaQBS$, and $({-})_{\bot}$ is the lifting monad.
We define a strong graded monad lifting of $\ProbMonad$ along the $\Omega$-valued predicate fibration over $\omegaQBS$ by adapting a $\top\top$-lifting for $\QBS$ used in \cite{AguirreICFP2021}.
Then, we combine it with a lifting of the lifting monad $({-})_{\bot}$.
Details are given in \referappendix{Section}{D.4}{sec:expectation-logic-details}.
By Theorem~\ref{thm:soundness}, we obtain the soundness of the expectation logic.

\begin{toappendix}
\begin{definition}\label{def:expectation-graded-monad-lifting}
	We define a strong $[0, \infty]$-graded monad lifting of the probability monad $\ProbMonad$ along $\mathrm{pred}^{[0, \infty]^{\op}}_{\omegaQBS} : \mathbf{Pred}_{[0, \infty]^{\op}}(\omegaQBS) \to \omegaQBS$.
	\begin{align}
		\ExpectationGradedMonad_{\delta} (X, P) &= \lambda \zeta : \ProbMonad X. \sup \{ S(\delta + \delta')(x \gets \zeta; f(x)) \mid f \in \omegaQBS(X, \ProbMonad [0, \infty]), \\
		&\qquad\qquad \sup \{S(\delta')(f(i)) \mid P(i) < S(\delta')(f(i)) \} < S(\delta + \delta')(x \gets \zeta; f(x)) \}
		\label{eq:expectation-graded-monad-lifting}
	\end{align}
	where $S : [0, \infty] \to \mathbf{Pred}_{\Omega}(\omegaQBS)_{\ProbMonad [0, \infty]}$ is a parameter for the graded $\top\top$-lifting defined as follows.
	\[ S(\delta) \coloneqq \lambda \zeta. \max \{ 0, \mathbb{E}_{x \gets \zeta}[x] - \delta \} : \ProbMonad [0, \infty] \to [0, \infty] \]
	Here, $[0, \infty] \in \omegaQBS$ is defined by $([0, \infty], \mathbf{Meas}(\mathbb{R}, [0, \infty]), {\le})$ because we later need to extend $f \in \omegaQBS(X, [0, \infty])$ to $[f, 0] \in \omegaQBS(X_{\bot}, [0, \infty])$.
	The subtraction by $\infty$ is defined by $x - \infty = 0$ so that we have $x \le y + z$ if and only if $x - y \le z$.
\end{definition}

We combine the graded monad lifting $\ExpectationGradedMonad_{\delta}$ with a lifting of the lifting monad $({-})_{\bot}$.

\begin{lemma}
	If $g \in \omegaQBS(X, [0, \infty])$ is such that $g \le P$, then $\ExpectationGradedMonad_{\delta} (X, P) \ge \lambda \zeta. \mathbb{E}_{x \gets \zeta}[g(x)] - \delta$.
\end{lemma}
\begin{proof}
	Take $\delta' = 0$ and $f = \eta^{\ProbMonad} \comp g$ in~\eqref{eq:expectation-graded-monad-lifting}.
	Then, we have $S(\delta + \delta')(x \gets \zeta; f(x)) = \max \{ 0, \mathbb{E}_{x \gets \zeta}[g(x)] - \delta \}$.
\end{proof}

We define a strong monad lifting of the lifting monad $({-})_{\bot}$ along $\mathrm{pred}^{[0, \infty]^{\op}}_{\omegaQBS} : \mathbf{Pred}_{[0, \infty]^{\op}}(\omegaQBS) \to \omegaQBS$ as follows.
\[ \PartialCorrectnessMonad (X, P) \quad\coloneqq\quad (X_{\bot}, [P, 0]) \]
where $[P, 0] : \omegaQBS(1, X_{\bot}) \cong \omegaQBS(1, X) + 1 \to [0, \infty]$ is the co-tupling of $P : \omegaQBS(1, X) \to [0, \infty]$ and $0 : 1 \to [0, \infty]$.

\begin{lemma}\label{lem:expectation-graded-monad-lifting}
	The composite $\delta, (X, P) \mapsto \ExpectationGradedMonad_{\delta} \PartialCorrectnessMonad (X, P)$ is a strong $[0, \infty]$-graded monad lifting of $\ProbMonad ({-})_{\bot}$ along $\mathrm{pred}^{[0, \infty]^{\op}}_{\omegaQBS} : \mathbf{Pred}_{[0, \infty]^{\op}}(\omegaQBS) \to \omegaQBS$.
	Moreover, it is admissible.
\end{lemma}
\begin{proof}[Proof of Lemma~\ref{lem:expectation-graded-monad-lifting}]
	We show the existence of a distributive law by Lemma~\ref{lem:partial-correctness-distributive-law}.
	It suffices to show $0 \ge \ExpectationGradedMonad_{\delta} \PartialCorrectnessMonad (X, P) (\eta^{\ProbMonad}(\bot))$.
	We prove by contradiction.
	Note that we have $x \gets \eta^{\ProbMonad}(\bot); f(x) = f(\bot)$.
	Suppose that there exist $\delta' \in [0, \infty]$ and $f \in \omegaQBS(X_{\bot}, \ProbMonad [0, \infty])$ such that $S(\delta + \delta')(f(\bot)) > 0$ and
	\begin{equation}
		\sup \{S(\delta')(f(i)) \mid [P, 0](i) < S(\delta')(f(i)) \} < S(\delta + \delta')(f(\bot)).
		\label{eq:expectation-graded-monad-lifting-proof}
	\end{equation}
	Then, it must be the case that $S(\delta')(f(\bot)) > 0$ by $S(\delta + \delta')(f(\bot)) < S(\delta')(f(\bot))$.
	However, by~\eqref{eq:expectation-graded-monad-lifting-proof}, we also have $S(\delta')(f(\bot)) < S(\delta + \delta')(f(\bot))$, which is a contradiction.
	As for the admissibility, the condition for the least element follows from the above argument.
	So, it suffices to show that for any $\omega$-chain $\{ \zeta_n \in \ProbMonad X_{\bot} \}_n$ such that $\ExpectationGradedMonad_{\delta} \PartialCorrectnessMonad (X, P)(\zeta_n) \le u$ for some $u \in [0, \infty]$, we have $\ExpectationGradedMonad_{\delta} \PartialCorrectnessMonad (X, P)(\sup_n \zeta_n) \le u$.
	This follows from the fact that $S(\delta + \delta')(x \gets \zeta; f(x))$ is Scott-continuous in $\zeta$.
\end{proof}
\end{toappendix}

\begin{corollary}[soundness of expectation logic]\label{cor:expectation-logic-soundness}
	Suppose that $\dot{\Gamma} \vdash M : T_{\mathcal{E}} \{ x : b \mid \phi \}$ is well-typed.
	If $\interpret{\phi}(\gamma, {-}) : \interpret{b}_s \to [0, \infty]$ is a morphism in $\omegaQBS$ for each $\gamma$, then
	\[ \forall \gamma \in \interpret{\underlying{\dot{\Gamma}}},\quad \interpret{\dot{\Gamma}}(\gamma) + \interpret{\mathcal{E}}(\gamma) \quad\ge\quad \mathbb{E}_{x \gets \interpret{M}_s (\gamma)}[\interpret{\phi}(\gamma, x)] \]
	Here, we implicitly extend the domain of $\interpret{\phi}(\gamma, {-})$ by $\interpret{\phi}(\gamma, \bot) = 0$.
	Note that if $\interpret{b}_s$ is a measurable space with discrete order, then any measurable function $\interpret{\phi}(\gamma, {-})$ is a morphism in $\omegaQBS$.
	\qed
\end{corollary}

\subsubsection{Example Program}

\begin{figure}[tb]
	\scriptsize
\begin{verbatim}
let unfair_cowboy_duel t a b =
  let weight t c = if t then return a + c else return b in
  let rec aux t c =
    let w = weight t c in
    let s = Bern w in         (* bound: t => c * b / (a + b - a * b), not t => 0 *)
    if s then return t
    else aux (not t) (c / 2)  (* bound:     t => 1 / 3 * c * b / (a + b - a * b),
                                        not t => 2 / 3 * c * b / (a + b - a * b) *)
  in aux t ((1 - a) / 3)
\end{verbatim}
\caption{The unfair duelling cowboys. Two cowboys A ($t = \mathtt{true}$) and B ($t = \mathtt{false}$) are duelling. Cowboy A hits cowboy B with probability $a$, and cowboy B hits cowboy A with probability $b$.
However, Cowboy A is given an unfair advantage, which increases the probability by $c$.
The program returns the winner of the duel.}
\label{fig:unfair-cowboy-duel}
\end{figure}

We consider a program (Fig.~\ref{fig:unfair-cowboy-duel}) that returns a winner of a duel between two cowboys \cite{McIver2005}.
We modified the original program so that Cowboy A is given an unfair advantage, which increases the probability of hitting Cowboy B.
The advantage is given by a parameter $c$, which is halved at each round of the duel.
If there is no unfair advantage (i.e., $c = 0$), then the ``fair'' winning probability of Cowboy A is $p^{\mathrm{fair}}_A \coloneqq a / (a + b - a \cdot b)$ if Cowboy A shoots first, and $p^{\mathrm{fair}}_B \coloneqq (1 - b) a / (a + b - a \cdot b)$ if Cowboy B shoots first.
We would like to bound the difference between the fair winning probability and the unfair winning probability.
Using our dependent effect system, we can derive the following type for $\mathtt{aux}$, the main loop of the duel.
\begin{align}
	&\mathtt{aux} : (t {:} \{ u {:} \UnitType \mid p^{\mathrm{fair}}_A \} \!+\! \{ u {:} \UnitType \mid p^{\mathrm{fair}}_B \}) \times (c {:} \{ c {:} \RealType \mid \langle 0 \le c \land a + c \le 1 \rangle \}) \\
	&\qquad\qquad\to T_{[t] \cdot \frac{4 c b}{3 (a + b - ab)} + [\lnot t] \cdot \frac{2 c b}{3 (a + b - ab)}} (\{ u : \UnitType \mid 1 \} + \{ u : \UnitType \mid 0 \})
\end{align}
This type reads as follows.
The post-expectation is $1$ if Cowboy A wins, and $0$ if Cowboy B wins.
The pre-expectation for $t$ is the fair winning probability of Cowboy A.
The difference between the fair winning probability and the unfair winning probability is bounded by $4 / 3 \cdot c b / (a + b - ab)$ if Cowboy A shoots first, and $2 / 3 \cdot c b / (a + b - ab)$ if Cowboy B shoots first.
Here, $[{-}]$ is the Iverson bracket, which is defined as $[t] = 1$ if $t$ is true, and $[t] = 0$ otherwise.

\begin{toappendix}
The following code shows the refinement type of each variable used in $\mathtt{cowboy\_duel}$.
Here, we use \hyperlink{rule:VT-VarSelf}{\textsc{VT-VarSelf}}, and its extension to $\mathtt{bool} = \UnitType + \UnitType$.
We also slightly extend the syntax of value terms to allow case analysis so that we can naturally write predicates on $\mathtt{bool} = \UnitType + \UnitType$.

{\scriptsize
\begin{verbatim}
not : bool -> bool
not : {u : unit | A} + {u : unit | B} -> {u : unit | B} + {u : unit | A}

let cowboy_duel t a b =
let weight t = if t then return a else return b in
let rec aux t =
  (*  t : { u : unit | a / (a + b - a * b) } + { u : unit | (1 - b) a / (a + b - a * b) }  *)

  let w = weight t in
  (* w : { w : real | <t = true> /\ <w = a> \/ <t = false> /\ <w = b> } *)

  (* .. |- w : { w' : real | <w' = w> } *)
  (* .. |- w : { w' : real | <0 <= w' <= 1>/\
    (<t = true> /\ w' + (1 - w') (1 - b) a / (a + b - a * b)
    \/ <t = false> /\ (1 - w') a / (a + b - a * b)) } *)
  let s = Bern w in
  (* s : {u : unit | <t = true> /\ 1 \/ <t = false> /\ 0}
         + {u : unit | <t = true> /\ (1 - b) a / (a + b - a * b) \/ <t = false> /\ a / (a + b - a * b)} *)

  if s then
    (* s : {u : unit | <t = true> /\ 1 \/ <t = false> /\ 0} *)
    (* .. |- t : { u : unit | <t = true> } + { u : unit | <t = false> }  *)
    (* .. |- t : { u : unit | 1 } + { u : unit | 0 }  *)
    return t
  else
    (* s : {u : unit | <t = true> /\ (1 - b) a / (a + b - a * b) \/ <t = false> /\ a / (a + b - a * b)} *)
    (* .. |- t : { u : unit | <t = true> } + { u : unit | <t = false> }  *)
    (* .. |- t : { u : unit | (1 - b) a / (a + b - a * b) } + { u : unit | a / (a + b - a * b) } *)
    aux (not t)
in
aux t
\end{verbatim}
}

The following code shows effects and the refinement type of each variable used in $\mathtt{unfair\_cowboy\_duel}$, the unfair version of the cowboy duel.
{\scriptsize
\begin{verbatim}
let unfair_cowboy_duel t a b =
let weight t c = if t then return a + c else return b in
let rec aux t c =
  (*  t : { u : unit | a / (a + b - a * b) } + { u : unit | (1 - b) * a / (a + b - a * b) }  *)

  let w = weight t c in
  (* w : { w : real | <t = true> /\ <w = a + c> \/ <t = false> /\ <w = b> } *)

  (* .. |- w : { w' : real | <w' = w> } *)
  (* .. |- w : { w' : real | <0 <= w' <= 1>/\
    (<t = true> /\ w' + (1 - w') * (1 - b) * a / (a + b - a * b)
    \/ <t = false> /\ (1 - w') * a / (a + b - a * b)) } *)
  let s = Bern w in  (* effect: t => c * b / (a + b - a * b), not t => 0 *)
  (* s : {u : unit | <t = true> /\ 1 \/ <t = false> /\ 0}
         + {u : unit | <t = true> /\ (1 - b) * a / (a + b - a * b) \/ <t = false> /\ a / (a + b - a * b)} *)

  if s then
    (* s : {u : unit | <t = true> /\ 1 \/ <t = false> /\ 0} *)
    (* .. |- t : { u : unit | <t = true> } + { u : unit | <t = false> }  *)
    (* .. |- t : { u : unit | 1 } + { u : unit | 0 }  *)
    return t
  else
    (* s : {u : unit | <t = true> /\ (1 - b) * a / (a + b - a * b) \/
                       <t = false> /\ a / (a + b - a * b)} *)
    (* .. |- t : { u : unit | <t = true> } + { u : unit | <t = false> }  *)
    (* .. |- t : { u : unit | (1 - b) * a / (a + b - a * b) } + { u : unit | a / (a + b - a * b) } *)
    aux (not t) (c / 2)  (* effect:     t => 1 / 3 * c * b / (a + b - a * b),
                                    not t => 2 / 3 * c * b / (a + b - a * b) *)
in
aux t ((1 - a) / 3)

(* aux t (c : { real | <0 <= c /\ a + c <= 1>})
    : T (    t => 4 / 3 * c * b / (a + b - a * b),
         not t => 2 / 3 * c * b / (a + b - a * b)) *)
\end{verbatim}
}

Note that the semantic validity for $[0, \infty]^{\op}$-valued predicates is given as in the following example.
\begin{example}
	Let $f, g, h_1, h_2$ be $[0, \infty]^{\op}$-valued predicates.
	\[ x : \{ x : \RealType \mid f(x) \},\quad y : \{ y : \RealType \mid g(x, y) \} \quad\vDash\quad h_1(x, y) \implies h_2(x, y) \]
	The above semantic validity is equivalent to the following condition.
	\[ \forall x, y \in \mathbb{R},\quad \max \{ f(x), g(x, y),  h_1(x, y) \} \ge h_2(x, y) \]
\end{example}

\end{toappendix}

\ifthenelse{\boolean{longversion}}{
\subsection{Temporal Safety}\label{sec:temporal-safety}

\begin{toappendix}
	\subsection{Temporal Safety}
\end{toappendix}

We consider temporal safety verification \cite{NanjoLICS2018,SekiyamaPOPL2023}.
Given a program and a language $L \subseteq \Sigma^{*}$ where $\Sigma$ is a finite set of event symbols, we want to verify that the sequence of events emitted by the program is in $L$.
For this purpose, we consider a monoid $(\PowerSet{\Sigma^{*}}, \{ \varepsilon \}, {\cdot}, {\subseteq})$ of languages over $\Sigma$ where $\varepsilon$ is the empty word and $({\cdot})$ is the concatenation operator.
We use $\Omega = \mathbf{2}$ as a complete Heyting algebra.

\subsubsection{Generic Effects}

Let $\mathtt{event}$ be a base type for event symbols in $\Sigma$.
We consider a generic effect $\mathtt{Emit}(e)$ that emits an event symbol $e \in \Sigma$.
The refinement type signature is given by
$\mathtt{Emit} : (e : \mathtt{event}) \stackrel{\{ e \}}{\rightarrowtriangle} \UnitType$.

\subsubsection{Model}

We define a simple FGCBV model as $(\omegaCPO, \WriterLiftingMonad, \interpret{{-}}_s)$ and assume that all base types are interpreted as discrete $\omega$cpos.
The strong monad $\WriterLiftingMonad$ is defined as follows.

\begin{definition}
	We consider an algebraic theory for a generic effect $\mathtt{write}_a$ for each $a \in \Sigma$ and the bottom element $\bot$ with one equational axiom: $\bot \le x$ for any $x$ (see, e.g, \cite{HylandTheoreticalComputerScience2006} for algebraic theories in $\omegaCPO$).
	Then, we define $\WriterLiftingMonad X$ as the free algebra generated by $X$, which yields a pseudo-lifting strong monad $\WriterLiftingMonad$ on $\omegaCPO$.
	Concretely, $\WriterLiftingMonad X$ is given by
	$\WriterLiftingMonad X \coloneqq (\Sigma^{*} \times X + \Sigma^{*} \times \{ \bot \} + \Sigma^{\omega} \times \{ \bot \}, {\le}_{\WriterLiftingMonad X})$
	where ${\le}_{\WriterLiftingMonad X}$ is defined as the minimum relation such that (1) for any $(s, x), (t, y) \in \Sigma^{*} \times X$, $(s, x) \le_{\WriterLiftingMonad X} (t, y)$ if $s = t$ and $x \le y$, (2) for any $s, t \in \Sigma^{*} + \Sigma^{\omega}$, $(s, \bot) \le_{\WriterLiftingMonad X} (t, \bot)$ if $s$ is a prefix of $t$, and (3) for any $s \in \Sigma^{*}$ and $(t, x) \in \Sigma^{*} \times X$, $(s, \bot) \le_{\WriterLiftingMonad X} (t, x)$ if $s$ is a prefix of $t$.
	The unit and the multiplication are defined similarly to the standard writer monad.
	\qed
\end{definition}

\begin{toappendix}
\begin{definition}[language monoid]
	Let $\Sigma$ be a set of alphabet symbols and $\Sigma^{*}$ be the set of finite words over $\Sigma$.
	The power set $\PowerSet{\Sigma^{*}}$ has the following preordered monoid structure.
	The multiplication is given by the concatenation operator extended to sets of words.
	\[ X \cdot Y \coloneqq \{ x \cdot y \mid x \in X, y \in Y \} \]
	The unit is given by the singleton set $\{ \varepsilon \}$ where $\varepsilon$ is the empty word.
	The order relation is given by the subset relation ${\subseteq}$.
	We call the preordered monoid $(\PowerSet{\Sigma^{*}}, \{ \varepsilon \}, ({\cdot}), {\subseteq})$ the \emph{language monoid} over $\Sigma$.
\end{definition}
\end{toappendix}

\begin{definition}
	We define a $\PowerSet{\Sigma^{*}}$-graded monad lifting $\TemporalSafetyGradedMonad$ of $\WriterLiftingMonad$ along the predicate fibration $\mathrm{pred}_{\omegaCPO} : \mathbf{Pred}(\omegaCPO) \to \omegaCPO$ as follows. For each $(X, P) \in \mathbf{Pred}(\omegaCPO)$,
	\[ \TemporalSafetyGradedMonad_L P \quad\coloneqq\quad L \times P + \Sigma^{*} \times \{ \bot \} + \Sigma^{\omega} \times \{ \bot \} \quad\subseteq\quad \Sigma^{*} \times X + \Sigma^{*} \times \{ \bot \} + \Sigma^{\omega} \times \{ \bot \}. \]
	Note that the definition of $\TemporalSafetyGradedMonad$ means that if a computation terminates, then the word produced by the computation is in $L$ and the value produced by the computation is in $P$.
	\qed
\end{definition}

\begin{toappendix}
\begin{lemma}
	For any well-formed type $\dot{\Gamma} \vdash \dot{A}$ and $L \in \PowerSet{\Sigma^{*}}$, $\TemporalSafetyGradedMonad_L \interpret{\dot{\Gamma}, x : \dot{A}}$ is admissible.
\end{lemma}
\begin{proof}
	Since the least element is given by $(\varepsilon, \bot)$, we have $(\varepsilon, \bot) \in \TemporalSafetyGradedMonad_L \interpret{\dot{\Gamma}, x : \dot{A}}$.
	Suppose that $P$ is closed under $\sup$ of $\omega$-chains.
	We show that $\TemporalSafetyGradedMonad_L P$ is also closed under $\sup$ of $\omega$-chains.
	Let $\{ (s_n, x_n) \in \TemporalSafetyGradedMonad_L P \}_{n \in \mathbb{N}}$ be an $\omega$-chain.
	If there exists $N$ such that $(s_N, x_N) \in L \times P$, then we have $s_n = s_N$ for all $n \ge N$ and $\sup_n (s_n, x_n) = (s_N, \sup_n x_n)$.
	Since $P$ is closed under $\sup$, we have $\sup_n x_n \in P$.
	If there is no such $N$, then we have $\sup_n (s_n, x_n) \in \Sigma^{*} \times \{ \bot \} + \Sigma^{\omega} \times \{ \bot \}$.
	By Lemma~\ref{lem:omega-cpo-type-admissible}, it follows that $\TemporalSafetyGradedMonad_L \interpret{\dot{\Gamma}, x : \dot{A}}$ is admissible.
\end{proof}
\end{toappendix}

\begin{corollary}[soundness of temporal safety]\label{cor:temporal-safety-soundness}
	Suppose that $\dot{\Gamma} \vdash M : T_{\mathcal{E}} \dot{A}$ is well-typed.
	For any $\gamma \in \interpret{\dot{\Gamma}}$, let $(s, r) = \interpret{M}_s(\gamma)$.
	If $x \neq \bot$, i.e., the computation terminates, then $s \in \interpret{\mathcal{E}}(\gamma)$ and $\interpret{\dot{\Gamma}, x : \dot{A}}(\gamma, r)$ hold.
	\qed
\end{corollary}

\subsubsection{Example Program}
We consider call stack analysis as an example of temporal safety verification.
Let $\Sigma \coloneqq \{ \texttt{push}, \texttt{pop} \}$ be the set of event symbols representing pushing and popping a value onto/from a call stack.
By appropriately inserting $\mathtt{Emit}$ operations before and after function calls, we can track the history of the call stack in this setting.

\begin{figure}[tb]
	\scriptsize
\begin{verbatim}
(* fib : (n : nat) -> T { max_depth <= n + 1 /\ depth == 0 } nat *)
let rec fib n =
  Emit "push"               (* event: { push } *)
  if n <= 1 then
    Emit "pop"              (* event: { pop } *)
    return 1
  else
    let a = fib (n - 2) in  (* event: { max_depth <= n - 1 /\ depth == 0 } *)
    let b = fib (n - 1) in  (* event: { max_depth <= n /\ depth == 0 } *)
    Emit "pop"              (* event: { pop } *)
    return a + b
\end{verbatim}
\caption{The fibonacci function with events for pushing and popping the call stack. Here, \texttt{max\_depth <= n} is a shorthand for the language $L_{\mathrm{max\_depth}}(n)$ and \texttt{depth == n} is a shorthand for the language $L_{\mathrm{depth}}(n)$.}
\label{fig:fibonacci}
\end{figure}

For example, consider the maximum depth of the call stack of the fibonacci function in Fig.~\ref{fig:fibonacci}.
We would like to show that the maximum stack depth of the function $\mathtt{fib}$ is at most $n + 1$ where $n$ is the input argument.
Formally, we define the following languages over the alphabet $\Sigma$.
For simplicity, we omit the underflow check; it can be handled by additionally requiring $\#_{\texttt{push}} - \#_{\texttt{pop}} \ge 0$.
\begin{align}
	L_{\mathrm{depth}}(n)\ &\coloneqq\ \{ s \in \Sigma^{*} \mid \#_{\texttt{push}}(s) - \#_{\texttt{pop}}(s) = n \} \label{eq:depth-language} \\
	L_{\mathrm{max\_depth}}(n)\ &\coloneqq\ \{ s \in \Sigma^{*} \mid \text{for any prefix $s'$ of $s$}, \#_{\texttt{push}}(s') - \#_{\texttt{pop}}(s') \le n \} \label{eq:max-depth-language}
\end{align}
where $\#_{\texttt{push}}(s)$ is the number of occurrences of the symbol $\texttt{push}$ in $s$ and similarly for $\#_{\texttt{pop}}(s)$.
Intuitively, $L_{\mathrm{depth}}(n)$ means that the current depth of the call stack is $n$, and $L_{\mathrm{max\_depth}}(n)$ means that the maximum depth of the call stack is at most $n$ at any point in the past.
Then, we can show that the fibonacci function has the following type, which ensures that the maximum depth of the call stack is at most $n + 1$.
\[ \mathtt{fib} : (n : \NatType) \to T_{L_{\mathrm{depth}}(0) \cap L_{\mathrm{max\_depth}}(n + 1)} \NatType \]

\begin{toappendix}
To derive the type for the fibonacci function $\mathtt{fib}$, we use the following properties of the languages $L_{\mathrm{depth}}(n)$ and $L_{\mathrm{max\_depth}}(n)$.
\begin{align}
	\#_{\texttt{push}}(s_1 \cdot s_2) &= \#_{\texttt{push}}(s_1) + \#_{\texttt{push}}(s_2) \\
	L_{\mathrm{depth}}(n_1) \cdot L_{\mathrm{depth}}(n_2) &\subseteq L_{\mathrm{depth}}(n_1 + n_2) \\
	(L_{\mathrm{max\_depth}}(n_1) \cap L_{\mathrm{depth}}(n_2)) \cdot L_{\mathrm{max\_depth}}(n_3) &\subseteq L_{\mathrm{max\_depth}}(\max \{ n_1, n_2 + n_3 \})
\end{align}

We did not consider the underflow of the call stack, but it is easy to extend the above definition to handle exclude underflow by considering the following language.
\[ L_{\mathrm{no\_underflow}} \coloneqq \{ s \in \Sigma^{*} \mid \text{for any prefix $s'$ of $s$}, 0 \le \#_{\texttt{push}}(s') - \#_{\texttt{pop}}(s') \} \]
\end{toappendix}

\subsection{Union Bound Logic}\label{sec:union-bound-logic}

\begin{toappendix}
	\subsection{Union Bound Logic}\label{sec:union-bound-logic-detail}
\end{toappendix}

The union bound logic is a logic graded by (an upper bound of) the failure probability of a computation \cite{BartheICALP2016}.
The complete Heyting algebra and the grading monoid for the union bound logic are $\Omega = \mathbf{2}$ and $([0, \infty),\ 0,\ {+},\ {\le})$, respectively.

\subsubsection{Generic Effect}
We consider a generic effect $\mathtt{Lap}$, whose refinement type signature is given as
$\mathtt{Lap} : (\epsilon : \{ \epsilon : \RealType \mid \epsilon > 0\}) \times (m : \RealType ) \stackrel{\exp (-\epsilon t)}{\rightarrowtriangle} \{ y : \RealType \mid |y - m| \le t \}$ for each $t > 0$.
The generic effect $\mathtt{Lap}(\epsilon, m)$ samples a real number $y$ from the Laplace distribution with scale parameter $\epsilon$ and mean $m$.
The probability that the sampled value $y$ fails to satisfy $|y - m| \le t$ is upper bounded by $\exp(-\epsilon t)$.

\begin{remark}\label{rem:ghost-parameter}
	Instead of considering the refinement type signature parameterized by a constant real number $t$, it is also possible to consider a generic effect $\mathtt{Lap}^{\mathrm{ghost}}$ that takes an auxiliary (\emph{ghost}) parameter $t : \RealType$, which is not used in the actual computation.
	Ghost parameters are helpful for reasoning about programs, since they can be value terms that depend on the context.
	\begin{align}
		\mathtt{Lap}^{\mathrm{ghost}} :\ &(\epsilon : \{ \epsilon {:} \RealType \mid \epsilon > 0\})\ \times\ (m : \RealType)\ \times\ (t : \{ t {:} \RealType \mid t > 0 \}) \\
		&\qquad\stackrel{\exp (-\epsilon t)}{\rightarrowtriangle} \{ y {:} \RealType \mid |y - m| \le t \}
	\end{align}
\end{remark}

\subsubsection{Model}
We use the simple FGCBV model $(\omegaQBS, \ProbMonad ({-})_{\bot}, \interpret{-}_s)$ where $\omegaQBS$ is the category of $\omega$-quasi-Borel spaces \cite{VakarPOPL2019}, $\ProbMonad$ is the probabilistic powerdomain monad on $\omegaQBS$, and $({-})_{\bot}$ is the lifting monad.

Defining a graded monad lifting for the union bound logic is not straightforward.
We would like to define a strong graded monad lifting of the probabilistic powerdomain monad $\ProbMonad$ along $\mathrm{pred}_{\omegaQBS} : \mathbf{Pred}(\omegaQBS) \to \omegaQBS$ so that we can bound the failure probability of a predicate $P \subseteq X$ by $\beta \ge 0$, i.e., $\mathrm{Pr}[X \setminus P] \le \beta$.
However, a predicate $P$ in $\mathbf{Pred}(\omegaQBS)$ is not necessarily measurable, and thus the probability of $X \setminus P$ is not well-defined.
This issue is known to be solved by a graded $\top\top$-lifting \cite{KatsumataPOPL2014,AguirreICFP2021}, and we follow their definition.

\begin{definition}\label{def:union-bound-graded-monad-lifting}
	A $[0, \infty)$-graded monad lifting of the probability monad $\ProbMonad$ for union bound logic is defined as follows.
	\[ \UnionBoundGradedMonad_{\beta} (X, P) \coloneqq (\ProbMonad X, \{ \zeta \in \ProbMonad X \mid \forall B \in \BorelClosed{X}, P \subseteq B \Rightarrow \mathbb{P}_{x \leftarrow \zeta}[x \in X \setminus B] \le \beta \}) \]
	where $\BorelClosed{X}$ is the set of Borel-Scott closed subsets of $X$ and $\mathbb{P}_{x \leftarrow \zeta}[x \in X \setminus B]$ is the probability that $x$ drawn from the distribution $\zeta$ is in $X \setminus B$.
	Here, we say a subset of an $\omega$qbs is \emph{Borel-Scott closed} if it is Borel and Scott-closed \cite[Example~4.9]{VakarPOPL2019}.
	Details are explained in \referappendix{Appendix}{D.6}{sec:union-bound-logic-detail}.
	\qed
\end{definition}

We combine the graded monad lifting $\UnionBoundGradedMonad_{\beta}$ with a lifting of the lifting monad similarly to Section~\ref{sec:cost-analysis}.
Then, the following corollary follows from the soundness theorem.

\begin{corollary}[soundness of union bound]\label{cor:union-bound-soundness}
	Suppose that $\dot{\Gamma} \vdash M : T_{\mathcal{E}} \{ x : b \mid \phi \}$ is well-typed.
	If $\phi$ is a Borel-Scott closed predicate on $x : b$, then the probability that $M$ returns a value $x$ violating $\phi$ is upper bounded by $\mathcal{E}$ whenever the predicates in $\dot{\Gamma}$ are satisfied.
	\begin{align}
		\forall \gamma \in \interpret{\dot{\Gamma}},\quad &\text{$\interpret{\phi}(\gamma, {-}) \subseteq \interpret{b}_s$ is Borel-Scott closed} \\
		&\quad\implies\quad \mathbb{P}_{x \gets \interpret{M}_s(\gamma)} [x \in \interpret{b}_s \setminus \interpret{\phi}(\gamma, {-})] \le \interpret{\mathcal{E}}(\gamma)
	\end{align}
	Note that Borel-Scott closedness of $\interpret{\phi}(\gamma, {-})$ is automatically satisfied if $\interpret{b}_s$ is a measurable space with discrete order and $\interpret{\phi}(\gamma, {-})$ is a measurable predicate.
	\qed
\end{corollary}

\begin{toappendix}
	We will explain details of Definition~\ref{def:union-bound-graded-monad-lifting}.
	Following \cite{KatsumataPOPL2014,AguirreICFP2021,SatoPOPL2019}, we use a graded $\top\top$-lifting, but we change the base category from $\QBS$ to $\omegaQBS$.
	\begin{definition}
		Consider the following situation.
		\begin{itemize}
			\item Let $p : \category{P} \to \category{C}$ be a preordered fibration with fibred small products.
			\item The total and base category $\category{P}$ and $\category{C}$ are cartesian closed, and $p$ preserves the cartesian closed structure.
			\item Let $T$ be a strong monad on $\category{C}$.
			\item Let $\GradingMonoid = (\GradingMonoid, \GradeUnit, \GradeMult)$ be a preordered monoid.
			\item Let $S : \GradingMonoid \to \category{P}_{T b}$ be a functor, which is called a parameter for graded $\top\top$-lifting.
		\end{itemize}
		The \emph{graded $\top\top$-lifting} \cite{KatsumataPOPL2014} is defined as follows.
		\[ \bigwedge_{e \in \GradingMonoid} \dot{T}^{e'}_e X \]
		Here, $\dot{T}^{e'}_e X$ is defined by the following reindexing.
		\begin{equation}
			\begin{tikzcd}
				\category{P} \ar[d] & \dot{T}^{e'}_e X \ar[r] & \dotExponential{(\dotExponential{X}{S e'})}{S (e \GradeMult e')} \\
				\category{C} & T I \ar[r, "\Lambda (\beta)"] & \exponential{(\exponential{I}{T b})}{T b}
			\end{tikzcd}
		\end{equation}
		\begin{equation}
			\beta \qquad\coloneqq\qquad
			\begin{tikzcd}
				T I \times (\exponential{I}{T b}) \ar[d, "\braiding"] \\
				(\exponential{I}{T b}) \times T I \ar[d, "\strength^T"] \\
				T ((\exponential{I}{T b}) \times I) \ar[d, "T \eval"] \\
				T^2 b \ar[d, "\mu"] \\
				T b
			\end{tikzcd}
		\end{equation}
	\end{definition}

	\begin{example}[union bound]
		The parameter for the graded $\top\top$-lifting for union bound is given as follows.
		Fibration: $\mathbf{Pred}(\omegaQBS) \to \omegaQBS$.
		\[ S \beta = \{ \zeta \in D 2^{\op} \mid \zeta(\chi_{\{ 0 \}}) \le \beta \} \]
		Here, $\chi$ is the characteristic function, and $2^{\op} = (\{ 0, 1 \}, \mathbf{Meas}(\mathbb{R}, \{ 0, 1 \}), {\ge})$ is an $\omega$-qbs with the reverse order.
	\end{example}

	We can simplify graded $\top\top$-liftings.
	By abstracting the proof of \cite[Theorem 2.2]{SatoENTCS2016}, we obtain the following lemma.
	
	\begin{lemma}\label{lem:graded-TT-lifting-reduction-unit}
		Assume the following condition holds.
		\begin{itemize}
			\item There exists $h_e : T b \to T b$ such that $h_e : S e \dotTo S 1$.
			\item For each $e, e' \in \GradingMonoid$, $(h_{e'}^{\sharp})^{*} S e \le \mu^{*} S (e \GradeMult e')$ in $\category{P}_{T^2 b}$.
		\end{itemize}
		Then we have
		\[ \bigwedge_{e \in \GradingMonoid} \dot{T}^{e'}_e X \quad=\quad \dot{T}^{1}_e X \]
		for each $X \in \category{P}$.
	\end{lemma}
	\begin{proof}
		It suffices to prove $\dot{T}^{1}_e X \le \dot{T}^{e'}_e X$ for each $e'$, which is equivalent to $\beta : \dot{T}^{1}_e X \dotTimes (\dotExponential{X}{S e'}) \dotTo S (e \GradeMult e')$ by definition of $\dot{T}^{e'}_e X$.
		Consider the following composite.
		\begin{equation}
			\begin{tikzcd}
				\dot{T}^{1}_e X \dotTimes (\dotExponential{X}{S e'}) \ar[r] & \dot{T}^{1}_e X \dotTimes (\dotExponential{X}{S 1}) \ar[r] & S e \\
				T I \times (\exponential{I}{T b}) \ar[r, "T I \times (\exponential{I}{h_{e'}})"] & T I \times (\exponential{I}{T b}) \ar[r, "\beta"] & T b
			\end{tikzcd}
		\end{equation}
		It follows that we have
		\[ T \eval \comp \strength^T \comp \braiding : \dot{T}^{1}_e X \dotTimes (\dotExponential{X}{S e'}) \dotTo (h_{e'}^{\sharp})^{*} S e \]
		by the following equation.
		\begin{align}
			\beta \comp (T I \times (\exponential{I}{h_{e'}})) &= \mu \comp T \eval \comp \strength^T \comp \braiding \comp (T I \times (\exponential{I}{h_{e'}})) \\
			&= \mu \comp T \eval \comp \strength^T \comp ((\exponential{I}{h_{e'}}) \times T I) \comp \braiding \\
			&= \mu \comp T \eval \comp T ((\exponential{I}{h_{e'}}) \times I) \comp \strength^T \comp \braiding \\
			&= \mu \comp T h_{e'} \comp T \eval \comp \strength^T \comp \braiding \\
			&= h_{e'}^{\sharp} \comp T \eval \comp \strength^T \comp \braiding
		\end{align}
		Now, we have $\beta : \dot{T}^{1}_e X \dotTimes (\dotExponential{X}{S e'}) \dotTo S (e \GradeMult e')$ by considering the following composite.
		\begin{equation}
			\begin{tikzcd}
				\dot{T}^{1}_e X \dotTimes (\dotExponential{X}{S e'}) \ar[r] & (h_{e'}^{\sharp})^{*} S e \ar[r, "\le"] & \mu^{*} S(e \GradeMult e') \ar[r] & S(e \GradeMult e') \\
				T I \times (\exponential{I}{T b}) \ar[r, "T \eval \comp \strength^T \comp \braiding"] & T^2 b \ar[r, equal] & T^2 b \ar[r, "\mu"] & T b
			\end{tikzcd}
		\end{equation}
	\end{proof}
	\begin{lemma}\label{lem:probability-over-2-unit-interval}
		The mapping $\zeta \mapsto \zeta(\chi_{\{ 0 \}})$ defines an isomorphism $\ProbMonad 2^{\op} \cong [0, 1]$ in $\omegaQBS$ where $[0, 1] \subseteq \mathbb{W}$.
	\end{lemma}
	\begin{proof}
		Any $f : 2^{\op} \to \mathbb{W}$ can be rewritten as
		\[ f \quad=\quad f(1) + (f(0) - f(1)) \chi_{\{ 0 \}} \]
		and thus, we have
		\[ \zeta(f) \quad=\quad f(1) + (f(0) - f(1)) \zeta(\chi_{\{ 0 \}}). \]
	\end{proof}
	
	\begin{lemma}\label{lem:shift-function}
		For each $\beta \in [0, \infty)$, the function $f_{\beta} : [0, 1] \to [0, 1]$ defined by $f_{\beta}(x) \coloneqq \max \{ x - \beta, 0 \}$ is a morphism in $\omegaQBS$.
		Moreover, we have $x \le \beta$ if and only if $f_{\beta}(x) \le 0$.
	\end{lemma}
	
	\begin{lemma}
		Let $X$ be an $\omega$qbs.
		There is a bijective correspondence between
		\begin{itemize}
			\item Borel-Scott closed subsets of $X$ and
			\item characteristic functions $X \to 2^{\op}$.
		\end{itemize}
	\end{lemma}
	
	\begin{example}[graded $\top\top$-lifting for union bound]
		Let $h_{\beta} : D 2^{\op} \to D 2^{\op}$ be such that
		\[ h_{\beta}(\zeta)(\chi_{\{ 0 \}}) = \max \{ \zeta(\chi_{\{ 0 \}}) - \beta, 0 \}. \]
		By Lemma~\ref{lem:probability-over-2-unit-interval} and the measurability of $\lambda x. \max \{ x - \beta, 0 \} : [0, 1] \to [0, 1]$, $h_{\beta}$ is a morphism in $\omegaQBS$.
		Note that for any $x \in [0, 1]$ and $\beta \in [0, \infty)$, we have $\max \{ x - \beta, 0 \} \in [0, 1]$ and
		\[ x \le \beta \quad\iff\quad \max \{ x - \beta, 0 \} \le 0. \]
		It follows that if $\zeta \in S \beta$, then $h_{\beta}(\zeta) \in S 0$. Thus, we have $h_{\beta} : S \beta \dotTo S 0$.
		Now we prove $(h_{\beta'}^{\sharp})^{*} S \beta \subseteq \mu^{*} S (\beta + \beta')$.
		Let $\zeta \in (h_{\beta'}^{\sharp})^{*} S \beta$.
		Then, we have
		\begin{align}
			h_{\beta'}^{\sharp}(\zeta)(\chi_{\{ 0 \}}) &= D h_{\beta'}(\zeta)(\lambda \nu. \nu(\chi_{\{ 0 \}})) \\
			&= \zeta(\lambda \nu. (h_{\beta'}(\nu))(\chi_{\{ 0 \}})) \\
			&= \zeta(\lambda \nu. \max \{ \nu(\chi_{\{ 0 \}}) - \beta', 0 \}) \\
			&\le \beta.
		\end{align}
		By the monotonicity and the linearity of $\zeta$, we have
		\begin{align}
			\mu(\zeta)(\chi_{\{ 0 \}}) &= \zeta(\lambda \nu. \nu(\chi_{\{ 0 \}})) \\
			&\le \zeta(\lambda \nu. \max \{ \nu(\chi_{\{ 0 \}}) - \beta', 0 \} + \beta') \\
			&= \zeta(\lambda \nu. \max \{ \nu(\chi_{\{ 0 \}}) - \beta', 0 \}) + \beta' \\
			&= \beta + \beta'.
		\end{align}
		Therefore, $\zeta \in \mu^{*} S (\beta + \beta')$ holds.
	\end{example}

	We combine the graded monad lifting $\UnionBoundGradedMonad$ with the monad lifting for partial correctness, which is defined in the same way as~\eqref{eq:partial-correctness-monad}.
\begin{align}
	\UnionBoundGradedMonad_{\beta} \PartialCorrectnessMonad (X, P) &= (D X_{\bot}, \{ \zeta \in D X_{\bot} \mid \forall B \in \BorelClosed{X_{\bot}}, P \cup \{ \bot \} \subseteq B \implies \mathbb{P}_{x \leftarrow \zeta}[x \in X_{\bot} \setminus B] \le \beta \}) \\
	&= (D X_{\bot}, \{ \zeta \in D X_{\bot} \mid \forall B \in \BorelClosed{X}, P \subseteq B \implies \mathbb{P}_{x \leftarrow \zeta}[x \in X \setminus B] \le \beta \}) \label{eq:union-bound-lifting-graded-monad-lifting}
\end{align}

\begin{lemma}
	The composite~\eqref{eq:union-bound-lifting-graded-monad-lifting} is a strong $[0, \infty)$-graded monad lifting of $D ({-})_{\bot}$ along $\mathrm{pred}_{\omegaQBS} : \mathbf{Pred}(\omegaQBS) \to \omegaQBS$, and it is admissible.
\end{lemma}
\begin{proof}
	There exists a distributive law by Lemma~\ref{lem:partial-correctness-distributive-law}.
	The admissibility follows from the fact that $\mathbb{P}_{x \leftarrow \zeta}[x \in X \setminus B]$ is Scott-continuous in $\zeta$.
\end{proof}

\end{toappendix}

\subsubsection{Example Program}
We consider a program that computes a noisy cumulative distribution function (CDF) \cite{Lobo-VesgaTOPLAS2021}.
The program in Fig.~\ref{fig:noisy-cdf} computes, for each threshold in the list $\mathtt{buckets}$, the number of elements in $\mathtt{db}$ that are below the threshold, and adds Laplacian noise to the result.
In the program, we elide obvious type conversions such as $\NatType \to \RealType$ to simplify the presentation.
We assume that the $\mathtt{filter}$ function is already defined and has the following type for any real number $\theta$.
\begin{align}
	\mathtt{filter} :\ &((n : \RealType) \to T_0 \{ b {:} \mathtt{bool} \mid b = (n \le \theta) \})\ \times\ (l : \mathtt{list}) \\
	&\quad\to T_0\ \{ l' {:} \mathtt{list} \mid \mathrm{cdf}(\theta, l) = \mathtt{len}\ l' \}
\end{align}
Here, $\mathrm{cdf}(\theta, l)$ is the real cumulative distribution function of the list $l$ at the threshold $\theta$.
Then, we can derive the following type for $\mathtt{noisy\_cdf}$ for any $b > 0$.
\begin{align}
	&\mathtt{noisy\_cdf} : (e : \{ e : \RealType \mid e > 0 \})\ \times\ (\mathtt{bkts} : \mathtt{list}) \ \times\ (\mathtt{db} : \mathtt{list}) \\
	&\to T_{(\mathtt{len}\ \mathtt{bkts}) \times \exp(-e b)}\ \{ \texttt{ncdf} : \texttt{list} \mid \max_i |\texttt{ncdf}[i] - \mathrm{cdf}(\texttt{bkts}[i], \texttt{db})| \le b \}
\end{align}

\begin{figure}[tb]
	\scriptsize
	\begin{verbatim}
(* noisy_cdf : (e : { e : real | e > 0 }) -> (buckets : list) -> (db : list) ->
  T (len buckets * exp(-e * b)) { ncdf : list | len ncdf = len buckets /\ 
            forall i. |ncdf[i] - cdf(buckets[i], db)| <= b }    where b > 0 *)
let rec noisy_cdf e buckets db =
  if len buckets = 0 then return []
  else
    let l = filter (fun n => n <= (head buckets)) db in
    let hd = Lap(len l, e) in                  (* bound: exp(-e * b) *)
    let tl = noisy_cdf e (tail buckets) db in  (* bound: (len buckets - 1) * exp(-e * b) *)
    return hd :: tl
\end{verbatim}
\caption{A program that computes a noisy cumulative distribution function. The function $\mathtt{noisy\_cdf}$ computes the CDF of the input list $\mathtt{db}$ at each threshold in a list $\mathtt{buckets}$, and adds Laplacian noise to the result.}
\label{fig:noisy-cdf}
\end{figure}

\begin{toappendix}
\paragraph{Analysing $\mathtt{noisy\_cdf}$}
The following code shows the effect of $\mathtt{noisy\_cdf}$ and the refinement type of each variable.

\small
\begin{verbatim}
(* noisy_cdf : (e : { e : real | e > 0 }) -> (buckets : list) -> (db : list) ->
  T (len buckets * exp(-e * b)) { ncdf : list | len ncdf = len buckets /\ 
                        forall i. |ncdf[i] - cdf(buckets[i], db)| <= b }
    where b > 0 *)
let rec noisy_cdf e buckets db =
  if len buckets = 0 then
    return []  (* T 0 { ncdf : list | len ncdf = 0 } *)
  else
    (* len buckets > 0 *)
    let l = filter (fun n => n <= (head buckets)) db in
    (* l : {l : list | cdf(buckets[0], db) = len l} *)

    let cdf_head = Lap(len l, e) in
    (* effect: exp(-e * b) *)
    (* cdf_head : {r : real | |r - cdf(buckets[0], db)| <= b } *)

    let cdf_tail = noisy_cdf e (tail buckets) db in
    (* effect: (len buckets - 1) * exp(-e * b) *)
    (* cdf_tail : { cdf_tail : list | len cdf_tail = len buckets - 1 /\ 
                        forall i. |cdf_tail[i] - cdf(buckets[i + 1], db)| <= b } *)

    return cdf_head :: cdf_tail
\end{verbatim}
\normalsize
\end{toappendix}

}{}

\section{Conclusions and Future Work}
We have presented a semantic framework for dependent effect systems based on indexed preordered monoids and indexed graded monads, which are generalizations of preordered monoids and graded monads, respectively.
We have shown how to construct indexed graded monads from graded monad liftings for simple effect systems and provided several concrete instances.
As future work, we would like to apply our semantic framework to relational refinement types for, e.g., differential privacy verification~\cite{BarthePOPL2015} and implement such an effect system.
Improving the $\mathtt{let}$ rule is another important future work.

\begin{credits}
\subsubsection{\ackname}
We would like to thank the anonymous reviewers for their helpful comments and suggestions.
This study was supported by JST K Program Grant Number JPMJKP24U2; JST CREST Grant Number JPMJCR21M3; JSPS KAKENHI Grant Number JP25H00446, JP25K21183, JP24H00699, JP23K24820, JP20H05703, JP20H05703.
Marco Gaboardi’s work was partially supported by the National Science Foundation under Grant
No. 2314324.

\end{credits}

\ifthenelse{\boolean{longversion}}{}{\checkpagelimit{26}}

\bibliographystyle{splncs04}
\bibliography{dependent-effect}

\end{document}